\documentclass[12pt,fleqn,a4paper]{article} 
\usepackage{clshan-font}
\usepackage{epsfig,graphics,graphicx}
\usepackage{clshan-math,clshan-dm-dd}
\newcounter{amidascode}
\usecounter{amidascode}

\def \gsim {\:\raisebox{-0.7ex}{$\stackrel{\textstyle>}{\sim}$}\:}
\renewcommand{\_}        {\ttsym{95}}
\newcommand{\beenr}      {\begin{enumerate}\setlength{\itemsep}{0cm}}
\newcommand{\eeenr}      {\end{enumerate}}
\newcommand{\tabt}  [1]  {\begin{flushleft}
                          \footnotesize
                          {\bf #1} \\
                          \vspace{0.2cm}
                          \renewcommand{\arraystretch}{1.1}
                          \begin{tabular}{|p{16 cm}|}
                          \hline}
\newcommand{\tabc}       {\hline
                          \end{tabular}
                          \vspace{0.1cm}
                          \renewcommand{\arraystretch}{1.15}
                          \begin{tabular}{|p{16 cm}|}
                          \hline}
\newcommand{\tabb}       {\hline
                          \end{tabular}
                          \vspace{0.2cm}
                          \end{flushleft}}
\newcommand{\code}  [1]  {\refstepcounter{amidascode}
                          \tabt{Code \arabic{amidascode}: #1}}
\newcommand{\includefigure}   [1]  {\begin{center}
                                    \includegraphics[width = 17    cm]{#1}
                                    \end{center}}
\begin{document}
\thispagestyle{empty}
\begin{flushright}
 March 2014
\end{flushright}
\begin{center}
{\large\bf
 AMIDAS-II:                                                    
 Upgrade of the AMIDAS Package and Website                       \\ \vspace{0.2 cm}
 for Direct Dark Matter Detection Experiments and Phenomenology} \\
\vspace*{0.7cm}
 {\sc Chung-Lin Shan} \\
\vspace*{0.5cm}
 {\it Physics Division,
      National Center for Theoretical Sciences           \\
      No.~101, Sec.~2, Kuang-Fu Road,
      Hsinchu City 30013, Taiwan, R.O.C.}                \\~\\
 {\it Department of Physics,
      Hangzhou Normal University                         \\
      No.~16, Xuelin Street, Xiasha Higher Education Zone,
      Hangzhou 310036, Zhejiang, China}                  \\~\\
 {\it Kavli Institute for Theoretical Physics China,
      Chinese Academy of Sciences                        \\
      No.~55, Zhong Guan Cun East Street,
      Beijing 100190, China}                             \\~\\
\vspace{0.05cm}
 {\it E-mail:} {\tt clshan@phys.nthu.edu.tw}             \\
\end{center}
\vspace{1cm}
\begin{abstract}
 In this paper,
 we give a detailed user's guide
 to the \amidas\ (A Model--Independent Data Analysis System)
 package and website,
 which is developed for
 online simulations and data analyses
 for direct Dark Matter detection experiments and phenomenology.
 Recently,
 the whole \amidas\ package and website system
 has been upgraded to the second phase:
 \amidasii,
 for including the new developed Bayesian analysis technique.

 \amidas\ has the ability to do
 full Monte Carlo simulations
 as well as
 to analyze real/pseudo data sets
 either generated by another event generating programs
 or
 recorded in direct DM detection experiments.
 Moreover,
 the \amidasii\ package
 can include several ``user--defined'' functions into the main code:
 the (fitting) one--dimensional WIMP velocity distribution function,
 the nuclear form factors for
 spin--independent and spin--dependent cross sections,
 artificial/experimental background spectrum
 for both of simulation and data analysis procedures,
 as well as
 different distribution functions
 needed in Bayesian analyses.
\end{abstract}
\clearpage
\includefigure{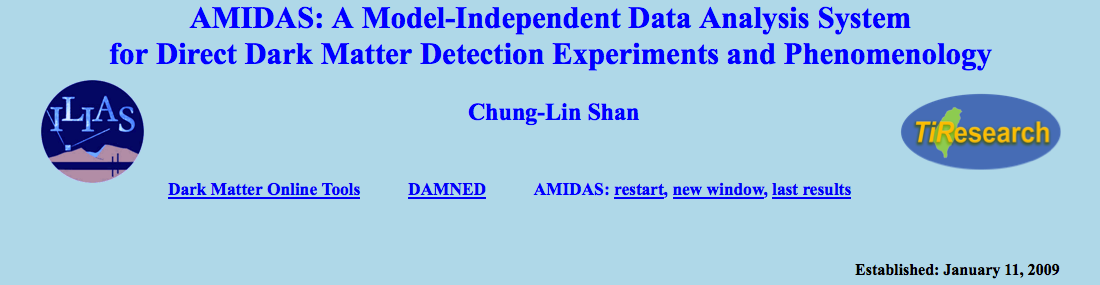}
\section{Introduction}

 Weakly Interacting Massive Particles (WIMPs) $\chi$
 arising in several extensions of
 the Standard Model of electroweak interactions
 are one of the leading candidates for Dark Matter (DM).
 Currently,
 direct DM detection experiments
 based on measuring recoil energy
 deposited in a low--background underground detector
 by elastic scattering of ambient WIMPs off target nuclei
 are one of the most promising methods
 for understanding
 DM properties,
 identifying them among new particles
 produced (hopefully in the near future) at colliders,
 as well as studying the (sub)structure of our Galactic halo
 (for reviews,
  see Refs.~\cite{SUSYDM96, Bertone05, Bergstrom12}).

 Since 2007
 we develop a series of new methods
 for analyzing data,
 i.e.~measured recoil energies,
 from 
 direct detection experiments
 as model--independently as possible.
 Up to now
 we could in principle
 reconstruct the (moments of the) one--dimensional
 velocity distribution function of halo WIMPs
 \cite{DMDDf1v, DMDDf1v-Bayesian},
 as well as
 determine the WIMP mass
 \cite{DMDDmchi}
 and (ratios between) different WIMP couplings/cross sections on nucleons
 \cite{DMDDfp2, DMDDranap}.

 Following the development of
 these model--independent data analysis procedures,
 we combined the code for our simulations to a compact system:
 \amidas\ (A Model--Independent Data Analysis System).
 Meanwhile,
 we modified the code
 to be able to analyze external data sets
 either generated by another event generating programs
 or
 recorded in direct DM detection experiments.
 Under the collaboration with the ILIAS Project
 \cite{ILIAS}
 and the Dark Matter Network Exclusion Diagram
 (DAMNED, a.k.a.~Dark Matter Online Tools)
 \cite{DAMNED},
 an online system
 has also been established in January 2009
 \cite{AMIDAS-web, AMIDAS-web-TiResearch},
 in order to offer
 an easier, more convenient and user--friendly environment
 for simulations and data analyses
 for (in)direct DM detection experiments
 and phenomenology.

 In the first phase of the \amidas\ package and website,
 the options for target nuclei,
 for the velocity distribution function of halo WIMPs,
 as well as
 for the elastic nuclear form factors
 for spin--independent (SI) and spin--dependent (SD)
 WIMP--nucleus interactions
 are fixed and
 only some commonly used analytic forms
 have been defined
 \cite{AMIDAS-SUSY09}.
 Users can not choose different detector materials
 nor use different WIMP velocity distribution function
 nor nuclear form factors
 for their simulations and/or data analyses.
 In order to offer more flexible use
 of the WIMP velocity distribution
 as well as
 the nuclear form factors,
 the \amidas\ package has been extended
 to be more user--oriented
 and able to include {\em user--uploaded} files
 for defining their own functional forms
 in September 2009
 \cite{AMIDAS-f1vFQ}%
\footnote{
 Note that,
 since the \amidas\ code has been written
 in the C programming language,
 all user--defined functions
 to be included into the \amidas\ package
 must be given in the syntax of C.
 On the other hand,
 for drawing output plots,
 since the {\tt Gnuplot} package has been used in \amidas,
 the uploaded files for drawing
 e.g.~the predicted spectrum of measured recoil energy,
 must be given in the syntax of {\tt Gnuplot}
 \cite{gnuplot}.
 More detailed descriptions
 will be given in Secs.~3 and 5
 as well as
 in Appendix.
}.

 Besides of the purposed model--independent reconstructions of
 different WIMP properties
 \cite{DMDDf1v, DMDDf1v-Bayesian,
       DMDDmchi,
       DMDDfp2, DMDDranap}%
\footnote{
 Note that
 our model--independent methods
 are developed for reconstructing WIMP properties
 with {\em positive} signals
 (and probably small fractions of unrejected background events
  in the analyzed data sets
  \cite{DMDDbg-mchi, DMDDbg-f1v, DMDDbg-fp2, DMDDbg-ranap}).
 With however {\em negative} results or
 before enough (${\cal O}(50)$ to ${\cal O}(500)$) WIMP signals
 could be accumulated,
 other well--built model--independent methods
 developed in Refs.~%
 \cite{Fox10a, Fox10b, Fox14,
       McDermott11,
       DelNobile13a, DelNobile13b, Cirelli13, NRopsDD,
       DelNobile14a,
       Feldstein14, Cherry14}
 would be useful for
 e.g.~giving exclusion limits from
 or comparing sensitivities of
 different detectors/experiments.
},
 one minor contribution of the \amidas\ package
 to direct DM detection experiments
 would be to be helpful
 in future detector design and material search.
 Firstly,
 by running Monte Carlo simulations with \amidas\
 one can estimate the required
 experimental exposures
 for reconstructing WIMP properties
 with acceptable statistical uncertainties
 and/or systematic biases.
 Secondly and more importantly,
 it has been discussed that,
 for determining the WIMP mass
 one frequently used heavy target nucleus
 needs to be combined with a light one
 \cite{DMDDmchi};
 for determining ratios
 between different WIMP--nucleon cross sections
 not only the conventionally used targets
 but also those with non--zero
 expectation values of both of the proton and neutron group spins
 are required
 \cite{DMDDranap}.
 In addition,
 different combinations of chosen detector materials
 with different nuclear masses and/or
 total nuclear spins and
 (ratios between)
 the nucleon group spins
 not only can affect the statistical uncertainties
 \cite{DMDDmchi, DMDDranap}
 but also are available
 in different range of interest
 of the reconstructed properties
 \cite{DMDDranap}.

 In this paper,
 we give a detailed user's guide
 to the \amidas\
 package and website.
 In Sec.~2,
 we list the \amidas\ functions and
 different
 simulation and data analysis
 modes
 for these functions.
 The use of intrinsically saved element data
 to set information of users' favorite target nuclei
 will be described in detail.
 In Sec.~3,
 we describe the meanings and options of
 all input parameters/factors
 for running Monte Carlo simulations.
 The preparation of uploaded files
 for defining
 the one--dimensional WIMP velocity distribution function,
 the elastic nuclear form factors
 for SI and SD WIMP--nucleus cross sections
 as well as
 artificial/experimental background spectrum
 will be particularly described.
 In Sec.~4,
 the preparation of the real/pseudo data files
 and the uploading/analyzing procedure
 on the \amidas\ website
 will be given.
 The new developed Bayesian analysis technique
 \cite{DMDDf1v-Bayesian}
 will be talked separately
 in Sec.~5.
 We conclude
 in Sec.~6.
 Some technical detail
 in the \amidas\ package
 will be given in Appendix.

\section{\amidas\ functions, working modes, and target nuclei}

 In this section,
 we list the functions of the \amidas\ package and website
 and different
 simulation and data analysis
 modes
 for these functions.
 The default options for the target nuclei
 used for different \amidas\ functions
 as well as
 the use of intrinsically saved element data
 to set information of users' favorite target nuclei
 will also be described.

\subsection{\amidas\ functions}

 Based on our works
 on the model--independent data analysis methods
 for extracting properties of Galactic WIMP DM particles
 \cite{DMDDf1v, DMDDf1v-Bayesian, DMDDmchi, DMDDfp2, DMDDranap},
 \amidas\ has so far the following functions:
\includefigure{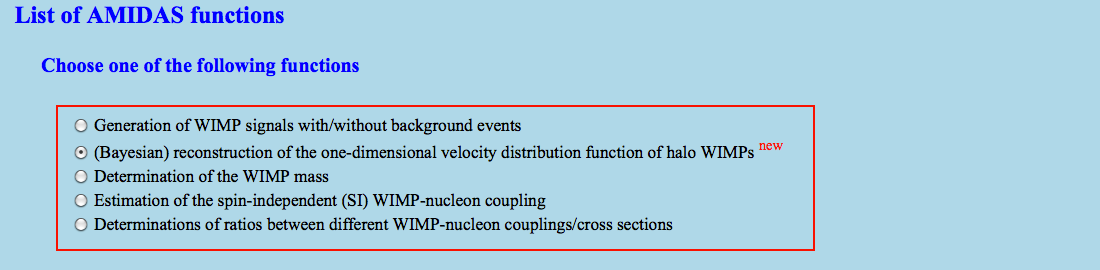}
\subsection{Reconstruction modes}

 Corresponding to each of the above listed \amidas\ functions,
 there are several different reconstruction modes
 for users to choose.

\subsubsection{For the WIMP--nucleus interaction}

 For generating WIMP signals with/without background events,
 one needs to choose
 either the SI scalar WIMP--nucleus interaction
 or the SD axial--vector one
 is dominated:
\includefigure{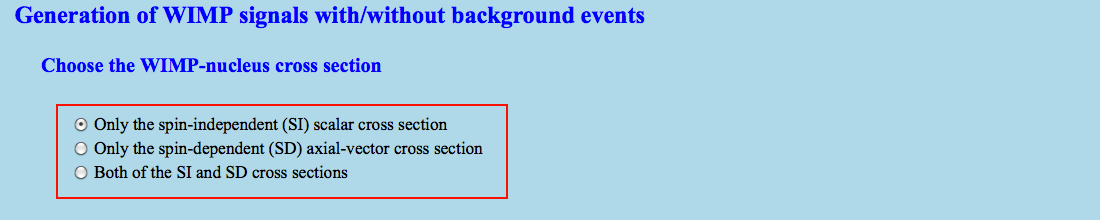}
\subsubsection{For the needed WIMP mass}

 For (Bayesian) reconstructing
 the one--dimensional WIMP velocity distribution function
 $f_1(v)$
 and estimating the SI WIMP--nucleon coupling $|\frmp|^2$,
 one needs the WIMP mass $\mchi$
 as an input parameter
 \cite{DMDDf1v, DMDDf1v-Bayesian, DMDDfp2}.
 This information could be obtained
 either from e.g.~collider experiments
 or from two (other) direct DM detection experiments
 \cite{DMDDmchi}.
 For these two cases,
 \amidas\ has three options of the input WIMP mass:
\includefigure{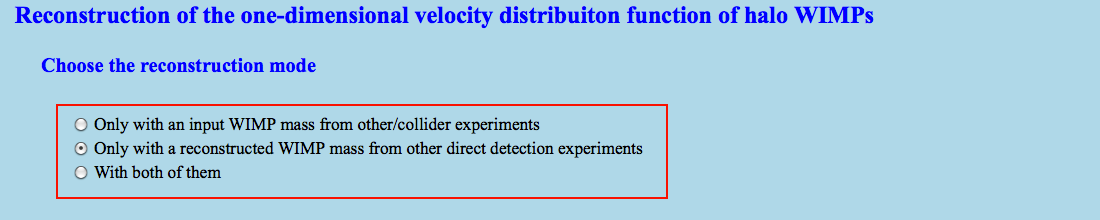}
\subsubsection{For the reconstruction of the WIMP mass}

 In addition,
 \amidas\ offers two modes
 for reconstructing the required WIMP mass $\mchi$
 \cite{DMDDmchi}:
\includefigure{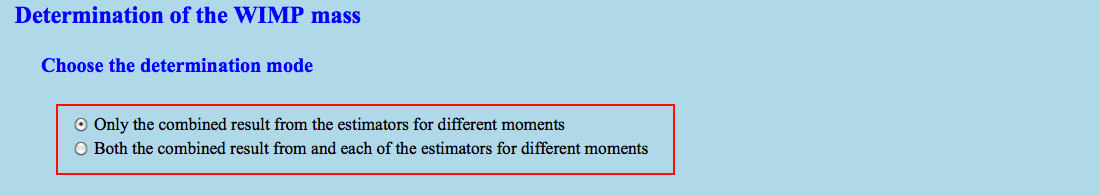}
\subsubsection{For the reconstruction of the ratio
               between two SD WIMP--nucleon couplings}

 Based on the assumption about the dominated WIMP--nucleus interaction,
 we obtained two expressions
 for reconstructing the ratio
 between two SD WIMP--nucleon couplings $\armn / \armp$
 \cite{DMDDranap}:
\includefigure{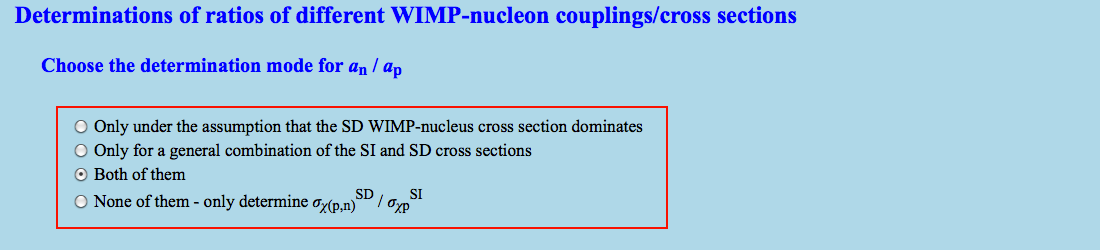}
\subsubsection{For the reconstructions of the ratios
               between SD and SI WIMP--nucleon cross sections}

 Similarly,
 by using different detector materials
 with different spin sensitivities with
 protons or neutrons,
 we have also two ways
 for reconstructing the ratios
 between the SD and SI WIMP--nucleon cross sections
 $\sigma_{\chi {\rm (p, n)}}^{\rm SD} / \sigmapSI$
 \cite{DMDDranap}:
\includefigure{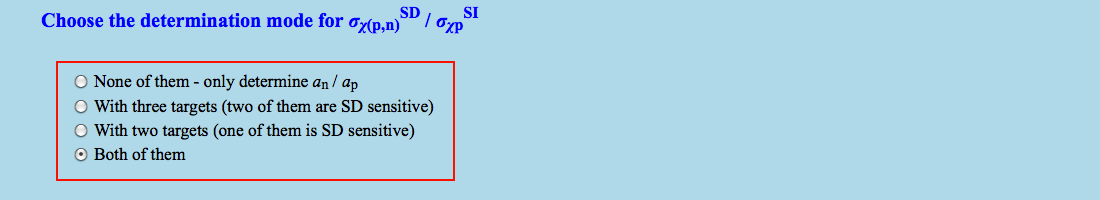}
\subsection{Target nuclei}

 Four frequently used detector materials:
 $\rmXA{Si}{ 28}$,
 $\rmXA{Ge}{ 76}$,
 $\rmXA{Ar}{ 40}$ and
 $\rmXA{Xe}{136}$
 are given as default options of target nuclei
 for simulations and data analyses
 by using \amidas\ package.
 Meanwhile,
 since September 2009
 it is achieved to let users set the target nuclei freely,
 with corresponding atomic number $Z$ and atomic mass number $A$
 as well as
 the total nuclear spin $J$ and
 the expectation values of the proton and neutron group spins
 $\expv{S_{\rm p, n}}$
 for each target nucleus.

\subsubsection{Only one required target nucleus}

 For generating WIMP signals
 as well as
 (Bayesian) reconstructing the one--dimensional
 WIMP velocity distribution function
 and estimating the SI WIMP--nucleon coupling
 with an {\em input} WIMP mass (from other/collider experiments),
 only one target nucleus is required and
 users can choose it from the four default options
 or type the element symbol of the target nucleus,
 and give the corresponding atomic information:
 $Z$, $A$, $J$, $\expv{S_{\rm p, n}}$
 by hand on the website directly.
 When considering only the SI scalar WIMP--nucleus interaction,
 one has%
\footnote{
 A similar panel will be needed
 for the third required target nucleus
 for reconstructing the ratio between
 two SD WIMP--nucleon couplings
 under the consideration of
 a general combination of the SI and SD WIMP--nucleus cross sections:
\includefigure{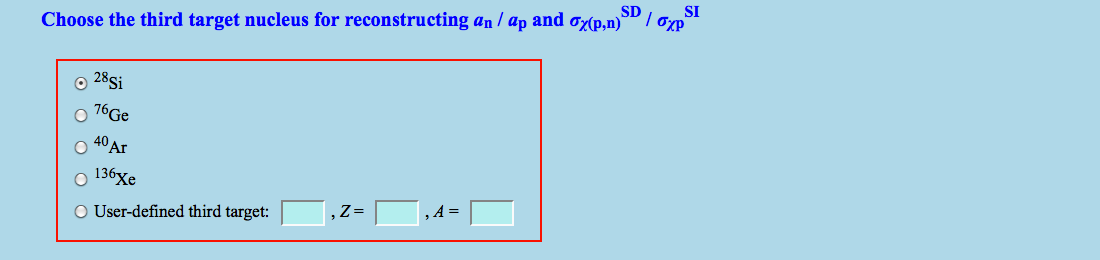}
}
\includefigure{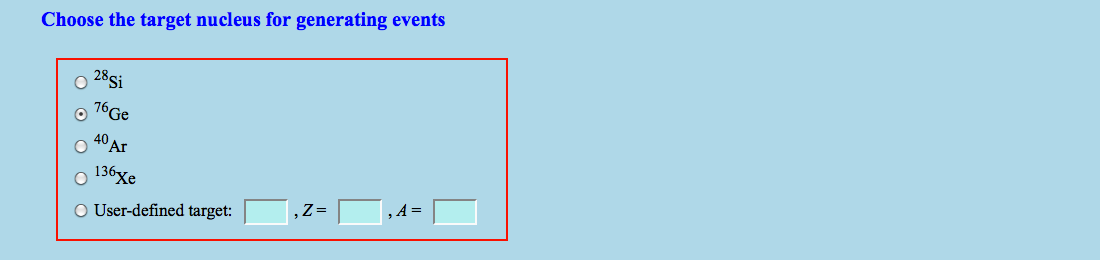}
 Or,
 once the SD interaction is taken into account,
 one will get
\includefigure{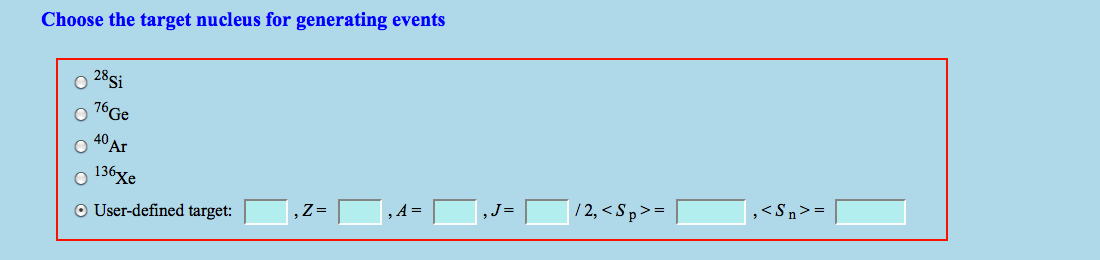}
 Note that,
 firstly,
 only for the \amidas\ function of generating WIMP signals
 with either only the SD axial-vector cross section
 or both of the SI and SD cross sections,
 the typing cells for
 the total nuclear spin $J$ and
 the expectation values of the proton and neutron group spins
 $\expv{S_{\rm p, n}}$
 for each target nucleus
 will appear,
 in order to remind users that
 these data are required.
 Secondly,
 the given value for the total nuclear spin $J$
 is only the {\em numerator} of the actually value.
 This means that,
 for instance,
 for the use of $\rmXA{I}{127}$
 one needs only give ``5'' in the typing cell between ``$J =$'' and ``$/ 2$''
 for assigning its $J$ value of $5 / 2$
 (see Table \ref{tab:default_nuclei}).

\begin{table}[t!]
\small
\begin{center}
\renewcommand{\arraystretch}{1.3}
\begin{tabular}{|| c   c   c   c   c   c   c ||}
\hline
\hline
 \makebox[1.8 cm][c]{Isotope} &
 \makebox[1.5 cm][c]{$Z$}     & \makebox[1.5 cm][c]{$A$}     & \makebox[1.5 cm][c]{$J$} &
 \makebox[2   cm][c]{$\Srmp$} & \makebox[2   cm][c]{$\Srmn$} & \\
\hline
\hline
 $\rmXA{Li}{  7}$ &  3 &   7 & 3/2 &                   0.497  &                   0.004 & \\
\hline
 $\rmXA{O} { 17}$ &  8 &  17 & 5/2 &                   0.0    &                   0.495 & \\
\hline
 $\rmXA{F} { 19}$ &  9 &  19 & 1/2 &                   0.441  & \hspace{-1.8ex}$-$0.109 & \\
\hline
 $\rmXA{Na}{ 23}$ & 11 &  23 & 3/2 &                   0.248  &                   0.020 & \\
\hline
 $\rmXA{Al}{ 27}$ & 13 &  27 & 5/2 &                   0.343  &                   0.030 & \\
\hline
 $\rmXA{Si}{ 29}$ & 14 &  29 & 1/2 & \hspace{-1.8ex}$-$0.002  &                   0.130 & \\
\hline
 $\rmXA{Cl}{ 35}$ & 17 &  35 & 3/2 & \hspace{-1.8ex}$-$0.059  & \hspace{-1.8ex}$-$0.011 & \\
\hline
 $\rmXA{Cl}{ 37}$ & 17 &  37 & 3/2 & \hspace{-1.8ex}$-$0.058  &                   0.050 & \\
\hline
 $\rmXA{K} { 39}$ & 19 &  39 & 3/2 & \hspace{-1.8ex}$-$0.180  &                   0.050 & \\
\hline
 $\rmXA{Ge}{ 73}$ & 32 &  73 & 9/2 &                   0.030  &                   0.378 & \\
\hline
 $\rmXA{Nb}{ 93}$ & 41 &  93 & 9/2 &                   0.460  &                   0.080 & \\
\hline
 $\rmXA{Te}{125}$ & 52 & 125 & 1/2 &                   0.001  &                   0.287 & \\
\hline
 $\rmXA{I} {127}$ & 53 & 127 & 5/2 &                   0.309  &                   0.075 & \\
\hline
 $\rmXA{Xe}{129}$ & 54 & 129 & 1/2 &                   0.028  &                   0.359 & \\
\hline
 $\rmXA{Xe}{131}$ & 54 & 131 & 3/2 & \hspace{-1.8ex}$-$0.009  & \hspace{-1.8ex}$-$0.227 & \\
\hline
 $\rmXA{Cs}{133}$ & 55 & 133 & 7/2 & \hspace{-1.8ex}$-$0.370  &                   0.003 & \\
\hline
 $\rmXA{W}{183}$  & 74 & 183 & 1/2 &                   0.0    & \hspace{-1.8ex}$-$0.031 & \\
\hline
\hline
\end{tabular}
\end{center}
\caption{
 List of the intrinsically defined nuclear spin data
 in the \amidasii\ package.
 More details can be found in
 e.g.~Refs.~\cite{SUSYDM96, Tovey00, Giuliani05, Girard05}.
}
\label{tab:default_nuclei}
\end{table}
\subsubsection{Intrinsically defined nuclear spin data}

 In the upgraded \amidasii\ package,
 we have further saved the basic atomic data for all 118 elements
 as well as
 the nuclear spin data ($J$ and $\expv{S_{\rm p, n}}$ values)
 for 17 frequently used detector nuclei
 (listed in Table \ref{tab:default_nuclei}).
 Hence,
 for setting user--required target nuclei for
 e.g.~generating WIMP--nucleus scattering events,
 users need only simply to give
 either the atomic symbol or the atomic number $Z$,
 since these two items are one--to--one corresponding.
 Then all other atomic and nuclear spin information
 about this chosen nucleus
 will be shown automatically
 in the following typing cells.

 Moreover,
 according to the users' assumption about
 the WIMP--nucleus interaction (SI, SD or both),
 the \amidas\ website will choose automatically
 the isotope of the chosen nucleus
 without or with spin sensitivity
 with the highest natural abundance.
 For example,
 when one types
 either the defined target ``Xe'' or the atomic number $Z$ ``54'',
 in the cell of the atomic mass number $A$
 ``132'' or ``129'' will appear
 under the choice of the SI or SD (or both) WIMP--nucleus interaction(s).
 For the latter case,
 once the nuclear spin data of the chosen nucleus
 is given intrinsically in the \amidasii\ package
 (listed in Table \ref{tab:default_nuclei}),
 the rest required information about the $J$ and $\expv{S_{\rm p, n}}$ values
 will also appear directly:
\includefigure{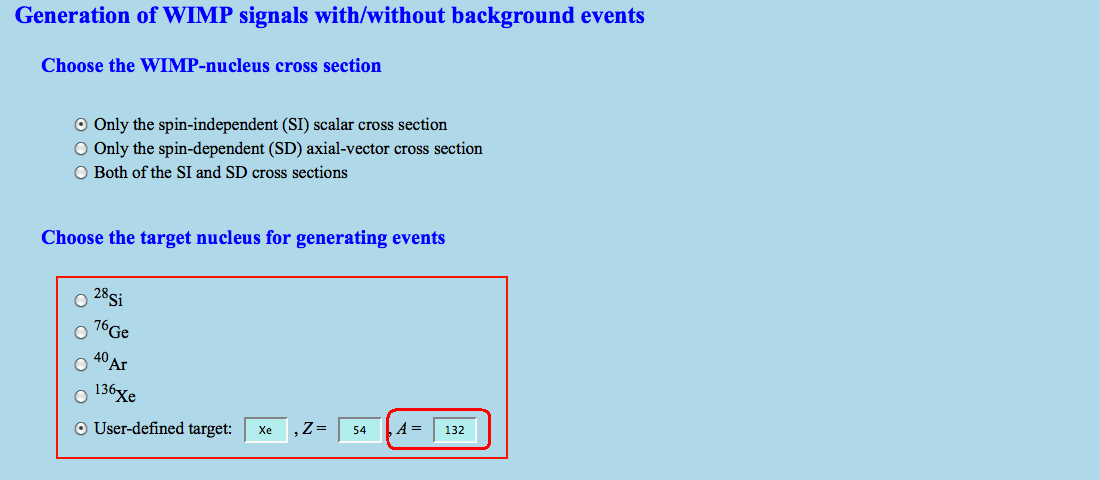}
\includefigure{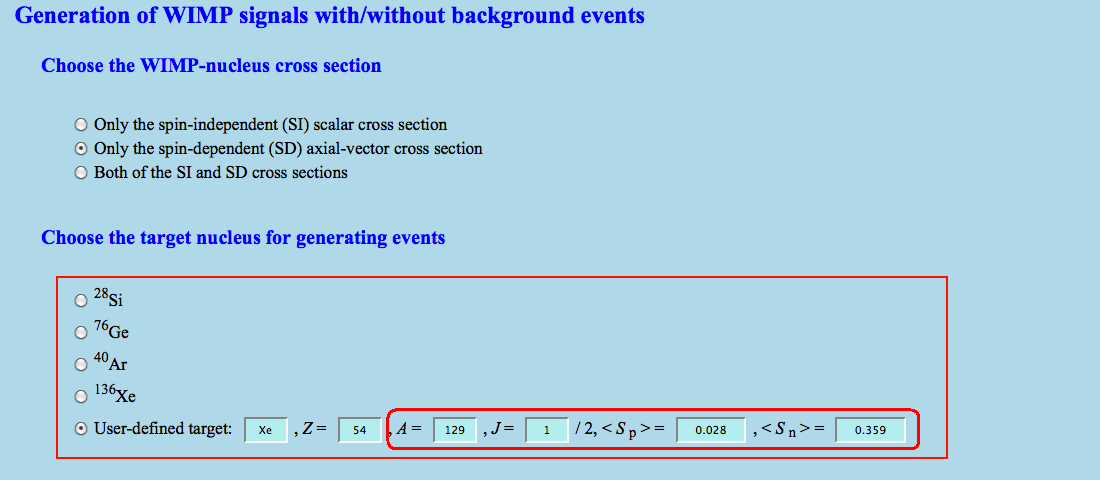}

 Note that,
 firstly,
 while the one--to--one corresponding
 atomic symbol and atomic number $Z$
 appear simultaneously,
 one could modify the atomic mass number $A$ by hand:
\includefigure{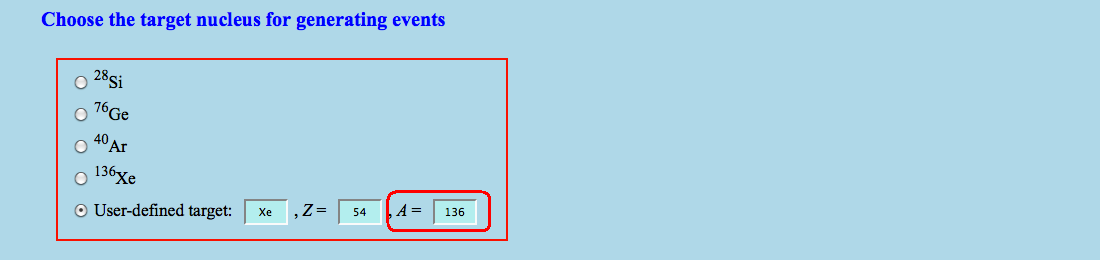}
\includefigure{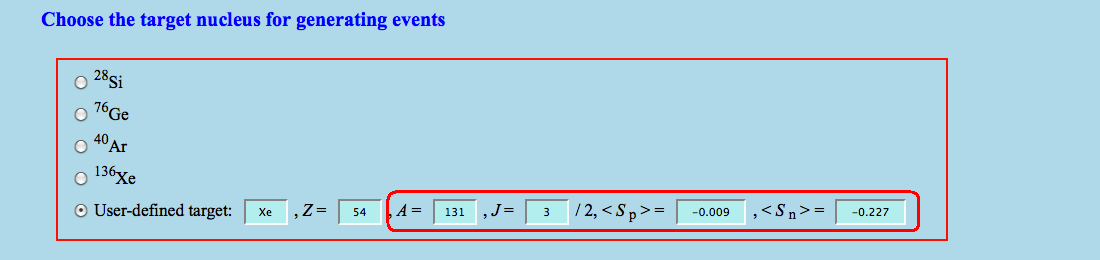}
 Secondly,
 for one target nucleus,
 the expected proton and neutron group spins are sometimes model--dependent:
 e.g.~for $\rmXA{Xe}{129}$ and $\rmXA{Xe}{131}$ nuclei,
 \mbox{$\Srmp_{\rmXA{Xe}{129}} = -0.002$},
 \mbox{$\Srmn_{\rmXA{Xe}{129}} =  0.273$} and
 \mbox{$\Srmp_{\rmXA{Xe}{131}} = -0.0007$},
 \mbox{$\Srmn_{\rmXA{Xe}{131}} = -0.125$}
 are also often used
 \cite{Ressell97, Toivanen09, Garny12, Klos13}.
 These model--dependent values
 are however not saved in the \amidas\ package yet
 and thus have to be given by hand:
\includefigure{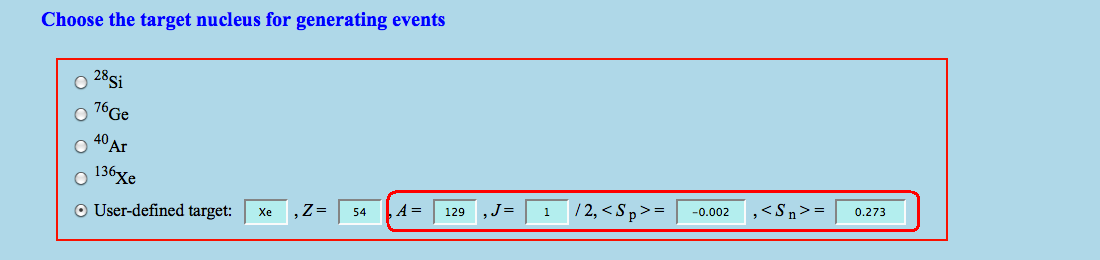}
\includefigure{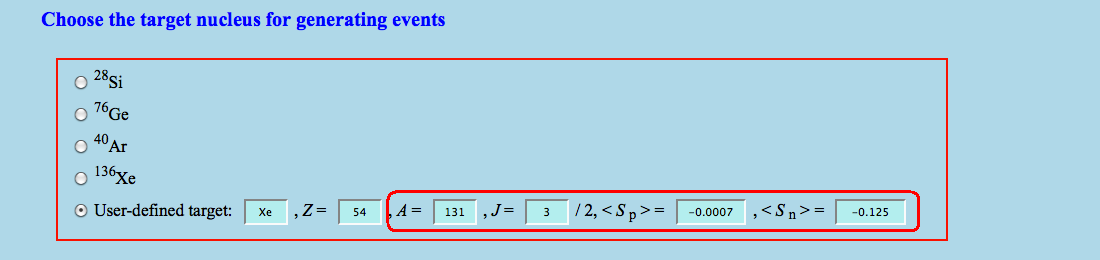}
 Moreover,
 once the $J$ and $\expv{S_{\rm p, n}}$ values of the chosen nucleus
 are {\em not} defined intrinsically in the \amidasii\ package,
 ``undefined'' would be shown in these cells:
\includefigure{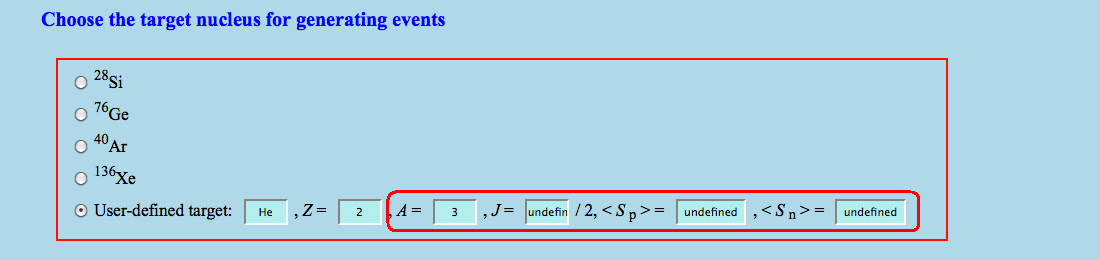}
\subsubsection{For the determination of the WIMP mass}

 For determining the WIMP mass $\mchi$,
 two combinations
 have been considered and programmed in the \amidas\ package
 \cite{DMDDmchi}:
\includefigure{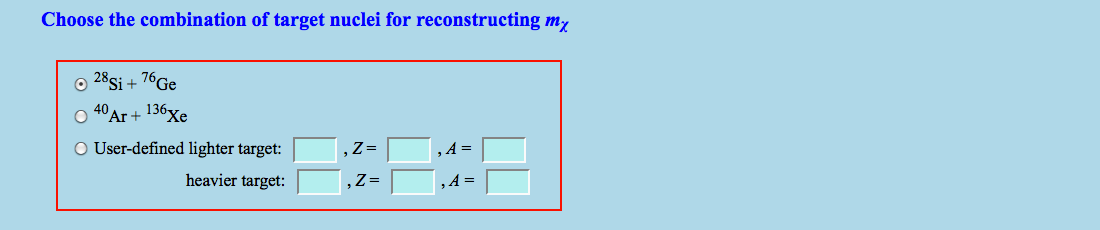}
 Note here that
 the {\em lighter} ({\em heavier}) nucleus of
 the user--chosen target combination
 should be given in the {\em first} ({\em second}) line of
 the item ``user--defined'' combination%
\footnote{
 Nevertheless,
 the \amidas\ website can rearrange the order of the chosen nuclei
 according to the given atomic mass numbers $A$.
}.
\subsubsection{For the determination of the ratio between
               two SD WIMP--nucleon couplings}

 For determining the ratio between
 two SD WIMP--nucleon couplings $\armn / \armp$,
 two combinations
 have been considered and programmed in the \amidas\ package
 \cite{DMDDranap}:
\includefigure{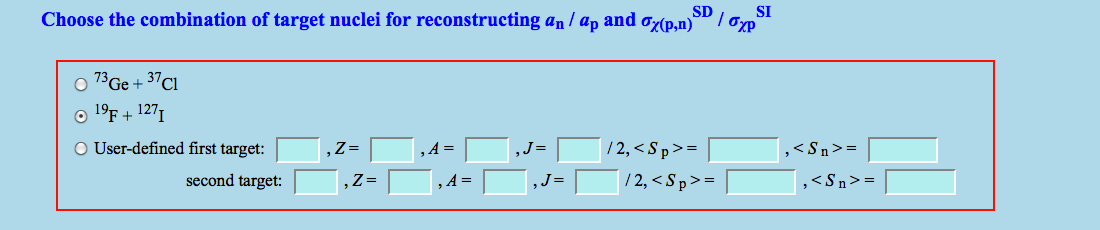}
 Remind here that
 two user--chosen target nuclei
 should be spin--sensitive
 and their total nuclear spin $J$
 and expectation proton and neutron group spins
 $\expv{S_{\rm (p, n)}}$
 are also required.
 As described in Sec.~2.3.2,
 the intrinsically defined nuclear spin data
 can be used here directly
 or modified by hand.

\subsubsection{For the determination of the ratio between
               SD and SI WIMP--proton cross section}

 For determining the ratio between
 SD and SI WIMP--proton cross section $\sigmapSD / \sigmapSI$,
 only one combination of target nuclei
 has been considered and programmed in the \amidas\ package
 \cite{DMDDranap}:
\includefigure{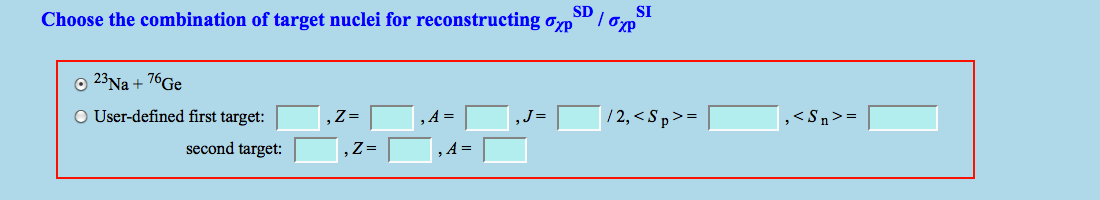}
 Note here that
 the {\em first} nucleus of
 the user--chosen target combination
 should be spin--sensitive
 with a {\em very small} or even {\em negligible}
 expectation value of the {\em neutron} group spin,
 whereas
 the {\em second} one
 should be {\em spin--nonsensitive}
 (without unpaired protons nor neutrons).

\subsubsection{For the determination of the ratio between
               SD and SI WIMP--neutron cross section}

 For determining the ratio between
 SD and SI WIMP--neutron cross section $\sigmanSD / \sigmapSI$,
 two combinations of target nuclei
 has been considered and programmed in the \amidas\ package
 \cite{DMDDranap}:
\includefigure{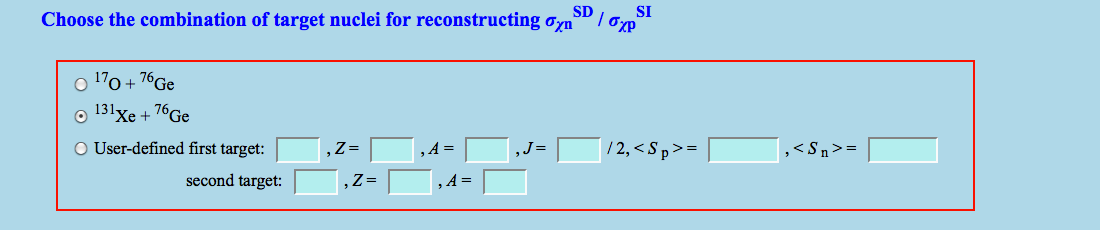}
 Note here that
 the {\em first} nucleus of
 the user--chosen target combination
 should be spin--sensitive
 with a {\em very small} or even {\em negligible}
 expectation value of the {\em proton} group spin,
 whereas
 the {\em second} one
 should be {\em spin--nonsensitive}
 (without unpaired protons nor neutrons).

\subsection{Data type}

 The probably most important and useful design
 of the \amidas\ package and website is
 the ability of
 {\em not only} doing simulations with self--generated events
 based on Monte Carlo method,
 {\em but also} analyzing user--uploaded real/pseudo data set(s)
 either generated by other event generators
 or
 recorded in direct DM detection experiments
 {\em without} modifying the source code.

\subsubsection{Data type}

 Corresponding to our design of the ability of
 doing Monte Carlo simulations
 as well as
 analyzing real/pseudo data set(s),
 users have two options for the data type:
\includefigure{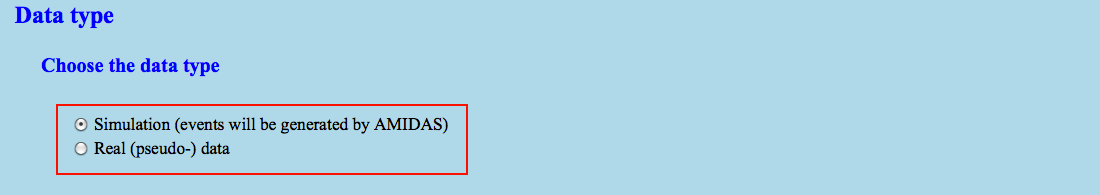}
 More detailed descriptions about
 the preparation of data files
 and the uploading/analyzing procedure
 on the \amidas\ website
 will be given in Sec.~4.

\subsubsection{Simulation mode}

 One can run numerical Monte Carlo simulations
 with all \amidas\ functions given in Sec.~2.1%
\footnote{
 Note that,
 considering the current experimental sensitivity
 and the required executing time for these simulations,
 the \amidas\ website offers full Monte Carlo simulations
 with maximal 2,000 experiments and
 maximal 5,000 events on average per one experiment.
}.
 Meanwhile,
 since the running time of the algorithmic procedure
 for the reconstruction of the WIMP mass $\mchi$,
 needed also for
 the (Bayesian) reconstruction of
 the one--dimensional WIMP velocity distribution function
 $f_1(v)$
 and the estimation of the SI WIMP--nucleon coupling $|\frmp|^2$,
 is pretty long,
 \amidas\ also offers users faster theoretical estimations
 as an alternative option:
\includefigure{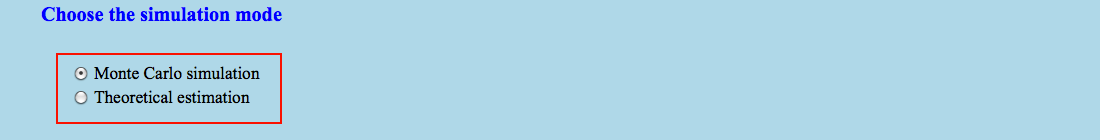}
 Here,
 instead of the summations of
 the (moments of the) measured recoil energies
 required in our model--independent data analysis procedures
 \cite{DMDDf1v, DMDDmchi, DMDDfp2, DMDDranap},
 numerical integrals over
 the theoretically predicted recoil spectrum
 will be used.
 Note however that,
 firstly,
 since for these estimations
 the statistical fluctuations
 have {\em not} been taken into account,
 these pure theoretically estimated results,
 especially for cases with only a few events,
 could be fairly {\em different} from
 results obtained by more realistic Monte Carlo simulations.
 Secondly,
 as the alternative option for Monte Carlo simulations
 with much shorter required executing time,
 the total event number used for these theoretical estimations
 is fixed%
\footnote{
 In contrast
 and more realistically,
 the actual number(s) of generated WIMP signals (and background events)
 for each Monte Carlo simulated experiment
 is Poisson--distributed around
 the expected value(s) set by users.
}
 and the calculations are limited to be done for only a few times.
 These restrictions could sometimes cause
 unexpected zigzags on the result curves.

 Note also here that,
 once residue background events
 are taken into account,
 users have to choose ``Monte Carlo simulation''!

\subsubsection{Output plots}

 In our work on
 the reconstruction of the WIMP mass, 
 we discussed also the statistical fluctuation
 of the reconstructed $\mchi$
 in the simulated experiments
 \cite{DMDDmchi}.
 Users can thus choose
 whether and for what WIMP mass
 they need the plot of the statistical fluctuation
 of the reconstructed WIMP mass:
\includefigure{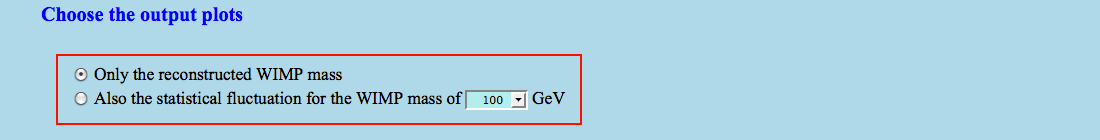}
\subsubsection{Simulation mode for
               \boldmath $\armn / \armp$ and
               $\sigma_{\chi {\rm (p, n)}}^{\rm SD} / \sigmapSI$}

 In our works on
 the determinations of ratios
 between different WIMP--nucleon couplings/cross sections,
 we considered two cases separately:
\includefigure{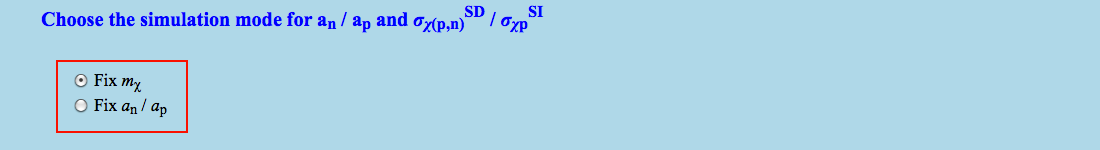}
\subsubsection{Background events}

 After the development of our model--independent methods
 for reconstructing different WIMP properties,
 we worked also on
 effects of residue background events
 in analyzed data sets
 \cite{DMDDbg-mchi, DMDDbg-f1v, DMDDbg-fp2, DMDDbg-ranap}.
 Hence,
 users have the option
 to do simulations
 with or without
 an intrinsically or user--defined
 background spectrum:
\includefigure{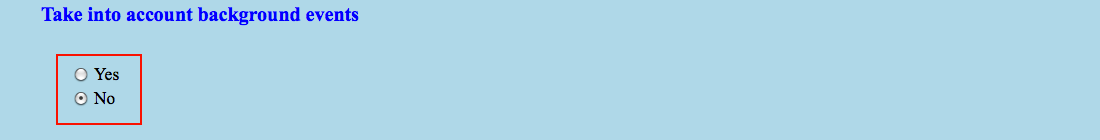}
 More detailed descriptions about
 intrinsically defined background spectra
 as well as
 the file preparation of
 users' own artificial/experimental background spectrum
 will be given in Sec.~3.6.

\subsubsection{Generated events}

 As an extra service,
 users can receive
 {\em all} WIMP--signal (and background) events
 generated by \amidas\
 in separate text files for different target nuclei:
\includefigure{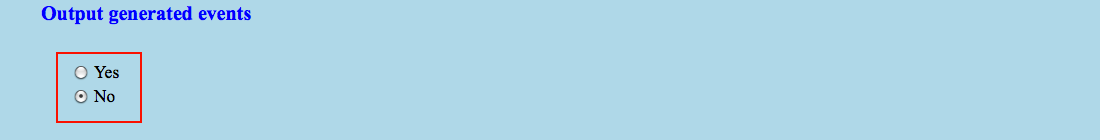}
 Note hare that,
 for the determination of the WIMP mass,
 the estimation of the SI WIMP--nucleon coupling
 as well as
 the determinations of ratios
 between different  WIMP--nucleon couplings/cross sections
 with several different input WIMP masses,
 the required value of the input WIMP mass
 has to be chosen:
\includefigure{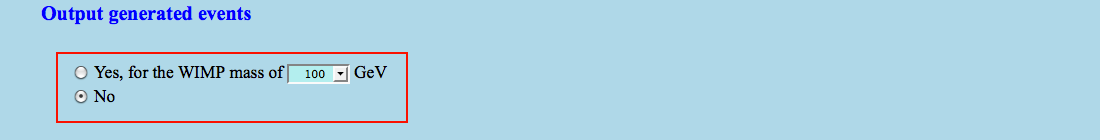}
 Similarly,
 for the determinations of ratios
 between different WIMP--nucleon couplings/cross sections
 with several different input SD WIMP--nucleon coupling ratios,
 users have to choose the required $\armn / \armp$ value:
\includefigure{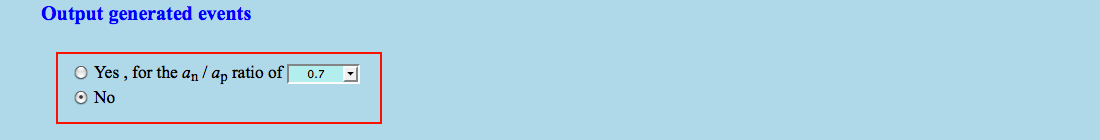}
\section{Running simulations}

 In this section,
 we describe the meanings and options of
 all input parameters/factors
 needed in principle {\em only}
 for predicting the recoil and background spectra
 for generating WIMP signals and residue background events
 and for drawing the output plots.
 Some commonly used and/or
 standard \amidas\ simulation/analysis values
 used in our works presented
 in Refs.~\cite{DMDDf1v, DMDDf1v-Bayesian, DMDDmchi, DMDDfp2, DMDDranap,
                DMDDbg-mchi, DMDDbg-f1v, DMDDbg-fp2, DMDDbg-ranap}
 have been given as default,
 but users can choose or modify
 all these parameters/options by hand.

 Moreover,
 the preparation of uploaded files
 for defining
 the one--dimensional WIMP velocity distribution function,
 the elastic nuclear form factors
 for SI and SD WIMP--nucleus cross sections
 as well as
 artificial/experimental background spectrum
 will be particularly described.

\subsection{Predicting event rate for elastic WIMP--nucleus scattering}

 For generally considering
 the SI and SD WIMP--nucleus interactions together,
 the basic expression for the differential event rate
 for elastic WIMP--nucleus scattering can be given as
 \cite{SUSYDM96, DMDDranap}:
\beq
   \dRdQ
 = \frac{\rho_0}{2 \mchi \mrN^2}
   \bbigg{\sigmaSI \FSIQ + \sigmaSD \FSDQ}
   \int_{\vmin}^{\vmax} \bfrac{f_1(v)}{v} dv
\~.
\label{eqn:dRdQ_SISD}
\eeq
 Here $R$ is the direct detection event rate,
 i.e.~the number of events
 per unit time and unit mass of detector material,
 $Q$ is the energy deposited in the detector,
 $\rho_0$ is the WIMP density near the Earth,
 $\sigma_0^{\rm (SI, SD)}$ are the SI/SD total cross sections
 ignoring the form factor suppression and
 $F_{\rm (SI, SD)}(Q)$ are the corresponding
 elastic nuclear form factors,
 $f_1(v)$ is the one--dimensional velocity distribution function
 of the WIMPs impinging on the detector,
 $v$ is the absolute value of the WIMP velocity
 in the laboratory frame.
 The reduced mass $\mrN$ is defined by
\beq
        \mrN
 \equiv \frac{\mchi \mN}{\mchi + \mN}
\~,
\label{eqn:mrN}
\eeq
 where $\mchi$ is the WIMP mass and
 $\mN$ that of the target nucleus.
 Finally,
 $\vmin$ is the minimal incoming velocity of incident WIMPs
 that can deposit the energy $Q$ in the detector:
\beq
   \vmin
 = \alpha \sqrt{Q}
\label{eqn:vmin}
\eeq
 with the transformation constant
\beq
        \alpha
 \equiv \sfrac{\mN}{2 \mrN^2}
\~,
\label{eqn:alpha}
\eeq
 and $\vmax$ is the maximal WIMP velocity
 in the Earth's reference frame,
 which is related to
 the escape velocity from our Galaxy
 at the position of the Solar system,
 $\vesc~\gsim~600$ km/s.

 The SI scalar WIMP--nucleus cross section
 can be expressed as
 \cite{SUSYDM96, Bertone05}:
\beqn
           \sigmaSI
 \=        \afrac{4}{\pi} \mrN^2
           \bBig{Z f_{\rm p} + (A - Z) f_{\rm n}}^2
\~.
\label{eqn:sigma0_scalar}
\eeqn
 Here $\mrN$ is the reduced mass defined in Eq.~(\ref{eqn:mrN}),
 $Z$ and $A$ are the atomic and atomic mass numbers
 of the target nucleus,
 respectively,
 $f_{\rm (p, n)}$ are the effective
 scalar couplings of WIMPs on protons and on neutrons,
 respectively.
 For the lightest supersymmetric neutralino
 (and for all WIMPs which interact primarily through Higgs exchange),
 the theoretical prediction
 that
 the scalar couplings are approximately the same
 on protons and on neutrons:
\( 
        f_{\rm n}
 \simeq f_{\rm p}
\)
 is often adopted.
 For this case,
 $\sigmaSI$
 can then be written as
\beq
        \sigmaSI
 \simeq \afrac{4}{\pi} \mrN^2 A^2 |f_{\rm p}|^2
 =      A^2 \afrac{\mrN}{\mrp}^2 \sigmapSI
\~,
\label{eqn:sigma0SI}
\eeq
 where $\mrp$ is the reduced mass
 of the WIMP mass $\mchi$ and the proton mass $m_{\rm p}$,
 and
\beq
   \sigmapSI
 = \afrac{4}{\pi} \mrp^2 |f_{\rm p}|^2
\label{eqn:sigmapSI}
\eeq
 is the SI WIMP--nucleon cross section.
 The tiny mass difference between a proton and a neutron
 has been neglected.

 On the other hand,
 the SD WIMP--nucleus cross section
 can be expressed as
 \cite{SUSYDM96, Bertone05}:
\beq
   \sigmaSD
 = \afrac{32}{\pi} G_F^2 \~ \mrN^2
   \afrac{J + 1}{J} \bBig{\Srmp \armp + \Srmn \armn}^2
\~.
\label{eqn:sigma0SD}
\eeq
 Here $G_F$ is the Fermi constant,
 $J$ is the total spin of the target nucleus,
 $\expv{S_{\rm (p, n)}}$ are the expectation values of
 the proton and neutron group spins,
 and $a_{\rm (p, n)}$ are the effective SD WIMP couplings
 on protons and on neutrons.
 Then,
 since
 for a proton or a neutron
 $J = \frac{1}{2}$ and $\Srmp$ or $\Srmn = \frac{1}{2}$,
 the SD WIMP--nucleon cross sections
 can be given as
\beq
   \sigma_{\chi {\rm (p, n)}}^{\rm SD}
 = \afrac{24}{\pi} G_F^2 \~ m_{\rm r, (p, n)}^2 |a_{\rm (p, n)}|^2
\~.
\label{eqn:sigmap/nSD}
\eeq
\subsection{WIMP properties}

 The following information on WIMP properties
 is required for predicting the recoil spectrum
 and/or analyzing user--uploaded data:
{\small
\begin{center}
\renewcommand{\arraystretch}{1.5}
\begin{tabular}{|| c | l | l ||}
\hline
\hline
 \makebox[ 1.4 cm][c]{Symbol}  &
 \makebox[12   cm][c]{Meaning} &
 \makebox[ 2.3 cm][c]{Remarks} \\
\hline
\hline
 $\mchi$                       &
 The input WIMP mass           &
 GeV$/c^2$                     \\
 $\sigma(\mchi)$               &
 An overall uncertainty on the input WIMP mass &
 [0, 1]                        \\
\hline
 $\sigmapSI$                   &
 The SI WIMP--proton cross section &
 pb                            \\
\hline
 $\frmn / \frmp$               &
 The ratio of the SI WIMP coupling on neutrons to that on protons &
                               \\
\hline
 $\armp$                       &
 The SD WIMP--proton coupling  &
                               \\
 $\armn / \armp$               &
 The ratio of the SD WIMP coupling on neutrons to that on protons &
                               \\
\hline
\hline
\end{tabular}
\end{center}
}
 ~
\includefigure{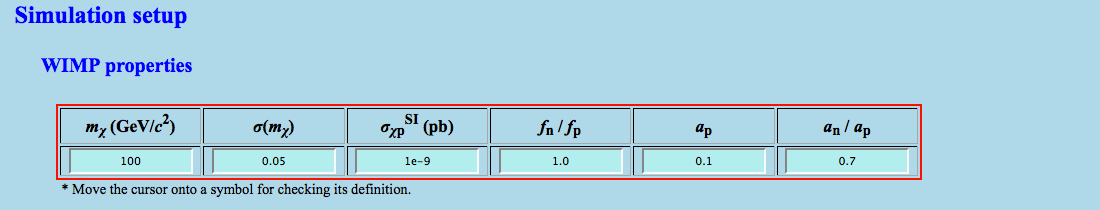}
 Note that,
 firstly,
 {\em not} all of these items
 are needed for every \amidas\ function.
 The website will give the needed items automatically
 according to users earlier options.
 Secondly,
 in case that any required datum is missed,
 this omission will be detected
 automatically after the submission and
 users will be reminded of that
 with a {\em red} block around the table.
 For this case
 {\em all data} in this setup table
 will be {\em reset} to the default values
 and should therefore be checked and modified once again
 to the users' own setup.
 Remind also that,
 the overall uncertainty on the input WIMP mass
 should be between 0 and 1.
 On the other hand,
 users can hover the curser onto an item symbol
 in the setup table
 for checking its definition.

\subsection{Astronomical setup}

 \amidas\ requires also information on
 the following astronomical parameters
 for predicting/fitting
 the velocity distribution function of halo WIMPs
 as well as
 for estimating the SI WIMP--nucleon coupling:
%
{\small
\begin{center}
\renewcommand{\arraystretch}{1.5}
\begin{tabular}{|| c | l | l ||}
\hline
\hline
 \makebox[ 1.4 cm][c]{Symbol}  &
 \makebox[12   cm][c]{Meaning} &
 \makebox[ 2.3 cm][c]{Remarks} \\
\hline
\hline
 $\rho_0$                        &
 The WIMP density near the Earth &
 GeV$/c^2/{\rm cm}^3$            \\
\hline
 $v_0$                         &
 The Sun's orbital speed around the Galactic center &
 km/s                          \\
 $\vmax$                       &
 The maximal cut--off on the 1-D WIMP velocity distribution function     &
 km/s                          \\
\hline
 $t_{\rm p}$                   &
 \begin{minipage}{12cm}
 \vspace{0.2 cm}
 The date on which the Earth's velocity relative to the WIMP halo is maximal
 \vspace{0.25cm}
 \end{minipage}                &
 day                           \\
 $t_{\rm expt}$                &
 The (middle of) the running date of the experiment   &
 day                           \\
\hline
\hline
\end{tabular}
\end{center}
}
 ~
\includefigure{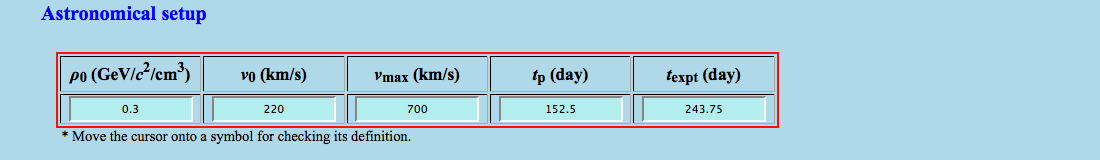}
 Remind that,
 in case that any required datum is missed,
 this omission will be detected
 automatically after the submission and
 users will be reminded of that
 with a {\em red} block around the table.
 For this case
 {\em all data} in this setup table
 will be {\em reset} to the default values
 and should therefore be checked and modified once again
 to the users' own setup.
 On the other hand,
 users can hover the curser onto an item symbol
 in the setup table
 for checking its definition.

\subsection{Velocity distribution function of halo WIMPs}

 For predicting the elastic WIMP--nucleus scattering spectrum
 in order to generate WIMP--induced signal events,
 one needs crucially
 the one--dimensional WIMP velocity distribution function $f_1(v)$.
 In the \amidasii\ package,
 we define intrinsically
 three most commonly used theoretical distributions.
 Meanwhile,
 users also have the option to define (by uploading/typing)
 an analytic form of
 their favorite one--dimensional velocity distribution:
\includefigure{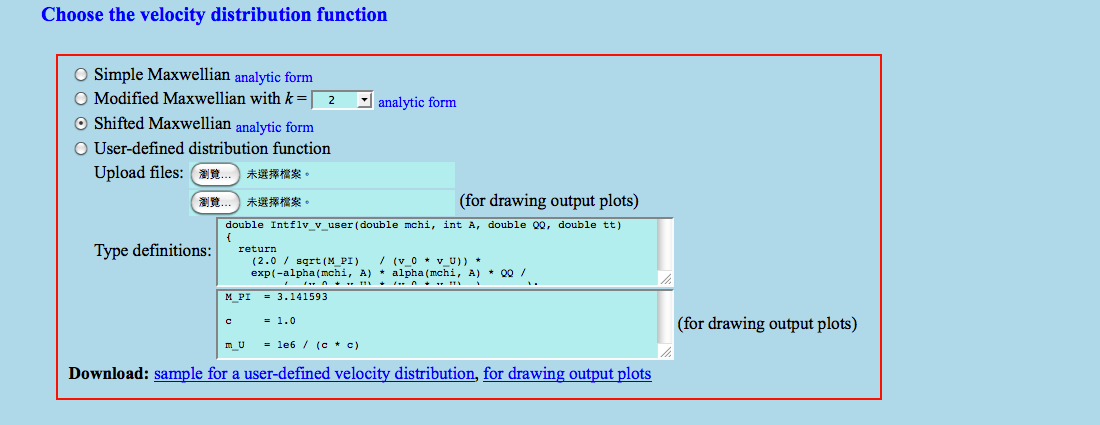}
 Note here that,
 firstly,
 for checking the analytic forms of
 the intrinsically defined velocity distributions,
 users can hover the cursor onto ``\verb+analytic form+'';
 users can also click the ``\verb+analytic form+''
 to open a new webpage
 with more detailed information and useful references.
 Secondly,
 as reminded on the website,
 the second uploaded file/typing area is
 {\em only} for drawing output plot(s) of
 the generated WIMP--signal (and background) events
 and,
 as a comparison to,
 the (Bayesian) reconstructed one--dimensional
 WIMP velocity distribution function.
 Moreover,
 considering the normal height of a browser window,
 we shrink the typing areas as default.
 However,
 once users click one of the two typing areas,
 the clicked one will be extended to show the full content;
 the extended typing area(s) will shrink automatically
 after users click
 one of the three default velocity distribution functions.
\includefigure{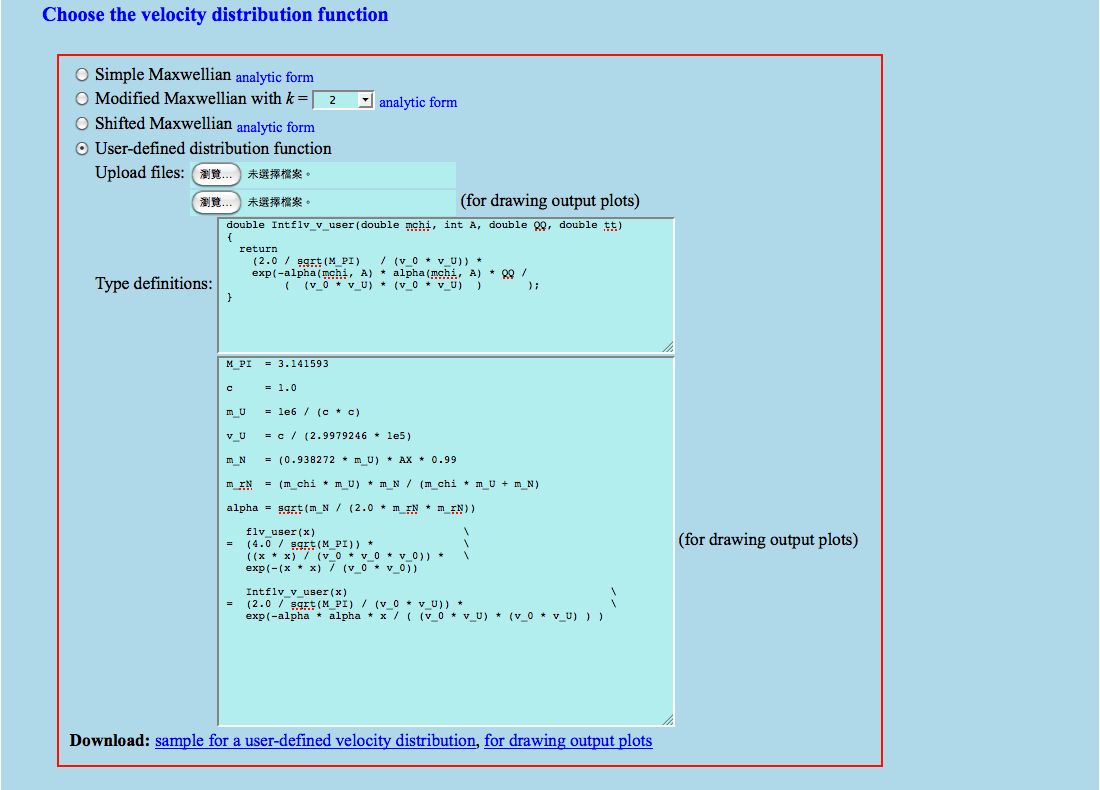}

 In this subsection,
 we give first the definitions of
 the three default velocity distribution functions
 in the \amidasii\ package.
 Then we will describe
 how to (modify these definitions to) define
 user's own velocity distribution.

\subsubsection{Default one--dimensional WIMP velocity distribution functions}

 So far users have three options
 for the one--dimensional WIMP velocity distribution function
 defined intrinsically in the \amidasii\ package
 \cite{DMDDf1v, DMDDf1v-Bayesian}:
\beenr
\item
 the simple Maxwellian velocity distribution function
 \cite{SUSYDM96}
\beq
    f_{1, \Gau}(v)
 =  \frac{4}{\sqrt{\pi}}
    \afrac{v^2}{v_0^3} e^{-v^2 / v_0^2}
\~;
\label{eqn:f1v_Gau}
\eeq
\item
 the modified Maxwellian velocity distribution function
 \cite{Lisanti10, YYMao12, YYMao13, Kuhlen13}
\beq
    f_{1, \Gau, k}(v)
 =  \frac{v^2}{N_{f, k}}
    \abrac{  e^{-v^2     / k v_0^2}
           - e^{-\vmax^2 / k v_0^2}  }^k
\~,
    ~~~~~~~~~~~~ 
    ({\rm for}~v \le \vmax),
\label{eqn:f1v_Gau_k}
\eeq
 for $k = 1,~2,~3,~4$,
 where
 $N_{f, k}$ is the normalization constant
 depending on the value of the power index $k$;
\item
 the shifted Maxwellian velocity distribution function
 \cite{SUSYDM96, Lewin96}
\beq
    f_{1, \sh}(v)
 =  \frac{1}{\sqrt{\pi}} \afrac{v}{v_0 \ve}
    \bBig{  e^{-(v - \ve)^2 / v_0^2}
          - e^{-(v + \ve)^2 / v_0^2}  }
\~,
\label{eqn:f1v_sh}
\eeq
 where the time--dependent Earth's velocity in the Galactic frame is given as
\beq
    \ve(t)
 =  v_0 \bbrac{1.05 + 0.07 \cos\afrac{2 \pi (t - t_{\rm p})}{1~{\rm yr}}}
\~.
\label{eqn:v_e}
\eeq
\end{enumerate}
\subsubsection{User--defining the one--dimensional WIMP velocity distribution function}

 For defining the one--dimensional WIMP velocity distribution
 for generating WIMP signals,
 users should in practice give
 the {\em integral} over the velocity distribution function:
\beq
    \int_{\vmin = \alpha \sqrt{Q}}^{\vmax} \bfrac{f_1(v, t)}{v} dv
\eeq
 with the name of ``\verb+Intf1v_v_user+''.
 Note here that
 the lower limit of the integral,
 the minimal incoming velocity of incident WIMPs
 that can deposit the energy $Q$ in the detector,
 $\vmin$,
 should be expressed as a function of the energy $Q$ through
 $\vmin = \alpha \sqrt{Q}$;
 the upper limit of the integral,
 $\vmax$,
 can be set as one of the required astronomical parameters
 (see Sec.~3.3).
 Meanwhile,
 since the transformation constant
\cheqnref{eqn:alpha}
\beq
         \alpha
 \equiv  \sfrac{\mN}{2 \mrN^2}
\eeq
\cheqnN{-1}
 is defined with the reduced mass of the WIMP mass $\mchi$
 and that of the target nucleus $\mN$
\cheqnref{eqn:mrN}
\beq
         \mrN
 \equiv  \frac{\mchi \mN}{\mchi + \mN}
\~,
\eeq
\cheqnN{-1}
 and $\mN$ is given as a function of the atomic mass number $A$
 in the \amidas\ package,
 the expression of the integral over the WIMP velocity distribution
 should thus be a function of $\mchi$ and $A$.
 Finally,
 in more general cases,
 the (integral over the) velocity distribution function
 should also be a function of time $t$.

 As an example,
 we give here
 the \amidasii\ code for
 the integral over the
 simple Maxwellian velocity distribution function.
 The codes for the (integrals over the)
 other intrinsically defined
 one--dimensional WIMP velocity distribution functions
 will be given in Appendix B.1
 for users' reference.

\code{Integral over the simple Maxwellian velocity distribution
      \boldmath$\int_{\alpha \sqrt{Q}}^{\infty} \bbig{f_{1, \Gau}(v) / v} dv$
      \label{code:Intf1v_v_user}}
\verb# double Intf1v_v_user(double mchi, int A, double QQ, double tt) # \\
\verb# {                                                              # \\
\verb#   return                                                       # \\
\verb#     ( 2.0 / sqrt(M_PI) / (v_0 * v_U) ) *                       # \\
\verb#     exp(-alpha(mchi, A) * alpha(mchi, A) * QQ /                # \\
\verb#          (  (v_0 * v_U) * (v_0 * v_U)  )       );              # \\
\verb# }                                                              # \\
\tabb
 Note that,
 firstly,
 \verb+v_U+ and
 the function \verb+alpha(mchi, A)+,
 where \verb+mchi+ and \verb+A+ stand for
 the WIMP mass $\mchi$ and
 the atomic mass number of the target nucleus, $A$,
 are defined intrinsically in the \amidas\ package%
\footnote{
 Relevant constants and functions
 defined in the \amidas\ package
 will be given in Appendix A.1 and A.2.
}.
 Secondly,
 \verb+v_0+,
 standing for
 $v_0$,
 is an input parameter
 which users can set separately on the website
 (see Sec.~3.3).
 Moreover,
 \verb+QQ+ and \verb+tt+ stand for
 the recoil energy $Q$
 and time $t$.
 Remind also here that,
 even for a {\em time--independent} velocity distribution
 or a numerical expression with
 a fixed experimental running date $t = t_{\rm expt}$,
 ``\verb+double tt+'' should always be declared
 as the {\em forth} parameter of the function
 \verb+Intf1v_v_user(mchi, A, QQ, tt)+.

 On the other hand,
 an {\em extra} file is required
 for defining the input theoretical
 one--dimensional velocity distribution function
 {\em itself}
 (not the integral over it!),
 for drawing this input velocity distribution,
 as a comparison,
 with the Bayesian reconstructed one
 together in the output plots.
 Meanwhile,
 the {\em integral} over
 this input velocity distribution
 is however needed
 for drawing the theoretical WIMP scattering spectrum,
 with the (binned) recorded WIMP--signals (and background) events
 together in the output plots.
 Since the {\tt Gnuplot} software
 has been adopted in the \amidas\ package
 for drawing output plots,
 the (integral over the) velocity distribution function
 defined in this file
 must be written in the syntax of {\tt Gnuplot}
 with the name of ``\verb+f1v_user(x)+'' or ``\verb+Intf1v_v_user(x)+''.
 Below we give our definitions for
 the (integral over the) simple Maxwellian velocity distribution
 as examples.
 The definitions for the (integral over the)
 other intrinsically defined
 one--dimensional WIMP velocity distribution function
 for \verb+Gnuplot+
 are given in Appendix B.1
 for users' reference.

\code{Simple Maxwellian velocity distribution
      \boldmath$f_{1, \Gau}(v)$
      (for Gnuplot)
      \label{code:f1v_x_user}}
\verb#   M_PI = 3.141593                        # \\
\verb#                                          # \\
\verb#    f1v_user(x)                         \ # \\
\verb# =  (4.0 / sqrt(M_PI))              *   \ # \\
\verb#    ( (x * x) / (v_0 * v_0 * v_0) ) *   \ # \\
\verb#    exp(-(x * x) / (v_0 * v_0))           # \\
\tabb
\code{Integral over the simple Maxwellian velocity distribution
      \boldmath$\int_{\alpha \sqrt{Q}}^{\infty} \bbig{f_{1, \Gau}(v) / v} dv$
      (for Gnuplot)
      \label{code:Intf1v_v_x_user}}
\verb# c   = 1.0                                                       # \\
\verb# m_U = 1e6 / (c * c)                                             # \\
\verb# v_U = c / (2.9979246 * 1e5)                                     # \\
\verb#                                                                 # \\
\verb# m_N  = (0.938272 * m_U) * AX * 0.99                             # \\
\verb# m_rN = (m_chi * m_U) * m_N / (m_chi * m_U + m_N)                # \\
\verb#                                                                 # \\
\verb# alpha = sqrt(m_N / (2.0 * m_rN * m_rN))                         # \\
\verb#                                                                 # \\
\verb#    Intf1v_v_user(x)                                           \ # \\
\verb# =  (2.0 / sqrt(M_PI) / (v_0 * v_U)) *                         \ # \\
\verb#    exp(-alpha * alpha * x / ( (v_0 * v_U) * (v_0 * v_U) ) )     # \\
\tabb
 Note here that,
 firstly,
 ``\verb+\+'' (backslash) must be used
 in order to let the including of the definition(s) given in this file
 into the other intrinsic commands correctly.
 Secondly,
 three flexible--kept parameters:
 \verb+AX+,
 \verb+m_chi+ and \verb+v_0+
 will be read directly from
 the users' initial simulation/data analysis setup
 set earlier on the website.
 The meanings of the variables used here
 can be found in Appendix A.1.

 Remind also that
 two sample files,
 one for the \amidas\ code and
 the other one for the {\tt Gnuplot} package,
 can be downloaded from the \amidas\ website.

\subsection{Elastic nuclear form factors for the WIMP--nucleus cross sections}

 For predicting the elastic WIMP--nucleus scattering spectrum
 in order to generate WIMP--induced signal events
 as well as
 reconstructing different WIMP properties:
 the one--dimensional velocity distribution function $f_1(v)$,
 the mass $\mchi$,
 the SI coupling on nucleons $|\frmp|^2$
 and the ratios between different WIMP--nucleon couplings/cross sections
 $\armn / \armp$ and $\sigma_{\chi {\rm (p, n)}}^{\rm SD} / \sigmapSI$,
 one needs crucially
 the elastic nuclear form factor for SI WIMP--nucleus interaction,
 $\FSIQ$.
 In the \amidas\ package,
 we define intrinsically
 four most commonly used elastic nuclear form factors
 for the SI cross section.
 Meanwhile,
 users also have the option to define (by uploading/typing)
 an analytic form of
 their favorite form factor:
\includefigure{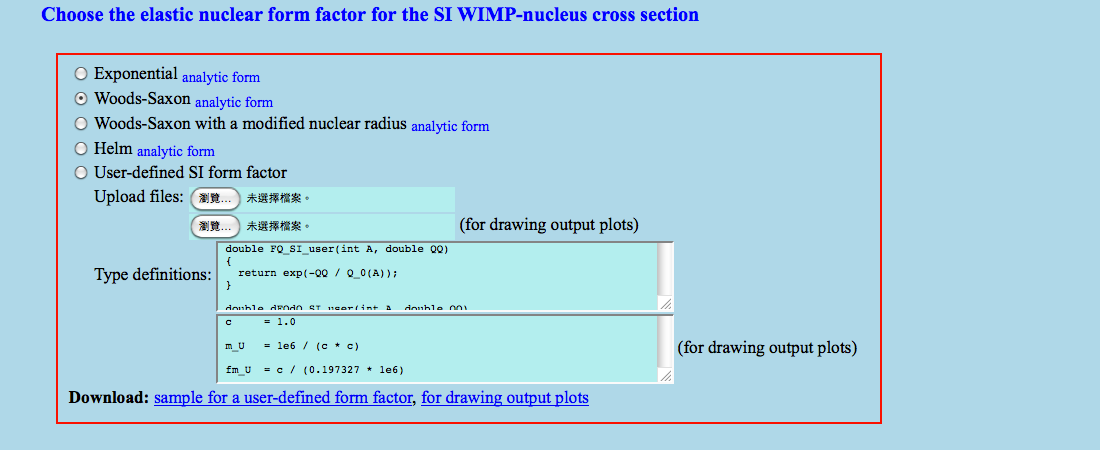}

 On the other hand,
 for
 generating WIMP--induced signal events
 (only) with SD WIMP--nucleus interaction
 as well as
 reconstructing the ratios
 between different WIMP--nucleon couplings/cross sections
 $\armn / \armp$ and $\sigma_{\chi {\rm (p, n)}}^{\rm SD} / \sigmapSI$,
 one needs crucially
 the elastic nuclear form factor for SD WIMP--nucleus interaction,
 $\FSDQ$.
 In the \amidas\ package,
 so far we define intrinsically only one elastic nuclear form factor
 for the SD cross section.
%
 As usual,
 users also have the option to define (by uploading/typing)
 an analytic form of
 their favorite form factor:
\includefigure{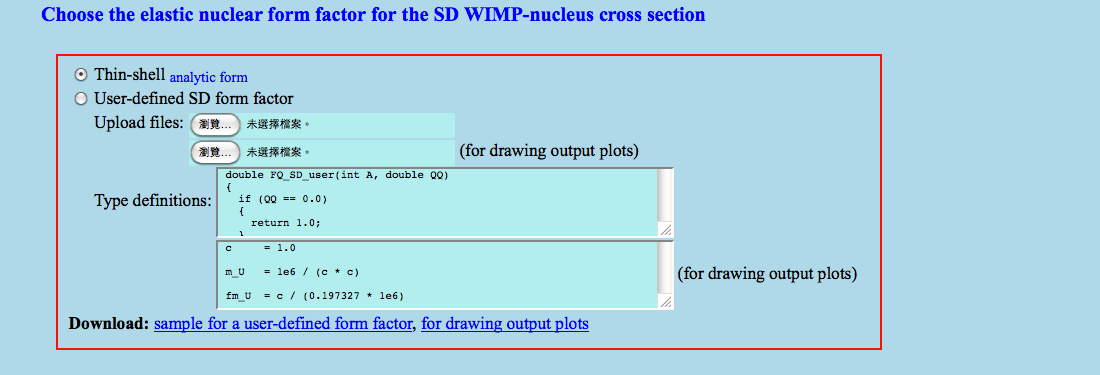}

 Note here that,
 firstly,
 for checking the analytic forms of
 the intrinsically defined SI and SD elastic nuclear form factors,
 users can hover the cursor onto ``\verb+analytic form+'';
 users can also click the ``\verb+analytic form+''
 to open a new webpage
 with more detailed information and useful references.
 Secondly,
 as reminded on the website,
 the second uploaded file/typing area is
 {\em only} for drawing output plot(s) of
 the generated WIMP--signal (and background) events.

 In this subsection,
 we give first the definitions of
 the four default elastic nuclear form factors
 for the SI WIMP--nucleus cross section
 and the unique one
 for the SD cross section
 in the \amidas\ package.
 Then we will describe
 how to (modify these definitions to) define
 user's own form factor.

\subsubsection{Default SI elastic nuclear form factors}

 So far users have four options
 for the SI elastic nuclear form factor
 defined intrinsically in the \amidas\ package
 \cite{
       DMDDf1v, DMDDmchi-NJP}:
\beenr
\item
 the exponential form factor
 \cite{Ahlen87, Freese88, SUSYDM96}
\beq
    F_{\rm ex}^2(Q)
 =  e^{-Q / Q_0}
\~,
\label{eqn:FQ_SI_ex}
\eeq
 where
\beq
    Q_0
 =  \frac{1.5}{\mN R_0^2}
\label{eqn:Q0}
\eeq
 is the nuclear coherence energy and
\beq
    R_0
 =  \bbrac{0.3 + 0.91 \afrac{\mN}{\rm GeV}^{1/3}}~{\rm fm}
\label{eqn:R0}
\eeq
 is the radius of the nucleus;
\item
 the Woods--Saxon form factor
 \cite{Engel91, SUSYDM96, Lewin96}
\beq
    F_{\rm WS}^2(Q)
 =  \bfrac{3 j_1(q R_1)}{q R_1}^2 e^{-(q s)^2}
\~,
\label{eqn:FQ_SI_WS}
\eeq
 where $j_1(x)$ is a spherical Bessel function,
\beq
    q
 =  \sqrt{2 m_{\rm N} Q}
\label{eqn:qq}
\eeq
 is the transferred 3-momentum,
\beq
    R_1
 =  \sqrt{R_A^2 - 5 s^2}
\label{eqn:R1}
\eeq
 is the effective nuclear radius
 with 
\beq
         R_A
 \simeq  1.2 \~ A^{1/3}~{\rm fm}
\~,
\label{eqn:RA}
\eeq
 and
\beq
         s
 \simeq  1~{\rm fm}
\label{eqn:ss}
\eeq
 is the nuclear skin thickness,
 $A$ is the atomic mass number of the nucleus;
\item
 the Woods--Saxon form factor
 with a modified nuclear radius
 \cite{Eder68, Lewin96},
 in which
\beq
         R_A
 \simeq  \abig{1.15 \~ A^{1/3} + 0.39}~{\rm fm}
\~;
\label{eqn:RA_Eder}
\eeq
\item
 the Helm form factor
 \cite{Helm56, Lewin96}
\beq
    R_1
 =  \sqrt{R_A^2 + {\T \afrac{7}{3}} \pi^2 r_0^2 - 5 s^2}
\label{eqn:R1_Helm}
\eeq
 is the effective nuclear radius with
\beq
         R_A
 \simeq  \abig{1.23 \~ A^{1/3} - 0.6}~{\rm fm}
\~,
\label{eqn:RA_Helm}
\eeq
\beq
         r_0
 \simeq  0.52~{\rm fm}
\~,
\label{eqn:r0_Helm}
\eeq
 and
\beq
         s
 \simeq  0.9~{\rm fm}
\label{eqn:ss_Helm}
\eeq
 is the nuclear skin thickness.
\end{enumerate}
\subsubsection{Default SD elastic nuclear form factor}

 Comparing to the SI case,
 the nuclear form factor
 for the SD WIMP--nucleus cross section
 is more complicated and target--dependent,
 due to its dependence on the SD WIMP--nucleon couplings
 as well as on the individual spin structure of target nuclei.
 Hence,
 so far
 in the \amidas\ package
 we define only one general analytic form
 for the SD elastic nuclear form factor:
\beenr
\item
 the thin--shell form factor
 \cite{Lewin96, Klapdor05}
\beq
    F_{\rm TS}^2(Q)
 =  \cleft{\renewcommand{\arraystretch}{1.6}
           \begin{array}{l l l}
            j_0^2(q R_1)              \~, & ~~~~~~ &
            {\rm for}~q R_1 \le 2.55~{\rm or}~q R_1 \ge 4.5 \~, \\
            {\rm const.} \simeq 0.047 \~, &        & {\rm for}~2.5 5 \le q R_1 \le 4.5 \~,
           \end{array}}
\label{eqn:FQ_SD_TS}
\eeq
 where $j_0(x)$ is a spherical Bessel function.
\end{enumerate}
\subsubsection{User--defining the elastic nuclear form factor}

 For defining users' favorite nuclear form factors,
 {\em not only} the definitions of
 the {\em squared} form factor $\FQ$
 {\em but also} those of the derivatives of $\FQ$
 with respect to the energy $Q$:
\beq
   \Dd{\FQ}{Q}
 = 2 F(Q) \bDd{F(Q)}{Q}
\label{eqn:dFQdQ}
\eeq
 (not $F(Q)$ itself nor $dF(Q) / dQ$)
 must be given {\em together} in {\em one} file
 with however the names of ``\verb+FQ_SI_user+'' (``\verb+FQ_SD_user+'')
 and ``\verb+dFQdQ_SI_user+'' (``\verb+dFQdQ_SD_user+'').

 As an example,
 we give here
 the \amidas\ code for
 the (derivative of the)
 exponential form factor.
 The codes for the (derivatives of the)
 other intrinsically defined
 elastic nuclear form factors
 will be given in Appendix B.2
 for users' reference.

\code{Squared exponential form factor
      \boldmath$F_{\rm ex}^2(A, Q)$
      \label{code:FQ_SI_user}}
\verb# double FQ_SI_user(int A, double QQ) # \\
\verb# {                                   # \\
\verb#   return exp(-QQ / Q_0(A));         # \\
\verb# }                                   # \\
\tabb
\code{Derivative of the squared exponential form factor
      \boldmath$d F_{\rm ex}^2(A, Q) / dQ$
      \label{code:dFQdQ_SI_user}}
\verb# double dFQdQ_SI_user(int A, double QQ)        # \\
\verb# {                                             # \\
\verb#   return -(1.0 / Q_0(A)) * FQ_SI_user(A, QQ); # \\
\verb# }                                             # \\
\tabb
 Note that
 the function \verb+Q_0(A)+
 is defined intrinsically in the \amidas\ package
 (see Appendix B.2).

 On the other hand,
 {\em two extra} files are required
 for defining the elastic nuclear form factors themselves
 ({\em without} their derivatives!)
 for the SI and/or SD WIMP--nucleus cross section(s) {\em separately},
 for drawing the theoretical WIMP scattering spectrum,
 as a comparison,
 with the (binned) recorded WIMP--signals (and background) events
 together in the output plots.
 Note that
 the form factors
 defined in these two files
 must be written in the syntax of {\tt Gnuplot}
 with the name of ``\verb+FQ_SI_user(x)+'' or `\verb+FQ_SD_user(x)+''.
 Below we give our definition for
 the exponential form factor
 as an example.
 The definitions for the other intrinsically defined
 elastic nuclear form factors
 are given in Appendix B.2
 for users' reference.

\code{Squared exponential form factor
      \boldmath$F_{\rm ex}^2(A, Q)$
      (for Gnuplot)
      \label{code:FQ_SI_x_user}}
\verb# c    = 1.0                                               # \\
\verb# m_U  = 1e6 / (c * c)                                     # \\
\verb# fm_U = c / (0.197327 * 1e6)                              # \\
\verb#                                                          # \\
\verb# m_N  = (0.938272 * m_U) * AX * 0.99                      # \\
\verb# m_rN = (m_chi * m_U) * m_N / (m_chi * m_U + m_N)         # \\
\verb#                                                          # \\
\verb# alpha = sqrt(m_N / (2.0 * m_rN * m_rN))                  # \\
\verb#                                                          # \\
\verb# R_0 = ( 0.3 + 0.91 * (m_N / m_U) ** (1.0 / 3.0) ) * fm_U # \\
\verb# Q_0 = 1.5 / (m_N * R_0 * R_0)                            # \\
\verb#                                                          # \\
\verb#    FQ_SI_user(x)   \                                     # \\
\verb# =  exp(-x / Q_0)                                         # \\
\tabb
 Remind that,
 firstly,
 ``\verb+\+'' (backslash) must be used
 in order to let the including of the definition(s) given in this file
 into the other intrinsic commands correctly.
 Secondly,
 two flexible--kept parameters:
 \verb+AX+ and \verb+m_chi+
 will be read directly from
 the users' initial simulation/data analysis setup
 set earlier on the website.
 The meanings of the variables used here
 can be found in Appendix A.1.

 Remind also that
 two sample files,
 one for the \amidas\ code and
 the other one for the {\tt Gnuplot} package,
 can be downloaded from the \amidas\ website.

\subsection{Background spectrum}

 As an extra consideration,
 users have the opportunity to
 take into account
 some unrejected background events
 in numerical simulations
 with the \amidasii\ package.

 So far
 in the \amidasii\ package,
 we defined intrinsically
 three analytic forms for the artificial background spectrum.
 Meanwhile,
 users also have the option to define (by uploading/typing)
 an analytic form of
 their favorite theoretical or experimental background spectrum:
\includefigure{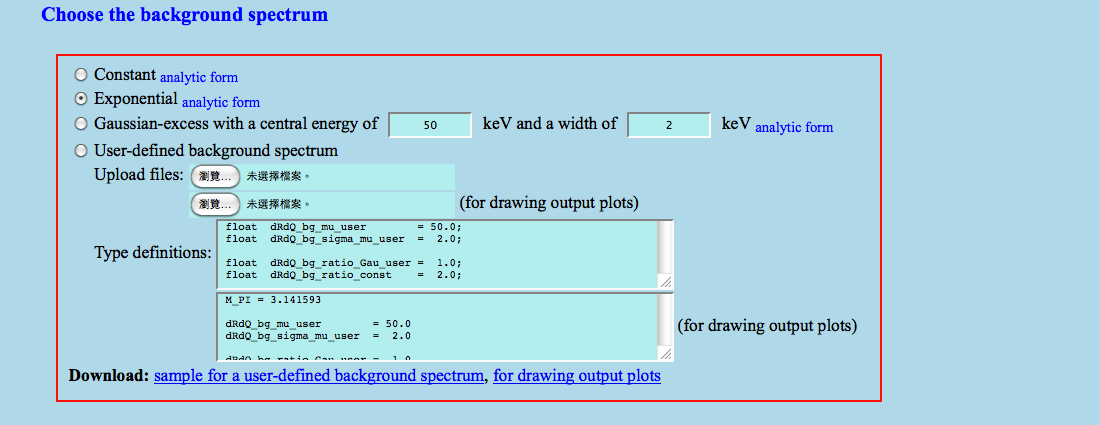}
 Note that,
 firstly,
 for checking the analytic forms of
 the intrinsically defined background spectra,
 users can hover the cursor onto ``\verb+analytic form+'';
 users can also click the ``\verb+analytic form+''
 to open a new webpage
 with more detailed information and useful references.
 Secondly,
 as reminded on the website,
 the second uploaded file/typing area is
 {\em only} for drawing output plot(s) of
 the generated WIMP--signal {\em and} background events.

 In this subsection,
 we give first the definitions of
 the three artificial background spectra
 in the \amidasii\ package.
 Then we will describe
 how to (modify these definitions to) define
 user's own theoretical or experimental background spectrum.

\subsubsection{Default background spectra}

 So far users have three options
 for the artificial background spectrum
 defined intrinsically in the \amidasii\ package
 \cite{DMDDbg-mchi, AMIDASbg-DSU2011}:
\beenr
\item
 constant background spectrum
 \cite{DMDDbg-mchi}
\beq
    \aDd{R}{Q}_{\rm bg, const}
 =  1
\~;
\label{eqn:dRdQ_bg_const}
\eeq
\item
 exponential background spectrum
 \cite{DMDDbg-mchi}
\beq
    \aDd{R}{Q}_{\rm bg, ex}
 =  \exp\abrac{-\frac{Q /{\rm keV}}{A^{0.6}}}
\~,
\label{eqn:dRdQ_bg_ex}
\eeq
 where $A$ is the atomic mass number of the target nucleus;
\item
 Gaussian--excess background spectrum
 \cite{AMIDASbg-DSU2011}
\beq
    \aDd{R}{Q}_{\rm bg, Gau}
 =  \frac{1}{\sqrt{2 \pi} \abrac{\sigma_{Q, {\rm bg}} / {\rm keV}}}
    \exp\bbrac{-\frac{(Q - Q_{\rm bg, peak})^2}{2 \sigma_{Q, {\rm bg}}^2}}
\~,
\label{eqn:dRdQ_bg_Gau}
\eeq
 where $Q_{\rm bg, peak}$ and $\sigma_{Q, {\rm bg}}$
 are the central energy and the width of the Gaussian background excess.
\end{enumerate}
\subsubsection{User--defining the background spectrum}

 For defining users' needed theoretical/experimental
 background spectrum
 for generating residue background events,
 users only have to give the analytic form
 of the spectrum
 {\em itself}
 as a function of the recoil energy $Q$
 (\verb+QQ+ in the \amidas\ code)
 in a file
 with the name of ``\verb+dRdQ_bg_user+''.

 As an example,
 we give here
 the \amidasii\ code for
 the target--dependent
 exponential background spectrum.
 The codes for the
 other intrinsically defined
 background spectra
 will be given in Appendix B.3
 for users' reference.

\code{Target--dependent exponential background spectrum
      \boldmath$\abrac{dR / dQ}_{\rm bg, ex}(A, Q)$
      \label{code:dRdQ_bg_user}}
\verb# double Q_0_bg_user(int A)             # \\
\verb# {                                     # \\
\verb#   return pow(A, 0.6);                 # \\
\verb# }                                     # \\
\verb#                                       # \\
\verb# double dRdQ_bg_user(int A, double QQ) # \\
\verb# {                                     # \\
\verb#   return exp(-QQ / Q_0_bg_user(A));   # \\
\verb# }                                     # \\
\tabb
 Note here that,
 since we take into account
 the target
 dependence of the
 background spectrum,
 the atomic mass number $A$
 is used here as the first function parameter
 (for detailed discussion see Ref.~\cite{DMDDbg-mchi}).

 On the other hand,
 an {\em extra} file is required
 for defining the background spectrum {\em itself}
 for drawing this extra part
 with the theoretical WIMP signal spectrum,
 as a comparison,
 with the (binned) recorded WIMP--signals and background events
 together in the output plots.
 Note that
 the integral over this background spectrum
 must also be given in the same file
 in order to {\em normalize}
 the background spectrum properly
 according to the user's required background ratio
 to the WIMP--induced signals.
 Remind that
 the (integral over the) background spectrum
 defined in this file
 must be written in the syntax of {\tt Gnuplot}
 with the name of ``\verb+dRdQ_bg_user(x)+'' and `\verb+IntdRdQ_bg_user(x)+''.
 Below we give our definition for
 the (integral over the) target--dependent exponential background spectrum
 as an example.
 The definitions for the other intrinsically defined
 (integral over the) background spectra
 are given in Appendix B.3
 for users' reference.

\code{Target--dependent exponential background spectrum
      \boldmath$\abrac{dR / dQ}_{\rm bg, ex}(A, Q)$
      (for Gnuplot)
      \label{code:dRdQ_bg_x_user}}
\verb#    dRdQ_bg_user(x)       \ # \\
\verb# =  exp(-x / AX ** 0.6)     # \\
\tabb
\code{Integral over the target--dependent exponential background spectrum
      \boldmath$\int \abrac{dR / dQ}_{\rm bg, ex}(A, Q) \~ dQ$
      (for Gnuplot)
      \label{code:IntdRdQ_bg_x_user}}
\verb#    IntdRdQ_bg_user(x)                  \ # \\
\verb# =- (AX ** 0.6) * exp(-x / AX ** 0.6)     # \\
\tabb
 Remind that,
 firstly,
 ``\verb+\+'' (backslash) must be used
 in order to let the including of the definition(s) given in this file
 into the other intrinsic commands correctly.
 Secondly,
 the flexible--kept parameters:
 \verb+AX+
 will be read directly from
 the users' initial simulation/data analysis setup
 set earlier on the website.
 The meanings of the variables used here
 can be found in Appendix A.1.

 Remind also that
 two sample files,
 one for the \amidas\ code and
 the other one for the {\tt Gnuplot} package,
 can be downloaded from the \amidas\ website.

\subsection{Experimental setup}

 Finally,
 one needs to set the following experimental information
 for both numerical
 simulations
 and data analyses.

\subsubsection{Experimental setup}

 For generating WIMP signals,
 running numerical simulations
 as well as
 analyzing uploaded data sets,
 the \amidas\ package needs
{\small
\begin{center}
\renewcommand{\arraystretch}{1.5}
\begin{tabular}{|| c | l | l ||}
\hline
\hline
 \makebox[ 1.4 cm][c]{Symbol}  &
 \makebox[12   cm][c]{Meaning} &
 \makebox[ 2.3 cm][c]{Remarks} \\
\hline
\hline
 $\Qmin$                        &
 The minimal cut--off energy    &
 keV                            \\
 $\Qmax$                        &
 The maximal cut--off energy    &
 keV                            \\
\hline
 $b_1$                          &
 The width of the first $Q-$bin &
 keV                            \\
\hline
 $N_{\rm tot}$                  &
 The expected total event number between $\Qmin$ and $\Qmax$ &
 Max. 5,000                     \\
\hline
 $B$                            &
 The number of $Q-$bin between $\Qmin$ and $\Qmax$ &
 [4, 10]                        \\
\hline
 $N_{\rm expt}$                 &
 The number of simulated experiments or uploaded data sets &
 Max. 2,000                     \\
\hline
\hline
\end{tabular}
\end{center}
}
 ~
\includefigure{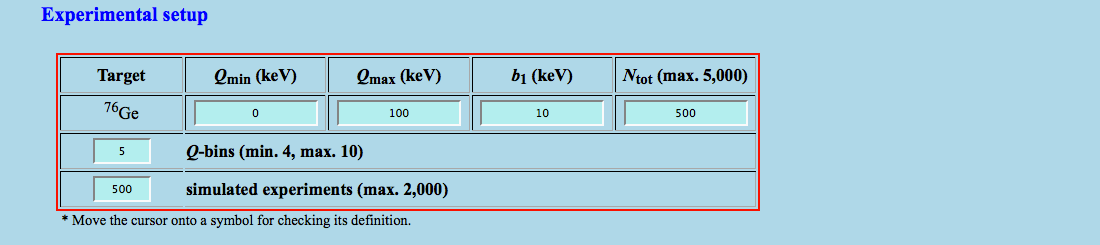}
 Remind that,
 firstly,
 the \amidas\ website offers full Monte Carlo simulations
 with maximal 2,000 experiments and
 maximal 5,000 events on average per one experiment;
 the number of $Q-$bin between $\Qmin$ and $\Qmax$
 should be between 4 and 10 bins.
 Secondly,
 users can hover the curser onto an item symbol
 in the setup table
 for checking its definition.

\subsubsection{Experimental setup for background events}

 Once users want to take into account background events
 for their numerical simulations,
 the following information should be given:
{\small
\begin{center}
\renewcommand{\arraystretch}{1.5}
\begin{tabular}{|| c | l | l ||}
\hline
\hline
 \makebox[ 1.4 cm][c]{Symbol}  &
 \makebox[12   cm][c]{Meaning} &
 \makebox[ 2.3 cm][c]{Remarks} \\
\hline
\hline
 $\Qminbg$                     &
 The lower bound of the background window &
 keV                           \\
 $\Qmaxbg$                     &
 The upper bound of the background window &
 keV                           \\
\hline
 $r_{\rm bg}$                  &
 The ratio of background events in the whole data set &
 [0, 1]                        \\
\hline
 $N_{\rm tot, bg}$             &
 The expected number of background events
 in the background window      &
 \begin{minipage}{2.3cm}
 \vspace{ 0.2 cm}
 Calculated automatically
 \vspace{-0.3 cm}
 \end{minipage}                \\
\hline
 $N_{\rm tot, sg}$             &
 The expected number of WIMP signals
 in the background window      &
 \begin{minipage}{2.3cm}
 \vspace{ 0.2 cm}
 Calculated automatically
 \vspace{-0.3 cm}
 \end{minipage}                \\
\hline
\hline
\end{tabular}
\end{center}
}
 ~
\includefigure{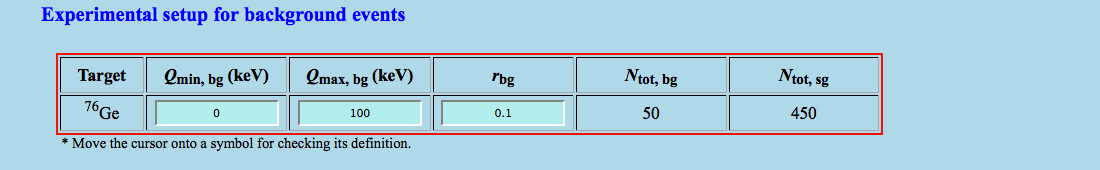}
 Remind that,
 firstly,
 the ratio of background events in the whole data set
 should be between 0 and 1.
 By setting the total event number $N_{\rm tot}$
 and the background ratio $r_{\rm bg}$,
 the average numbers of WIMP--signals and residue background events,
 $N_{\rm tot, sg}$ and $N_{\rm tot, bg}$,
 will be calculated automatically.
 Secondly,
 users can hover the curser onto an item symbol
 in the setup table
 for checking its definition.

\subsection{Running simulations}

 After giving all the required information
 for the aimed simulation,
 users have one more chance to check their choices,
 modify some of them,
 and then resubmit the whole setup,
 before they click the ``\verb+Simulation start+'' button
 to start the \amidas\ program.
\includefigure{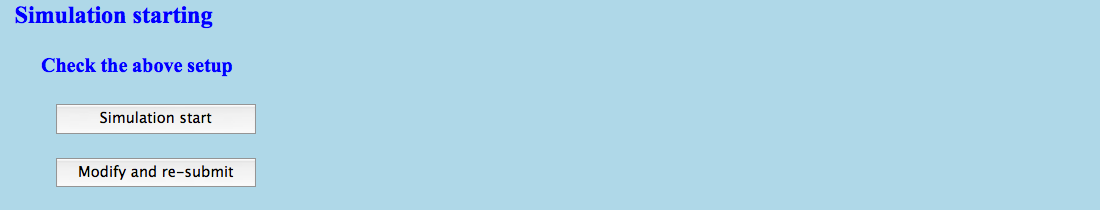}
 Remind that,
 in case that any required datum is missed,
 this omission will be detected
 automatically after the (re)submission and
 users will be reminded of that
 with a {\em red} block around the options/table.
 Note that
 {\em all data} in a table
 with missed information
 will be {\em reset} to the default values
 and should therefore be checked and modified once again
 to the users' own setup.

 Once all the required data
 have been checked,
 users have only to click
 the ``\verb+Simulation start+'' button
 and wait for the simulation results for a while%
\footnote{
 For the case that
 a simulation takes a long time
 and thus the results are not shown properly,
 users can check the following URL manually: \\
 {\tt http://pisrv0.pit.physik.uni-tuebingen.de/darkmatter/amidas/amidas-results.php}.
}.
\subsection{Output results (plots)}

 Simulation results will be presented
 in forms of plot(s) and occasionally of table(s).
 Considering users' different needs of plot formats,
 we offer {\em four} most commonly used file types:
 PostScript (PS),
 Encapsulated PostScript (EPS),
 Portable Document Format (PDF) and
 Portable Network Graphics (PNG).
 In order to let users
 understand the output results more clearly and
 use them more conveniently,
 each output plot or table
 will be accompanied with a short description.

 Meanwhile,
 for users' need of self--producing results
 for different kinds of presentations,
 the original TXT file(s) of the simulation results
 with users' personal simulation setup
 will also be given and downloadable on the website.
 Remind that
 it would be very grateful that
 a credit of the \amidas\ package and website
 could be given for using the output results.

 Below
 we give examples of output plots
 for simulations of reconstructions of different WIMP properties.
 The default \amidas\ simulation setup
 shown in Secs.~2.2 to 2.4 and Secs.~3.2 to 3.7
 has been generally used,
 with one common exception that
 the total event number
 is only 50 on average
 (Sec.~3.7.1);
 this means that
 the numbers of WIMP--signal and background events
 are $N_{\rm tot, sg} = 45$ and $N_{\rm tot, bg} = 5$,
 respectively
 (Sec.~3.7.2).

 \subsubsection{Generation of WIMP signals with/without background events}

 As a basic and minor function of the \amidas\ package,
 one can generate WIMP signals
 with even
 a fraction of artificial/experimental background events
 \cite{DMDDbg-mchi}:
\includefigure{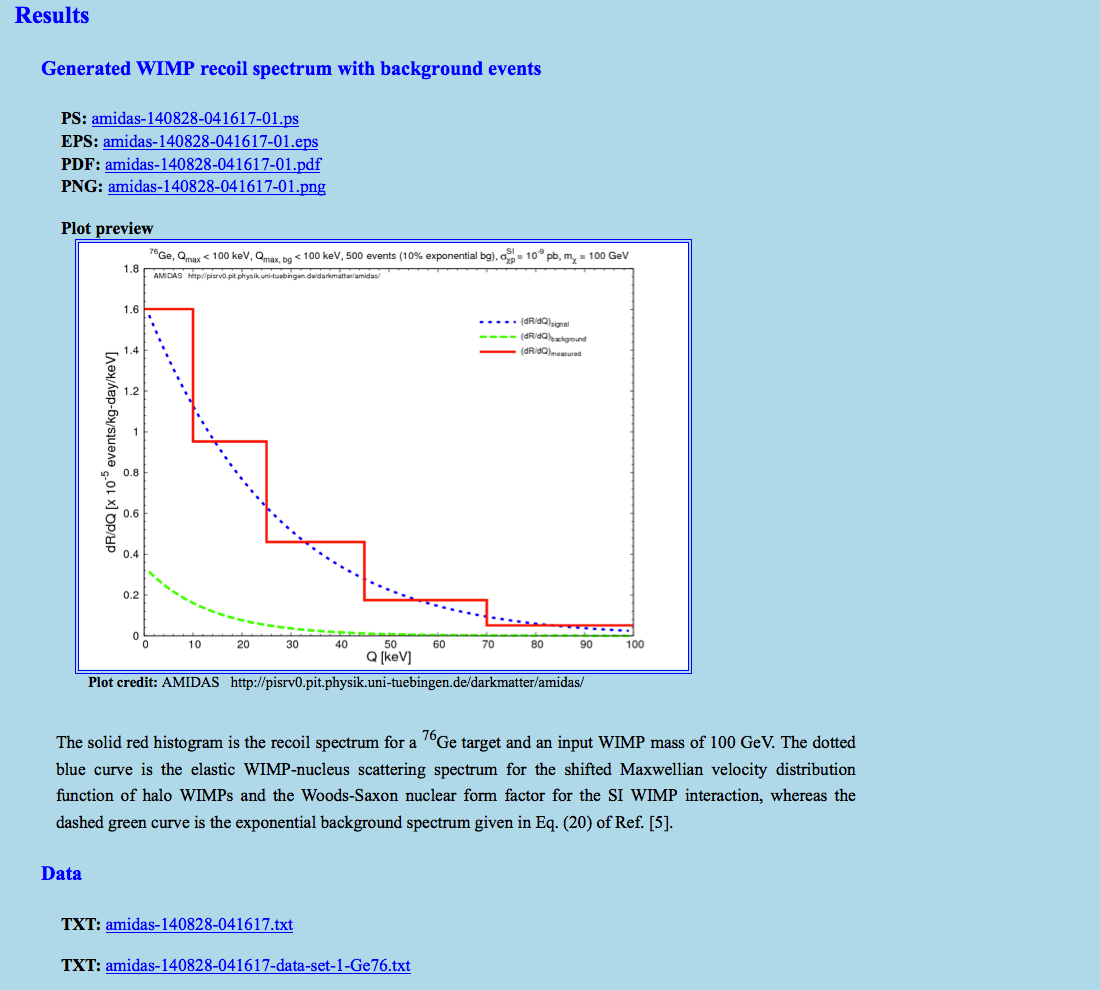}
 For each target nucleus,
 the theoretically predicted
 elastic WIMP--nucleus scattering spectrum,
 the artificial/experimental background spectrum,
 as well as
 the binned measured recoil events (histogram)
 will be given together in one plot.
 Moreover,
 as shown in the above figure,
 the generated WIMP--signal events
 (mixed with backgrounds)
 can also be offered
 in separate text files for different target nuclei
 (see Sec.~2.4.6).

 \subsubsection{Determination of the WIMP mass}

 The \amidas\ package
 provides one plot of the reconstructed WIMP mass
 with the 1$\sigma$ statistical uncertainty
 as functions of the input WIMP mass
 between 10 GeV and 1 TeV
 \cite{DMDDmchi}:
\includefigure{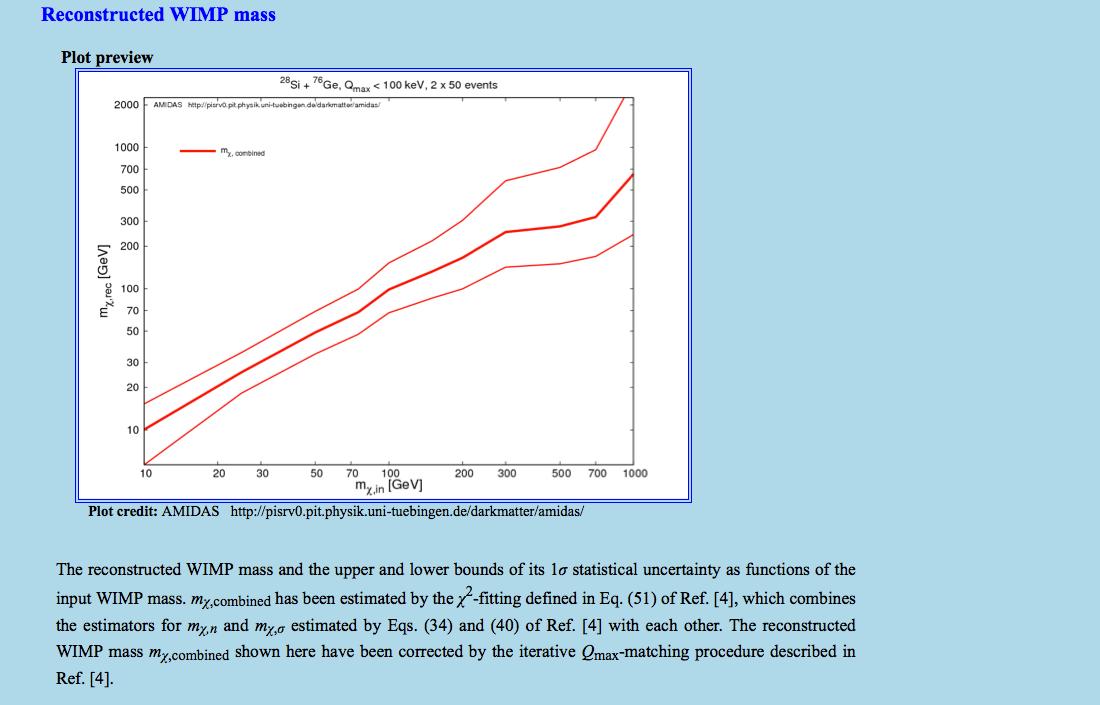}
 Meanwhile,
 as an optional output (Sec.~2.4.3),
 a plot of the statistical fluctuation of
 the reconstructed WIMP mass
 for the chosen input WIMP mass
 in all simulated experiments
 can also be given
 \cite{DMDDmchi}:
\includefigure{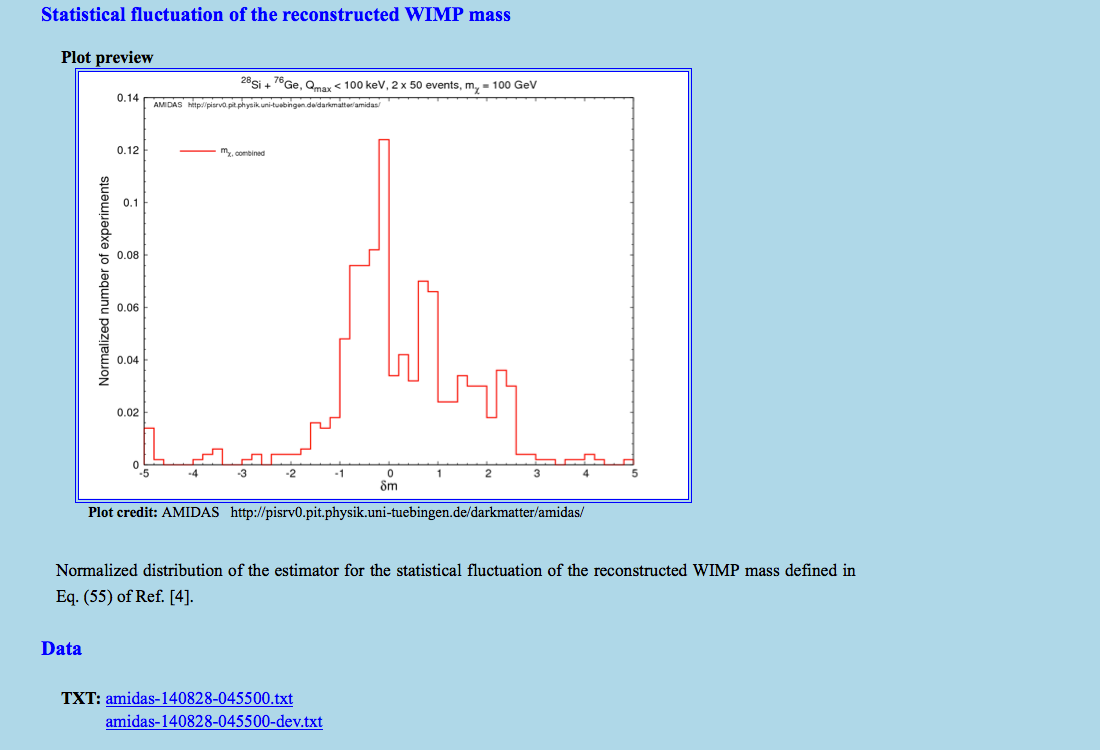}
 \subsubsection{Estimation of the SI WIMP--nucleon coupling}

 For the estimation of the SI WIMP--nucleon coupling
 \cite{DMDDfp2},
 the \amidas\ package provides two plots.
 One is
 the reconstructed squared SI WIMP coupling
 with the 1$\sigma$ statistical uncertainty
 as functions of the input WIMP mass
 between 10 GeV and 1 TeV:
\includefigure{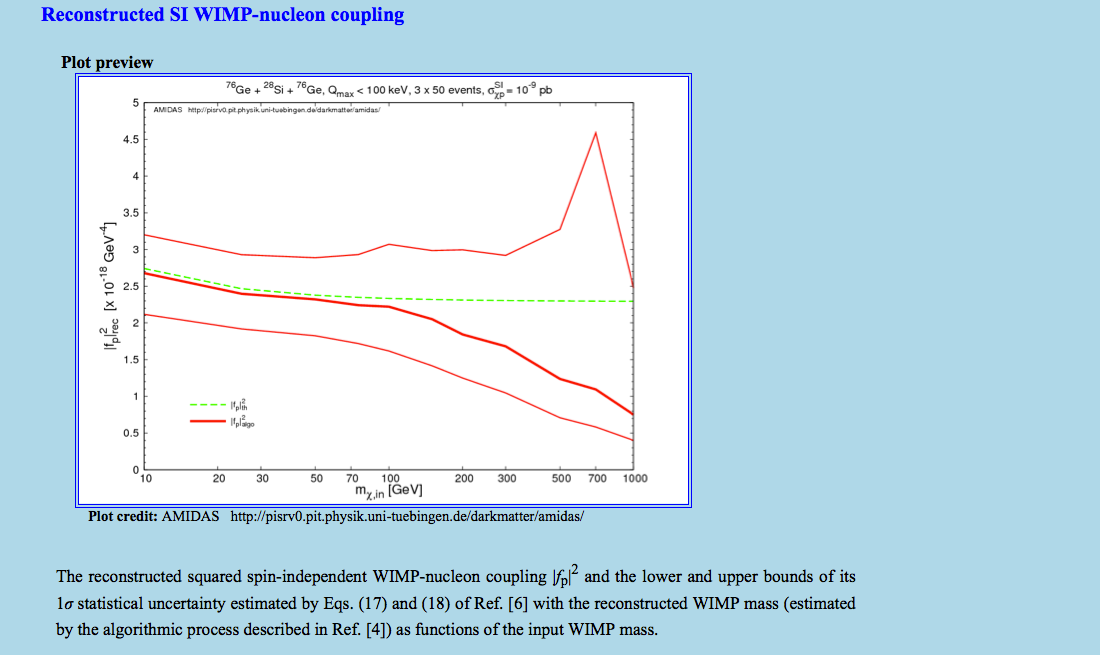}
 And
 the other one is
 the reconstructed squared SI WIMP couplings
 and the reconstructed WIMP masses
 with their 1$\sigma$ statistical uncertainties
 on the $\mchi - |\frmp|^2$ plane:
\includefigure{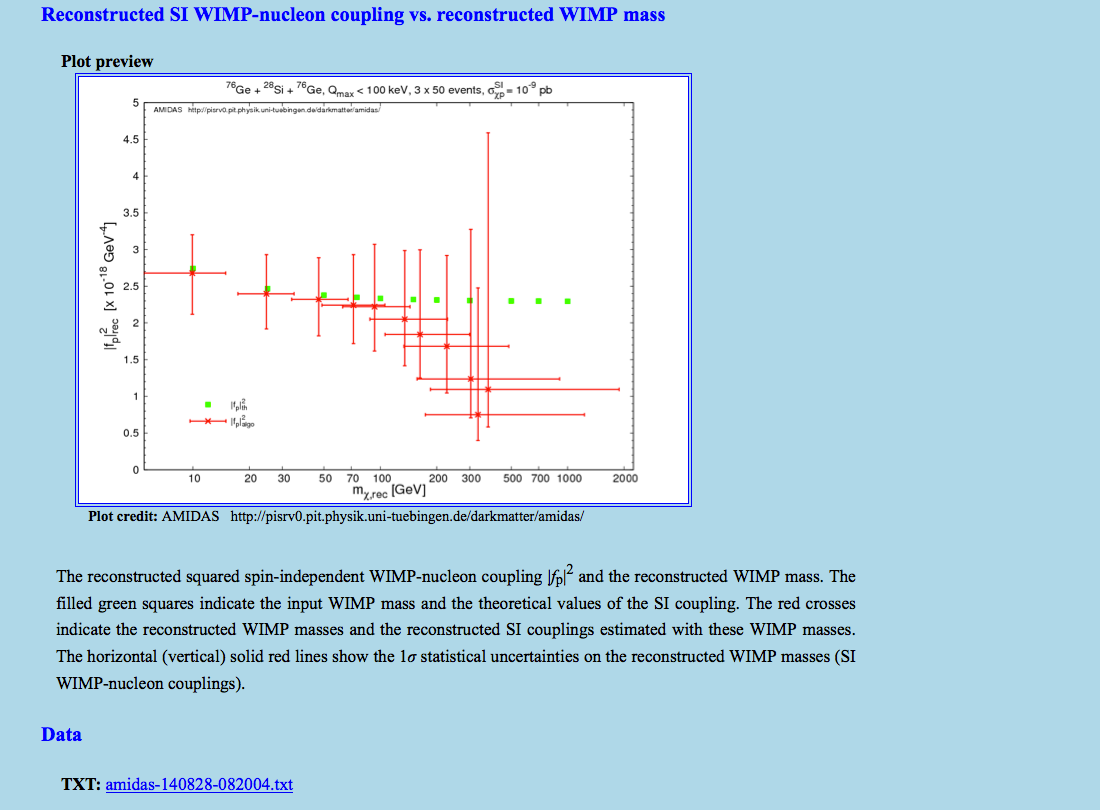}
 Note that
 here a reconstructed WIMP mass
 has been used
 (see Sec.~2.2.2).

 \subsubsection{Determinations of ratios
                between different WIMP--nucleon couplings/cross sections}

 As discussed in Sec.~2.4.4,
 for simulations of reconstructing ratios
 between different WIMP--nucleon couplings/cross sections
 \cite{DMDDranap},
 one needs to decide
 either fix the input WIMP mass
 or fix the input ratio of two SD WIMP--nucleon couplings,
 $\armn / \armp$.
 For each case,
 three plots of the ratios of
 $\armn / \armp$,
 $\sigmapSD / \sigmapSI$
 and $\sigmanSD / \sigmapSI$
 will be given as output results separately.

 For the case with a fixed WIMP mass,
 the reconstructed ratios
 with the 1$\sigma$ statistical uncertainties
 will be given
 as functions of the input $\armn / \armp$ ratio:
%
\includefigure{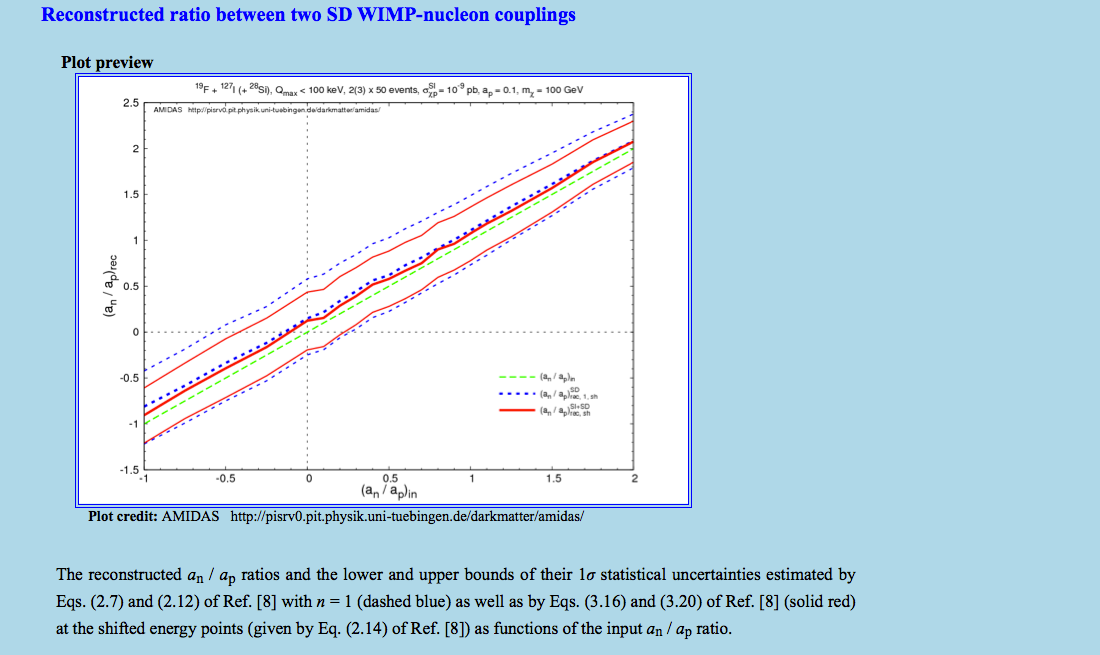}
\includefigure{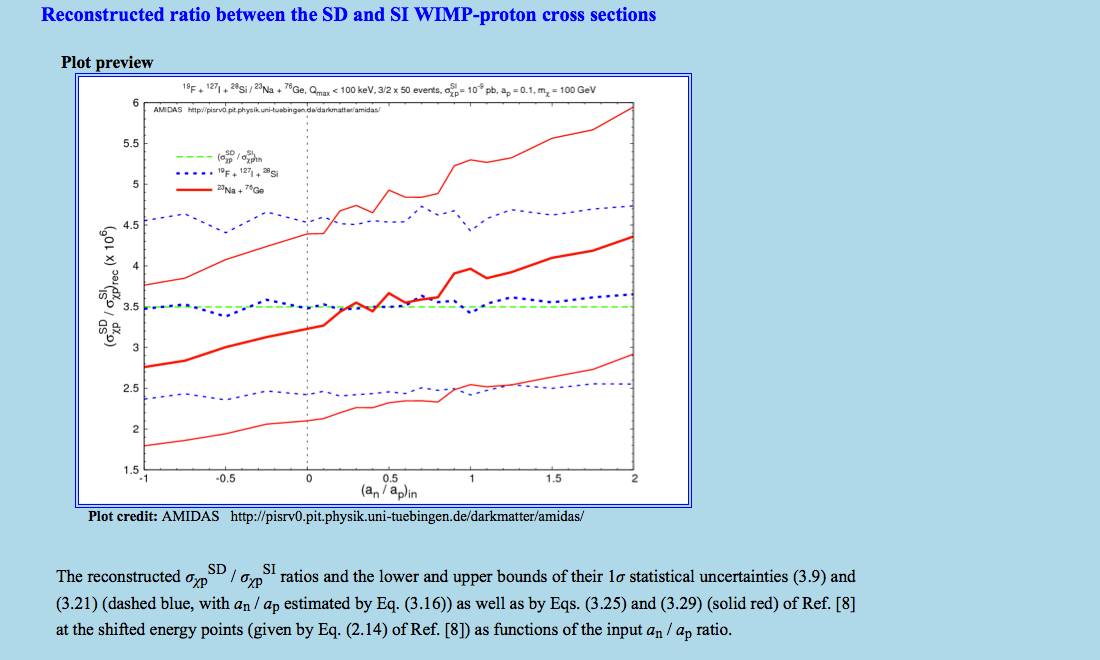}
\includefigure{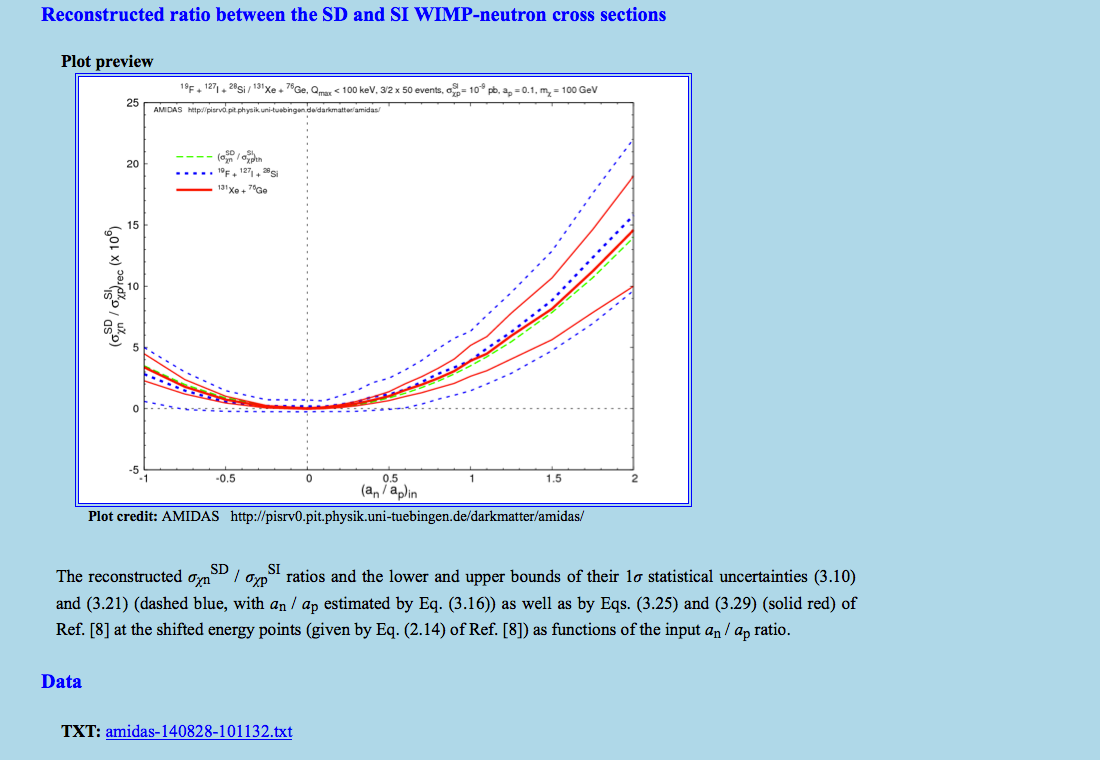}

 In contrast,
 for the case with a fixed $\armn / \armp$ ratio,
 the reconstructed ratios
 with the 1$\sigma$ statistical uncertainties
 will be given
 as functions of the input WIMP mass
 between 10 GeV and 1 TeV:
\includefigure{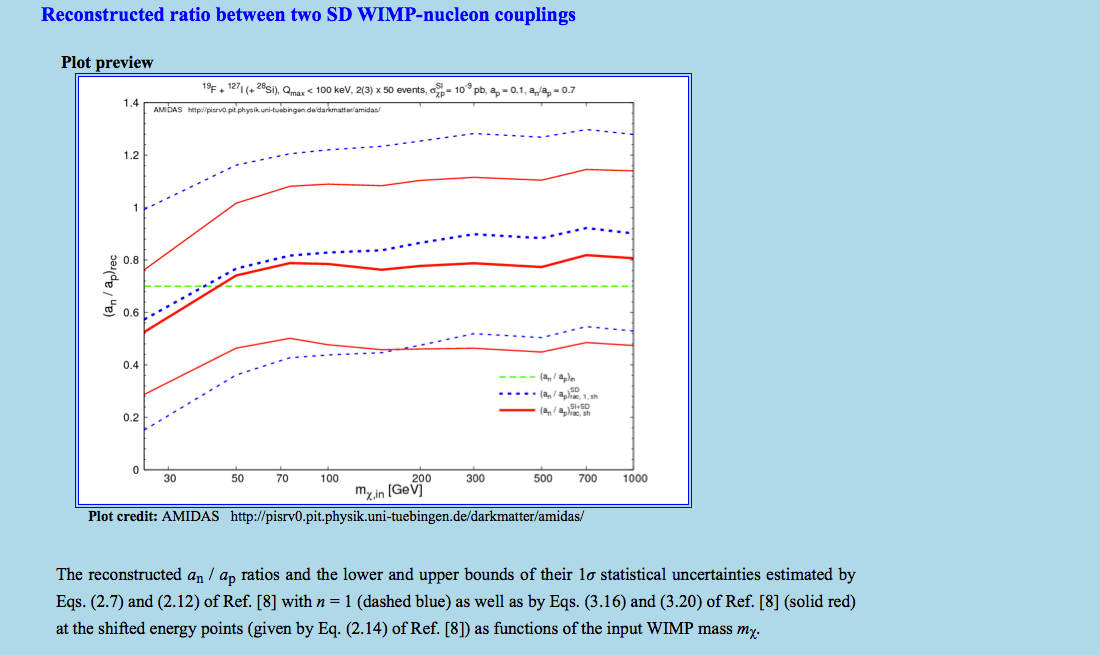}
\includefigure{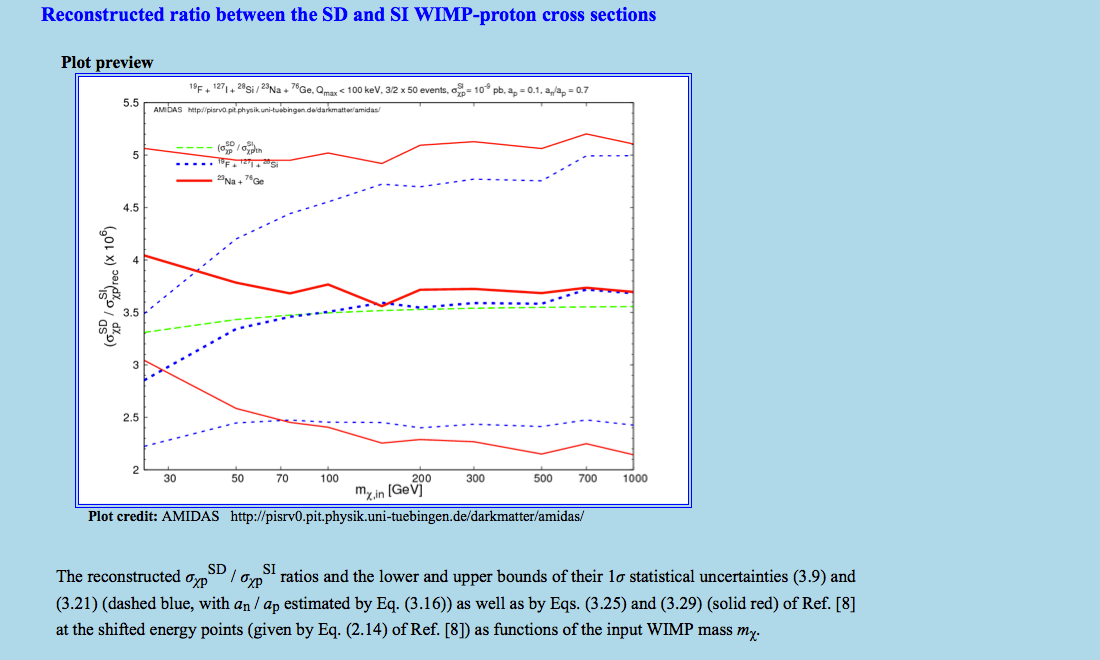}
\includefigure{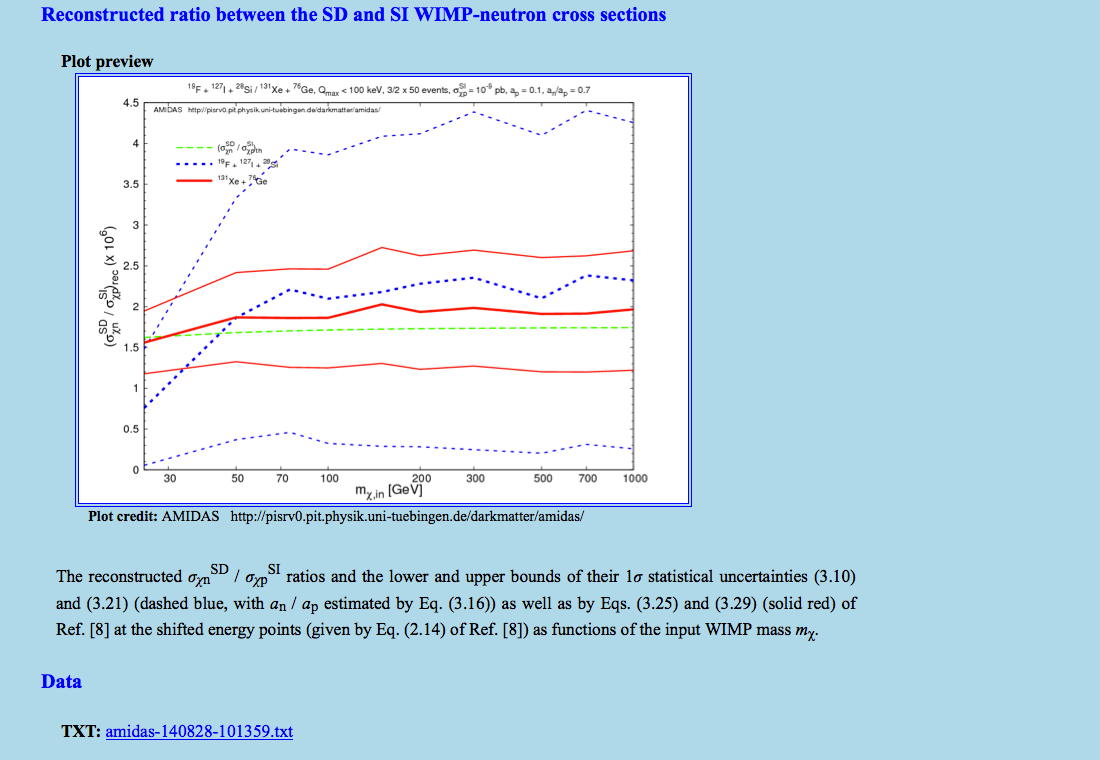}
\section{Analyzing real/pseudo data}

 The probably most important and useful design
 of the \amidas\ package and website is
 the ability of
 analyzing user--uploaded real/pseudo data set(s)
 recorded in direct DM detection experiments
 {\em without} modifying the source code.

 In this section,
 we describe
 the preparation of data files
 and the uploading/analyzing procedure
 on the \amidas\ website.
 A sample file for the uploaded data sets
 can be downloaded from the \amidas\ website.

\subsection{Preparing data set(s)}

 As mentioned above,
 on the \amidas\ website users can find and download
 a sample file for the uploaded data sets.
 Note that,
 for comments
 a ``0'' (zero) {\em has to be used} at the beginning,
 and all words in the comment lines
 must be connected by,
 e.g.~``\_'' (underscores).
 For instance,
\tabt{Example: an uploaded data file
      \label{code:uploaded_data_file}}
\verb# 0  m_[chi]_=_100_GeV                     # \\
\verb# 0  sigma_[chi,_p]^SI_=_1e-9_pb           # \\
\verb# 0  ......                                # \\
\verb#           :                              # \\
\verb#           :                              # \\
\verb#                                          # \\
\verb#   1 dataset, 58 events, 1628.565 kg-day: # \\
\verb#       1          1       0.068  keV      # \\
\verb#       1          2       8.325  keV      # \\
\verb#                  :                       # \\
\verb#                  :                       # \\
\verb#                                          # \\
\verb#   2 dataset, 48 events, 1628.565 kg-day: # \\
\verb#       2          1      11.743  keV      # \\
\verb#       2          2       1.824  keV      # \\
\verb#                  :                       # \\
\verb#                  :                       # \\
\verb#                                          # \\
\tabb
 Note that,
 as shown in the above example,
 it is {\em unnecessary}
 to order the generated/recorded recoil energies
 ascendingly or descendingly
 in your uploaded data file(s).
 The \amidas\ package will order the events in each data set
 after reading these events.
 However,
 the generated/recorded events
 with different target nuclei
 must be saved in {\em separate} files.

\subsection{Uploading data file(s)}

 Once users have chosen the data type
 as ``real (pseudo--) data''
 (Sec.~2.4.1),
 in the table for the experimental setup
 there will be one column
 for uploading users' data file(s)
 (cf.~Figure shown in Sec.~3.7.1).
\includefigure{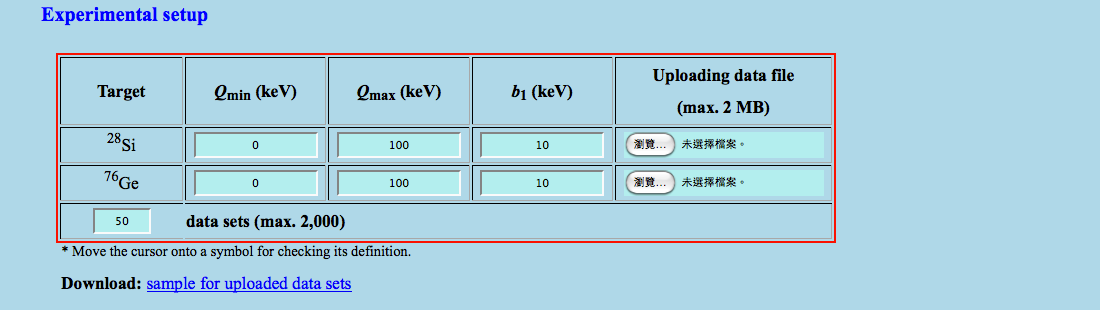}
 Users can upload their data file(s) as usual.
 Note only that
 the maximal size of {\em each} uploaded file is {\em 2 MB}.
 In addition,
 the numbers of data sets
 in {\em all} uploaded files
 must be {\em equal} or
 set as the {\em smallest} one%
\footnote{
 Events in the ``extra'' data sets
 will however be neglected
 in the analyses.
}.

 After that
 one or more data files
 have been uploaded and saved successfully,
 one extra column will appear
 in the experimental setup table
 to show the {\em original} name(s) of
 the uploaded data file(s).
\includefigure{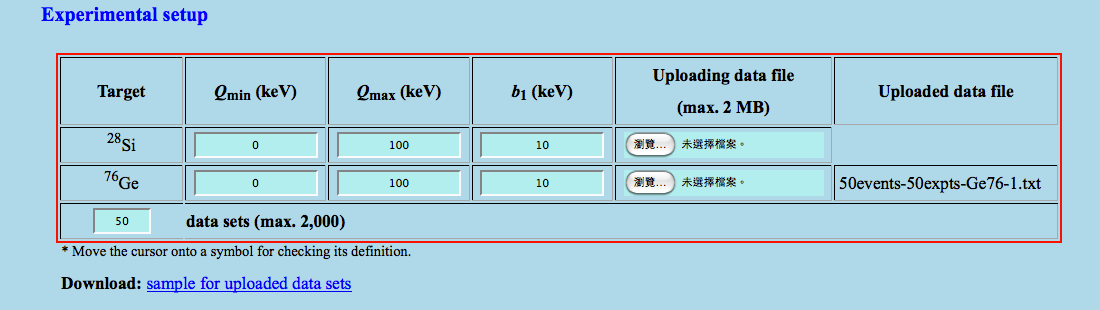}
 Users can check
 whether the correct files have been uploaded
 for the corresponding targets and,
 if necessary,
 upload the correct file(s) once again.

\subsection{Analyzing uploaded data}

 As for simulations,
 after giving all the required information
 for the aimed analysis,
 users have one more chance to check their choices and
 the {\em original} name(s) of their data file(s),
 modify some of them and
 replace the uploaded data file(s),
 and then resubmit the whole setup,
 before they click the ``\verb+Data analysis start+'' button
 to start the \amidas\ program.
\includefigure{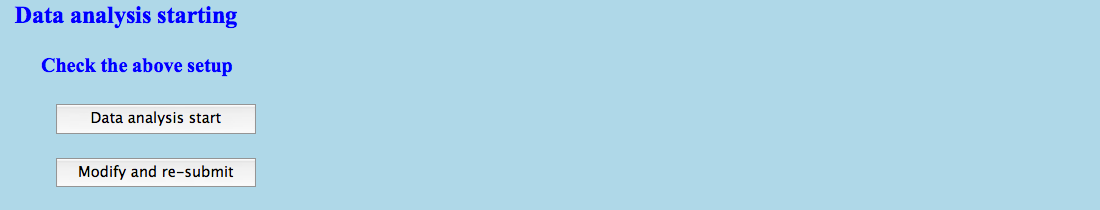}
 Remind that,
 in case that any required datum or {\em data file} is missed,
 this omission will be detected
 automatically after the (re)submission and
 users will be reminded of that
 with a {\em red} block around the options/table.
 Note that,
 while  {\em all data} in a table
 with missed information
 will be {\em reset} to the default values
 and should therefore be checked and modified once again,
 users {\em only} need to upload
 the {\em missed} data file(s)
 and/or the {\em replacement(s)}.

 Once all the required data and uploaded file(s)
 have been checked,
 users have only to click
 the ``\verb+Data analysis start+'' button
 and wait for the analyzed results for a while.

\subsection{Output results (tables)}

 Analyzed results will be presented
 in forms of table(s) and occasionally of plot(s).
 In order to let users
 understand the output results more clearly and
 use them more conveniently,
 each output plot or table
 will be accompanied with a short description.

 Meanwhile,
 the original TXT file(s) of the reconstructed results
 with users' personal experimental setup
 will also be given and downloadable on the website.
 Remind that
 it would be very grateful that
 a credit of the \amidas\ package and website
 could be given for using the output results.

 Below
 we give examples of output tables
 for data analyses of reconstructions of different WIMP properties.
 Detailed further analyses and uncertainty estimations
 can see Refs.~\cite{AMIDAS-CYGNUS2011, AMIDASbg-DSU2011}.
 The default \amidas\ simulation setup
 shown in Secs.~2.2 to 2.4 and Secs.~3.2 to 3.7
 has been generally used
 for generating pseudo data sets,
 with one common exception that
 the total number of WIMP--signal events
 is only 50 on average
 (Sec.~3.7.1).
 On the other hand,
 as the
 priorly required assumption
 in our data analyses
 \cite{DMDDmchi, DMDDfp2, DMDDranap},
 the elastic nuclear form factors
 for the SI and SD WIMP cross sections
 are set as the default \amidas\ options
 shown in Sec.~3.5.1.
 For each target nucleus,
 50 data sets
 accumulated in one uploaded file
 have been analyzed
 (Sec.~4.2).

 \subsubsection{Determination of the WIMP mass}

 By using two data sets
 with one light and one heavy target nuclei,
 one can reconstruct the WIMP mass directly
 \cite{DMDDmchi}:
\includefigure{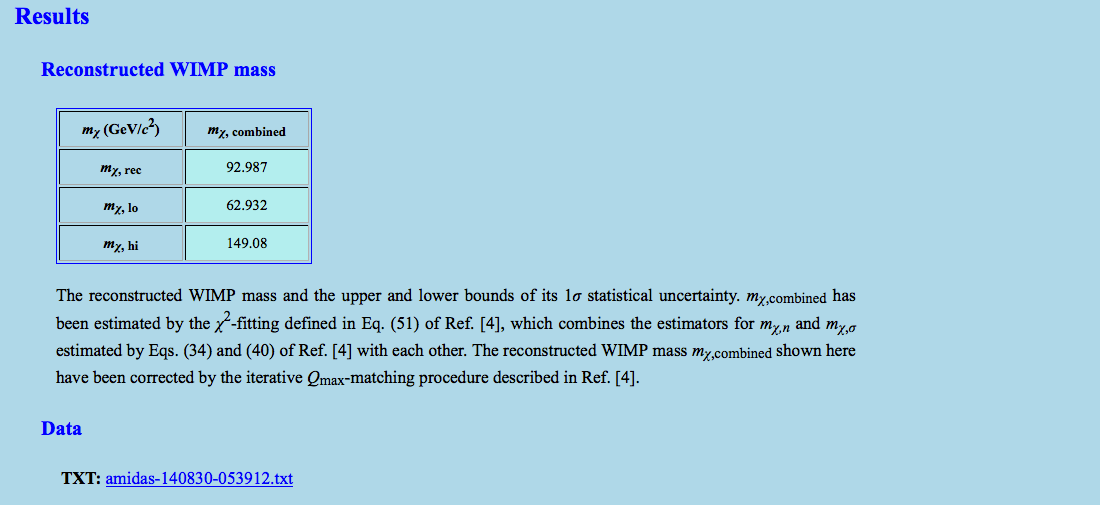}
 Remind that
 the input WIMP mass
 for generating the analyzed data events
 is $m_{\chi, {\rm in}} = 100$ GeV.

 \subsubsection{Estimation of the SI WIMP--nucleon coupling}

 By using two or three data sets
 and assuming the local Dark Matter density
 (the frequently used value of $\rho = 0.3~{\rm GeV/}c^2{\rm /cm^3}$
  has been adopted here),
 one can estimate the (squared) SI WIMP--nucleon coupling
 \cite{DMDDfp2}:
\includefigure{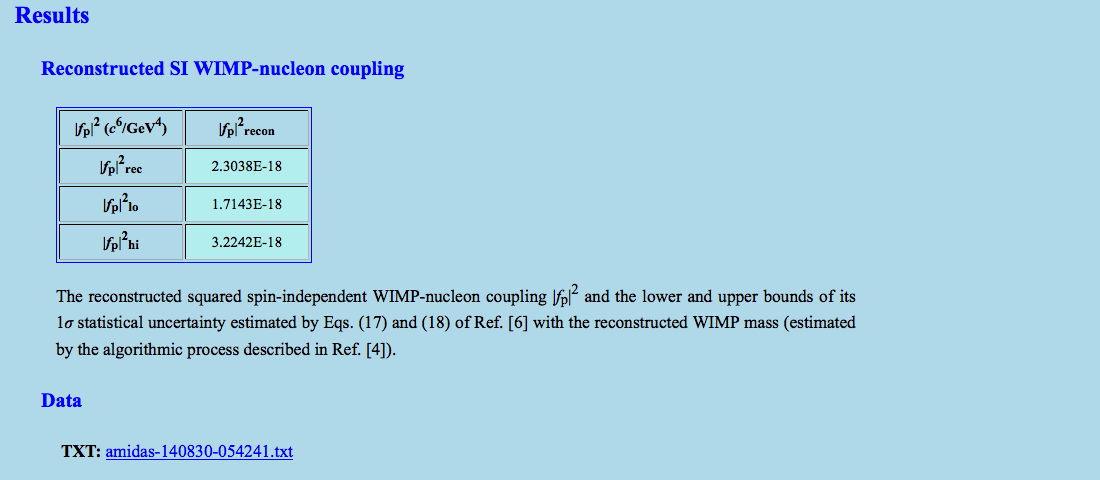}
 For readers' reference,
 the theoretical value of
 the squared SI WIMP--nucleon coupling
 for a WIMP mass of $\mchi = 100$ GeV
 and a SI WIMP--proton cross section of $\sigmapSI = 10^{-9}$ pb
 is $|\frmp|_{\rm th}^2 = 2.3344 \times 10^{-18}~{\rm GeV}^{-4}$,
 i.e.~$|\frmp|_{\rm th} = 1.5279 \times 10^{-9}~{\rm GeV}^{-2}$.

 \subsubsection{Determinations of ratios
                between different WIMP--nucleon couplings/cross sections}

 By using combinations of data sets
 with target nuclei having non--zero total nuclear spins,
 one can reconstruct different ratios
 between SD and SI WIMP--nucleon couplings/cross sections
 \cite{DMDDranap}:
\includefigure{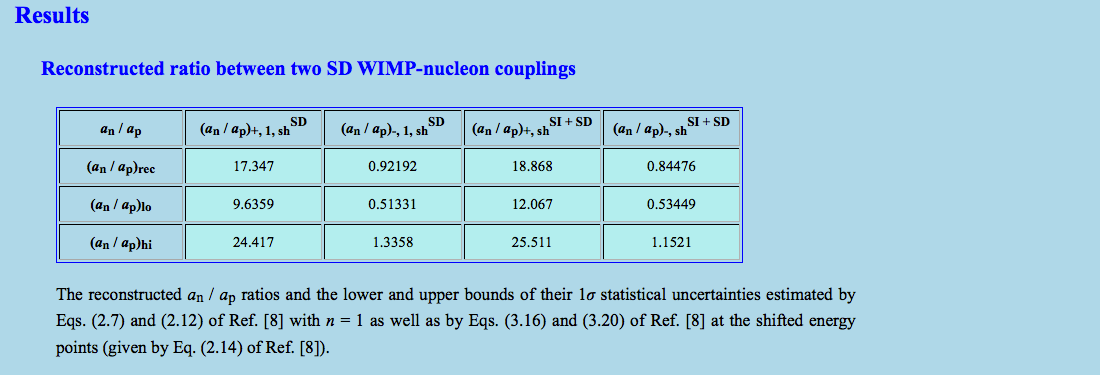}
\includefigure{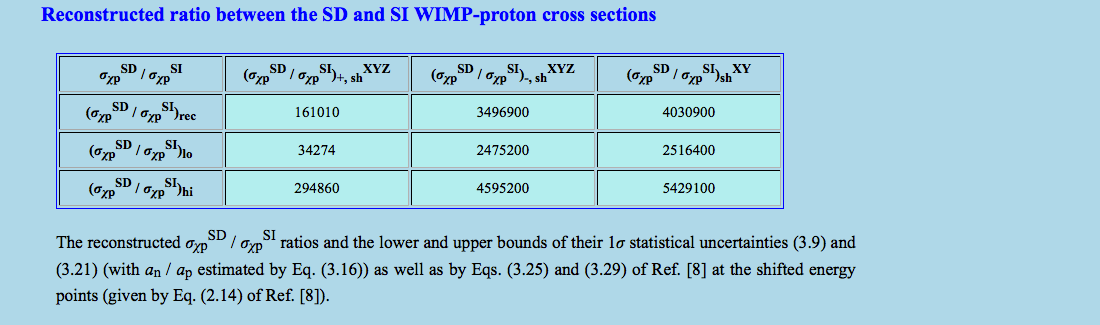}
\includefigure{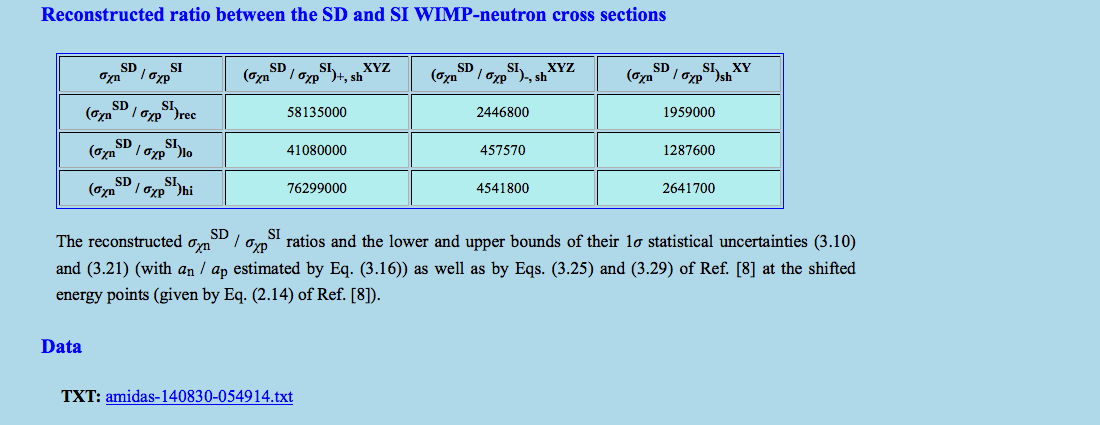}
 Remind that,
 as discussed in Ref.~\cite{DMDDranap},
 for reconstructing the ratio
 between two SD WIMP--nucleon couplings
 and thus the ratios
 between the SD and SI WIMP--nucleon cross sections
 with a third target nucleus,
 our model--independent methods
 will provide two (plus and minus) solutions. 
 Nevertheless,
 for our chosen combination of detector materials
 ($\rmXA{F}{19} + \rmXA{I}{127}$),
 one could 
 find clearly that
 the minus ($-$) solution should be the reasonable one.

 For readers' reference,
 the input ratio between two SD WIMP--nucleon couplings
 for generating the analyzed data events
 is $\abrac{\armn / \armp}_{\rm in} = 0.7$.
 And,
 for a WIMP mass of $\mchi = 100$ GeV
 and a SI WIMP--proton cross section of $\sigmapSI = 10^{-9}$ pb,
 the ratios between SD and SI WIMP--nucleon cross sections are
 $\abrac{\sigmapSD / \sigmapSI}_{\rm th} = 3.4967 \times 10^6$
 and
 $\abrac{\sigmanSD / \sigmapSI}_{\rm th} = 1.7134 \times 10^6$,
 respectively.

\section{Bayesian analyses}

 In Ref.~\cite{DMDDf1v-Bayesian},
 we applied Bayesian analysis technique
 to the reconstruction of the one--dimensional
 velocity distribution function of Galactic WIMPs.
 This newest development is released
 for both of simulation and real/pseudo data analysis
 on the \amidas\ website.

\subsection{Fitting velocity distribution function of halo WIMPs}

 As the most crucial part in Bayesian analyses,
 users need to choose
 one fitting one--dimensional WIMP velocity distribution function.
 In the \amidasii\ package,
 we offer so far five analytic forms
 for the fitting velocity distribution.
 Meanwhile,
 users also have the option to define (by uploading/typing)
 an analytic form of
 their favorite fitting one--dimensional velocity distribution:
\includefigure{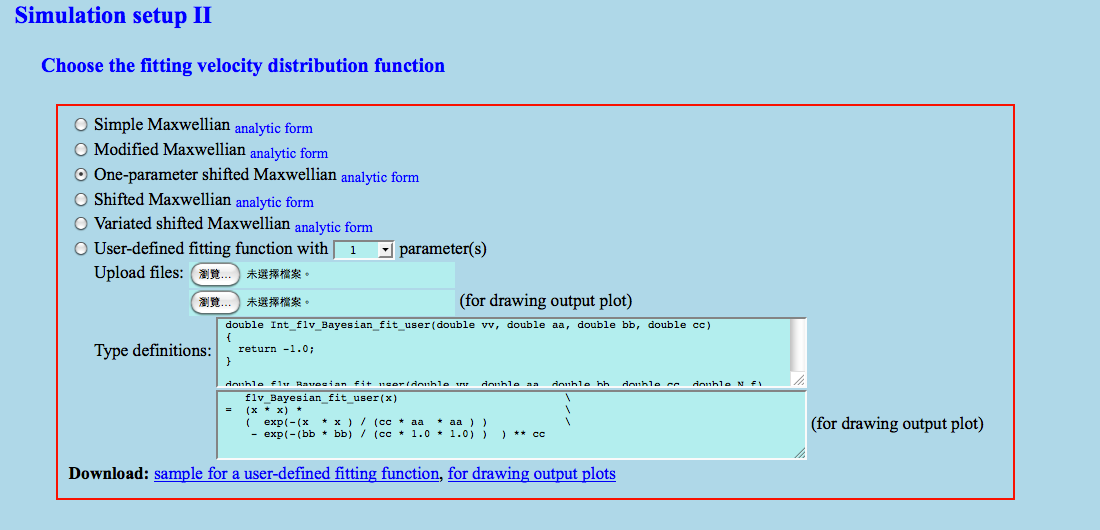}
 Note here that,
 firstly,
 for checking the analytic forms of
 the intrinsically defined fitting velocity distributions,
 users can hover the cursor onto ``\verb+analytic form+'';
 users can also click the ``\verb+analytic form+''
 to open a new webpage
 with more detailed information and useful references.
 Secondly,
 as reminded on the website,
 the second uploaded file/typing area is
 {\em only} for drawing output plot(s) of
 the Bayesian reconstructed one--dimensional
 WIMP velocity distribution function.
 Additionally,
 for defining users own fitting velocity distribution,
 the number of the fitting parameters
 must be set simultaneously on the website%
\footnote{
 So far the \amidasii\ package
 allows to fit with/scan maximal {\em three} fitting parameters.
}.

 In this subsection,
 we give first the definitions of
 the five default fitting one--dimensional WIMP velocity distribution functions
 in the \amidasii\ package.
 Then we will describe
 how to (modify these definitions to) define
 user's own fitting velocity distribution.

\subsubsection{Default fitting one--dimensional WIMP velocity distribution functions}

 Except of three default
 one--dimensional WIMP velocity distributions
 for generating WIMP signal events
 (see Sec.~3.4.1),
 in Ref.~\cite{DMDDf1v-Bayesian}
 two useful variations of
 the shifted Maxwellian velocity distribution
 have also been considered.
 Hence,
 so far
 users have five options for
 the fitting one--dimensional WIMP velocity distribution
 defined intrinsically in the \amidasii\ package
 \cite{DMDDf1v-Bayesian}:
\beenr
\item
 the simple Maxwellian velocity distribution function $f_{1, \Gau}(v)$
 \cite{SUSYDM96};
\item
 the modified Maxwellian velocity distribution function $f_{1, \Gau, k}(v)$
 \cite{Lisanti10, YYMao12, YYMao13, Kuhlen13};
\item
 the one--parameter shifted Maxwellian velocity distribution function
\beq
    f_{1, \sh, v_0}(v)
 =  \frac{1}{\sqrt{\pi}} \afrac{v}{v_0 \ve}
    \bBig{  e^{-(v - \ve)^2 / v_0^2}
          - e^{-(v + \ve)^2 / v_0^2}  }
\~,
\label{eqn:f1v_sh_v0}
\eeq
 with
\beq
    \ve
 =  1.05\~v_0
\label{eqn:v_e_ave}
\eeq
 is the time-–averaged
 Earth's velocity in the Galactic frame
 \cite{DMDDf1v-Bayesian};
\item
 the shifted Maxwellian velocity distribution function $f_{1, \sh}(v)$
 \cite{SUSYDM96, Lewin96};
\item
 the variated shifted Maxwellian velocity distribution function
 \cite{DMDDf1v-Bayesian}
\beq
    f_{1, \sh, \Delta v}(v)
 =  \frac{1}{\sqrt{\pi}} \bfrac{v}{v_0 \abrac{v_0 + \Delta v}}
    \cbigg{  e^{-\bbrac{v - \abrac{v_0 + \Delta v}}^2 / v_0^2}
           - e^{-\bbrac{v + \abrac{v_0 + \Delta v}}^2 / v_0^2}  }
\~,
\label{eqn:f1v_sh_Dv}
\eeq
 where
\beq
         \Delta v
 \equiv  \ve - v_0
\label{eqn:Delta_v}
\eeq
 is the difference between $v_0$ and
 the time--dependent Earth's velocity in the Galactic frame $\ve(t)$.
\end{enumerate}
\subsubsection{User--defining the fitting one--dimensional WIMP velocity distribution function}

 For defining users' favorite fitting velocity distribution function
 to include into the \amidasii\ package,
 one has to distinguish two cases:
 an analytic form for
 the {\em integral} over the fitting velocity distribution
 {\em exists} or {\em not}.

 As an example,
 we give here
 the \amidasii\ code for
 the (integral over the)
 simple Maxwellian velocity distribution function
 for the Bayesian fitting process.
 The codes for the (integral over the)
 other intrinsically defined
 ``fitting'' one--dimensional velocity distribution functions
 will be given in Appendix C.1
 for users' reference.

 For the case that
 an analytic form for
 the integral over the fitting velocity distribution
 exists,
 users have to define this integral
 with the name of ``\verb+Int_f1v_Bayesian_fit_user+''
 {\em first}:
\code{``Defined'' integral over the fitting simple Maxwellian velocity distribution
      \boldmath$\int f_{1, \Gau}(v) \~ dv$
      \label{code:Int_f1v_Bayesian_fit_user}}
\verb# double Int_f1v_Bayesian_fit_user(double vv, double aa, double bb, double cc) # \\
\verb# {                                                                            # \\
\verb#   return                                                                     # \\
\verb#       erf(vv / aa)                                                           # \\
\verb#     - (2.0 / sqrt(M_PI)) *                                                   # \\
\verb#       (vv / aa)          *                                                   # \\
\verb#       exp(-(vv * vv) / (aa * aa) );                                          # \\
\verb# }                                                                            # \\
\tabb
 And then one can define the fitting velocity distribution function
 with the name of \\ ``\verb+f1v_Bayesian_fit_user+''
 {\em itself}:
\code{Fitting simple Maxwellian velocity distribution
      \boldmath$f_{1, \Gau}(v)$
      \label{code:f1v_Bayesian_fit_user}}
\verb# double f1v_Bayesian_fit_user(double vv, double aa, double bb, double cc, double N_f) # \\
\verb# {                                                                                    # \\
\verb#   return                                                                             # \\
\verb#     (  (4.0 / sqrt(M_PI))                   *                                        # \\
\verb#        ( (vv * vv) / (aa * aa * aa) ) / v_U *                                        # \\
\verb#        exp(-(vv * vv) / (aa * aa) )          ) /                                     # \\
\verb#     (  Int_f1v_Bayesian_fit_user(v_max, aa, bb, cc)                                  # \\
\verb#      - Int_f1v_Bayesian_fit_user(0.0,   aa, bb, cc)  );                              # \\
\verb# }                                                                                    # \\
\tabb
 Note that,
 firstly,
 the function \verb+Int_f1v_Bayesian_fit_user+ for the integral
 has {\em always four} parameters:
 the velocity \verb+vv+ and
 {\em three} fitting parameters \verb+aa+, \verb+bb+ and \verb+cc+.
 It doesn't matter
 how many fitting parameters (only one or two or all three)
 are actually used.
 Meanwhile,
 the function \verb+f1v_Bayesian_fit_user+
 for the velocity distribution itself
 has always {\em five} parameters:
 except of \verb+vv+, \verb+aa+, \verb+bb+ and \verb+cc+,
 an extra parameter \verb+N_f+
 has to be included,
 which plays the role of the normalization constant
 needed {\em only} for the case that
 an analytic form for
 the integral over the fitting velocity distribution
 does {\em not} exist
 (see below).

 On the other hand,
 once it is too complicated or even impossible to give
 an analytic form for
 the integral over the fitting velocity distribution,
 one can {\em first} define it as ``$-1.0$'':
\code{``Undefined'' integral over the fitting simple Maxwellian velocity distribution
      \boldmath$\int f_{1, \Gau}(v) \~ dv$
      \label{code:Int_f1v_Bayesian_fit_user_N_f}}
\verb# double Int_f1v_Bayesian_fit_user(double vv, double aa, double bb, double cc) # \\
\verb# {                                                                            # \\
\verb#   return -1.0;                                                               # \\
\verb# }                                                                            # \\
\tabb
 And then one has to define the {\em main} ({\em velocity--dependent}) part of
 the fitting velocity distribution function as:
\code{Fitting simple Maxwellian velocity distribution
      \boldmath$f_{1, \Gau}(v)$
      \label{code:f1v_Bayesian_fit_user_N_f}}
\verb# double f1v_Bayesian_fit_user(double vv, double aa, double bb, double cc, double N_f) # \\
\verb# {                                                                                    # \\
\verb#   return                                                                             # \\
\verb#     (  ( (vv * vv) / (aa * aa * aa) ) *                                              # \\
\verb#        exp(-(vv * vv) / (aa * aa) )    ) /                                           # \\
\verb#     N_f;                                                                             # \\
\verb# }                                                                                    # \\
\tabb
 Then
 the \amidasii\ package will estimate
 the {\em fitting--parameter dependent} normalization constant $N_f$
 {\em point--by--point} numerically
 during the Bayesian fitting (scanning) procedure.

 Additionally,
 users have to set the symbol and unit of {\em each} fitting parameter,
 which will be used in the output result files and plots,
 with the names of ``\verb+ a_Bayesian_Symbol+''
 and ``\verb+ a_Bayesian_Unit+''
 in the {\em same} uploaded file:
\code{Symbol and unit of the (first) fitting parameter
      \label{code:symbol_unit_Bayesian_fit_user}}
\verb# a_Bayesian_Symbol = "v_0";  # \\
\verb# a_Bayesian_Unit   = "km/s"; # \\
\tabb

 Finally,
 as for drawing
 the generating velocity distribution
 (Sec.~3.4),
 an {\em extra} file is required
 for defining the ``fitting''
 one--dimensional velocity distribution function
 {\em itself}
 (the integral over it is not needed anymore)
 for drawing the Bayesian reconstructed/fitted velocity distribution
 in the output plots.
 Remind that,
 since the {\tt Gnuplot} package
 has been adopted in \amidas\
 for drawing output plots,
 the velocity distribution function
 defined in this file
 must be written in the syntax of {\tt Gnuplot}
 with the name of ``\verb+f1v_Bayesian_fit_user(x)+''.
 Below we give our definition for
 the simple Maxwellian velocity distribution
 with and without the known analytic form of
 the integral over it
 as examples.
 The definitions for the other intrinsically defined
 fitting one--dimensional WIMP velocity distribution function
 are given in Appendix C.1
 for users' reference.

 Once the analytic form of
 the integral over the velocity distribution function is known,
 we have
\code{Fitting simple Maxwellian velocity distribution
      \boldmath$f_{1, \Gau}(v)$
      (for Gnuplot)
      \label{code:f1v_Bayesian_fit_x_user}}
\verb#   M_PI = 3.141593                     # \\
\verb#                                       # \\
\verb#    f1v_Bayesian_fit_user(x)         \ # \\
\verb# =  (4.0 / sqrt(M_PI))           *   \ # \\
\verb#    ( (x * x) / (aa * aa * aa) ) *   \ # \\
\verb#    exp(-(x * x) / (aa * aa) )         # \\
\tabb
 In contrast,
 without a known analytic form of
 the integral over the velocity distribution function,
 one can simply define
\code{Fitting simple Maxwellian velocity distribution
      \boldmath$f_{1, \Gau}(v)$
      (for Gnuplot)
      \label{code:f1v_Bayesian_fit_x_user_N_f}}
\verb#    f1v_Bayesian_fit_user(x)         \ # \\
\verb# =  ( (x * x) / (aa * aa * aa) ) *   \ # \\
\verb#    exp(-(x * x) / (aa * aa) )       \ # \\
\tabb
 The value of the normalization constant
 needed ({\em inplicitly} in the \amidasii\ code) here
 (i.e.~$N_f$ in Code \ref{code:f1v_Bayesian_fit_user_N_f})
 will be estimated numerically
 by using the velocity distribution
 defined in Code \ref{code:f1v_Bayesian_fit_user_N_f}.%
\footnote{
 We set $N_f = 1$ at first
 and evaluate numerically the (reciprocal of the) integral over
 {\tt f1v\_Bayesian\_fit\_user(vv, aa, bb, cc, N\_f = 1.0)}
 by using the {\em Simpson's rule}.
}
 Remind that,
 firstly,
 ``\verb+\+'' (backslash) must be used
 in order to let the including of the definition(s) given in this file
 into the other intrinsic commands correctly.
 Secondly,
 the value of the parameter \verb+aa+
 will be given by the fitting result.
 Moreover,
 two sample files,
 one for the \amidas\ code and
 the other one for the {\tt Gnuplot} package,
 can be downloaded from the \amidas\ website.

\subsection{Distribution function for describing the statistical uncertainty}

 In this subsection,
 we discuss the second key factor
 required in Bayesian analyses:
 the distribution function for
 describing the statistical uncertainty
 on the analyzed/fitted data points,
 which will in turn be used in the likelihood function
 (for more details,
  see Ref.~\cite{DMDDf1v-Bayesian} and references therein).
 In the \amidasii\ package,
 we offer so far two analytic forms
 for the statistical--uncertainty distribution function.
 Meanwhile,
 users also have the option to define (by uploading/typing)
 an analytic form of
 their needed uncertainty distribution:
\includefigure{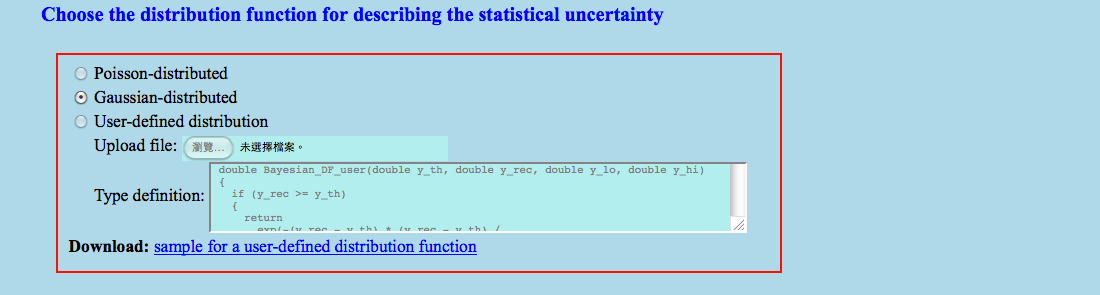}
 Note however here that,
 {\em only} the option for
 a ``Gaussian--distributed'' statistical uncertainty
 is considered for our Bayesian reconstruction of
 the one--dimensional WIMP velocity distribution function
 \cite{DMDDf1v-Bayesian}.
 Thus the other one option for
 a ``Poisson--distributed'' statistical uncertainty
 as well as
 the uploading/typing areas
 are ``unavailable'' (locked) currently.

 Nevertheless,
 as the users' guide,
 in this subsection,
 we still give first the definitions of
 the intrinsically defined
 uncertainty distribution functions
 in the \amidasii\ package.
 Then we will describe
 how to (modify this definition to) define
 user's own statistical--uncertainty distribution.

\subsubsection{Default statistical--uncertainty distribution functions}

 As the most commonly considered cases,
 in the \amidasii\ package
 we define two statistical--uncertainty distribution functions:
\beenr
\item
 Poisson--distributed
\beq
    {\rm Poi}(x_i, y_i; a_j, j = 1,~2,~\cdots,~N_{\rm Bayesian})
 =  \frac{f^{y_i}(x_i; a_j) \~ e^{-f(x_i; a_j)}}{y_i!}
\~,
\label{eqn:Bayesian_DF_Poi}
\eeq
 where $\abrac{x_i, y_i}$ for $i = 1,~2,~\cdots,~N$
 denote the $N$ analyzed/fitted data points
 and $f(x; a_j)$ is the theoretically predicted/fitting function
 with the $N_{\rm Bayesian}$ fitting parameters $a_j$,
 for $j = 1,~2,~\cdots,~N_{\rm Bayesian}$;
\item
 Gaussian--distributed
\beq
    {\rm Gau}(x_i, y_i, \sigma(y_i); a_j, j = 1,~2,~\cdots,~N_{\rm Bayesian})
 =  \frac{1}{\sqrt{2 \pi} \~ \sigma(y_i)} \~
    e^{-\bbig{y_i - f(x_i; a_j)}^2 / 2 \sigma^2(y_i)}
\~,
\label{eqn:Bayesian_DF_Gau}
\eeq
 where $\abrac{x_i, y_i \pm \sigma(y_i)}$ for $i = 1,~2,~\cdots,~N$
 denote the $N$ analyzed/fitted data points
 with the (measurement) uncertainties on $y_i$,
 $\sigma(y_i)$.
\end{enumerate}
\subsubsection{User--defining the statistical--uncertainty distribution function}

 As an example,
 we give here
 the \amidasii\ code for
 the (double--)Gaussian statistical--uncertainty distribution function
 used for our Bayesian reconstruction procedure
 \cite{DMDDf1v-Bayesian}
 with the name of ``\verb+Bayesian_DF_user+''.
 The code for the other intrinsically defined
 (i.e.~Poisson, so far)
 statistical--uncertainty distribution function
 will be given in Appendix C.2
 for users' reference.

\code{(Double--)Gaussian statistical--uncertainty distribution
      \boldmath${\rm Gau}(x_i, y_i, y_{i, {\rm (lo, hi)}};
                          y_{i, {\rm th}})$
      \label{code:Bayesian_DF_user}}
\verb# double Bayesian_DF_user(double y_th, double y_rec, double y_lo, double y_hi) # \\
\verb# {                                                                            # \\
\verb#   if (y_rec >= y_th)                                                         # \\
\verb#   {                                                                          # \\
\verb#     return                                                                   # \\
\verb#       exp(-(y_rec - y_th) * (y_rec - y_th)         /                         # \\
\verb#            (2.0 * (y_rec - y_lo) * (y_rec - y_lo))  ) /                      # \\
\verb#       (sqrt(2.0 * M_PI) * (y_rec - y_lo));                                   # \\
\verb#   }                                                                          # \\
\verb#                                                                              # \\
\verb#   else                                                                       # \\
\verb#   if (y_rec <  y_th)                                                         # \\
\verb#   {                                                                          # \\
\verb#     return                                                                   # \\
\verb#       exp(-(y_rec - y_th) * (y_rec - y_th)         /                         # \\
\verb#            (2.0 * (y_hi - y_rec) * (y_hi - y_rec))  ) /                      # \\
\verb#       (sqrt(2.0 * M_PI) * (y_hi - y_rec));                                   # \\
\verb#   }                                                                          # \\
\verb# }                                                                            # \\
\tabb
 Note here that
 we consider in the \amidasii\ package
 an {\em asymmetric} distribution of the statistical uncertainty.
 For the case that
 an analyzed/fitted data point is larger (smaller)
 than the theoretical value
 estimated from the fitting (WIMP velocity distribution) function,
 we take the 1$\sigma$ lower (upper) (statistical) uncertainty
 as the uncertainty 
 $\sigma(y_i)$
 in Eq.~(\ref{eqn:Bayesian_DF_Gau}).
 Remind that
 a sample file
 for the \amidas\ code
 can be downloaded from the \amidas\ website.

\subsection{Distribution function of each fitting parameter}

 In Bayesian analyses,
 one needs to give a probability distribution function
 for describing no/a prior knowledge
 about each fitting parameter.
 In the \amidasii\ package,
 we offer so far three probability distribution functions
 for each fitting parameter
 for describing the prior knowledge about it.
 Meanwhile,
 users also have the option to define (by uploading/typing)
 an analytic form of
 their favorite probability distribution:
\includefigure{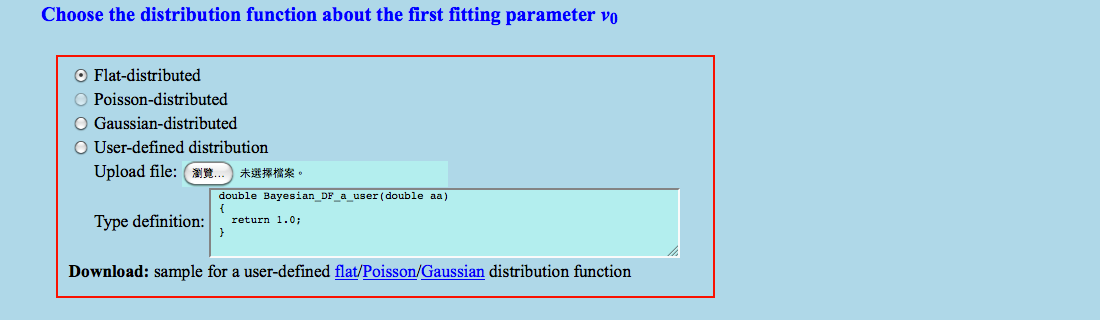}
 Note that,
 since the statistical uncertainties on
 the default fitting parameters
 used for our Bayesian reconstruction of $f_1(v)$
 are in principle {\em Gaussian},
 the option for
 a ``Poisson--distributed'' probability distribution function
 is ``unavailable'' (locked) here.
 It is however possible to use
 the intrinsically defined ``Poisson'' probability distribution
 (unlocked),
 once users choose to define
 their own fitting WIMP velocity distribution function
 (Sec.~5.1).

 In this subsection,
 we give first the definitions of
 the three intrinsically defined probability distribution functions
 for each fitting parameter
 in the \amidasii\ package.
 Then we will describe
 how to (modify these definitions to) define
 user's favorite probability distribution.

\subsubsection{Default distribution functions for the fitting parameters}

 As the most commonly considered cases,
 in the \amidasii\ package
 we define three probability distribution functions:
\beenr
\item
 Flat--distributed:
\beq
    {\rm p}_{i, {\rm flat}}(a_i)
 =  1
\~,
    ~~~~~~~~~~~~~~~~ 
    {\rm for~} a_{i, \rm min} \le a_i \le a_{i, \rm max},
\label{eqn:Bayesian_DF_a_flat}
\eeq
 where $a_{i, \rm min/max}$ denote
 the minimal and maximal bounds of the scanning interval of
 the fitting parameter $a_i$;
\item
 Poisson--distributed:
\beq
    {\rm p}_{i, {\rm Poi}}(a_i; \mu_{a, i})
 =  \frac{\mu_{a, i}^{a_i} \~ e^{-\mu_{a, i}}}{a_i!}
\~,
\label{eqn:Bayesian_DF_a_Poi}
\eeq
 with the expected value $\mu_{a, i}$ of
 the fitting parameter $a_i$;
\item
 Gaussian--distributed:
\beq
    {\rm p}_{i, {\rm Gau}}(a_i; \mu_{a, i}, \sigma_{a, i})
 =  \frac{1}{\sqrt{2 \pi} \~ \sigma_{a, i}} \~
    e^{-(a_i - \mu_{a, i})^2 / 2 \sigma_{a, i}^2}
\~,
\label{eqn:Bayesian_DF_a_Gau}
\eeq
 with the expected value $\mu_{a, i}$ of
 and the 1$\sigma$ uncertainty $\sigma_{a, i}$ on
 the fitting parameter $a_i$.
\end{enumerate}
\subsubsection{User--defining the distribution function for the fitting parameters}

 As an example,
 we give here
 the \amidasii\ code for
 the Gaussian probability distribution function
 used in our Bayesian analysis procedure
 \cite{DMDDf1v-Bayesian}
 with the name of ``\verb+Bayesian_DF_a_user+''.
 The codes for the other intrinsically defined
 probability distribution functions
 will be given in Appendix C.3
 for users' reference.

\code{Gaussian probability distribution for the fitting parameter
      \boldmath${\rm p}_{i, {\rm Gau}}(a_i; \mu_{a, i}, \sigma_{a, i})$
      \label{code:Bayesian_DF_a_user}}
\verb# double a_ave_Bayesian_user       = 2.0;             # \\
\verb# double sigma_a_ave_Bayesian_user = 0.5;             # \\
\verb#                                                     # \\
\verb# double Bayesian_DF_a_user(double aa)                # \\
\verb# {                                                   # \\
\verb#   return                                            # \\
\verb#     exp(-(aa - a_ave_Bayesian_user)    *            # \\
\verb#          (aa - a_ave_Bayesian_user)    /            # \\
\verb#          (2.0                       *               # \\
\verb#           sigma_a_ave_Bayesian_user *               # \\
\verb#           sigma_a_ave_Bayesian_user  )  ) /         # \\
\verb#     (sigma_a_ave_Bayesian_user * sqrt(2.0 * M_PI)); # \\
\verb# }                                                   # \\
\tabb
 Note that
 there is {\em only one} function parameter
 in the (user--defined) probability distribution function
 and the parameter name must be
 ``\verb+aa+'', ``\verb+bb+'' or ``\verb+cc+'' 
 corresponding to the first, second or the third fitting parameter.
 Remind also that,
 a sample file
 for the \amidas\ code
 can be downloaded from the \amidas\ website.

\subsubsection{Bounds of the scanning interval of each fitting parameter}

 For doing Bayesian fitting (scanning),
 users need to set the lower and upper bounds of the scanning range
 for each fitting parameter:
\includefigure{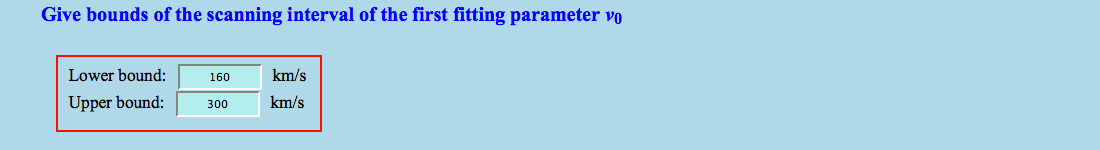}
 Moreover,
 once the probability distribution function
 of one fitting parameter
 has been chosen as,
 e.g.~Gaussian--distributed,
 it will be required automatically to set also
 the expected value and the standard deviation (uncertainty)
 of this parameter:
\includefigure{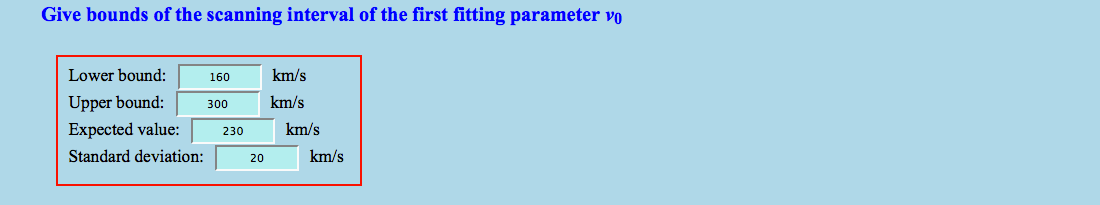}

 On the other hand,
 for the case that
 an user--defined probability distribution function
 has been used for one fitting parameter,
 users have also to give the notation and the unit of this parameter
 for the output files and plots:
\includefigure{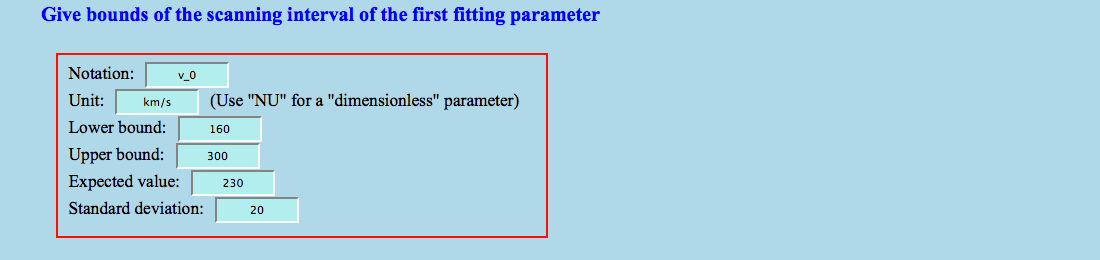}
 Note that,
 as remind on the website,
 for a ``dimensionless'' parameter,
 e.g.~the ratio $\ve / v_0$ and the power index $k$,
 it is required to use ``NU'' or ``nu'' in the input cell.

\subsection{Scanning the parameter space}

 Finally,
 as the last information for our Bayesian analysis procedure,
 users have the opportunity to choose
 different scanning method.

\subsubsection{Method for scanning the parameter space}

 So far in the \amidasii\ package
 we programmed {\em three} different scanning methods:
\includefigure{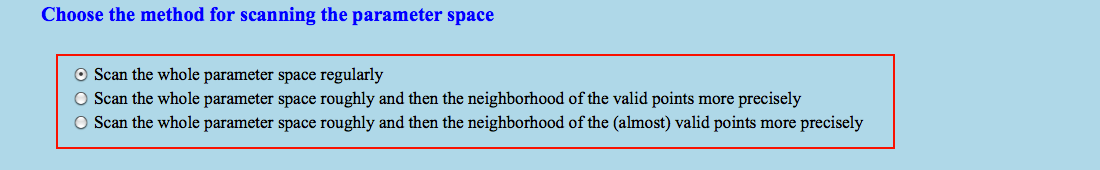}
 Here the second option
 ``scan the whole parameter space roughly and
 then the neighborhood of the valid points more precisely''
 means that
 \amidasii\ scans the whole parameter space
 to find the valid points;
 after that
 one valid point is found,
 \amidasii\ will scan {\em immediately} and {\em randomly}
 the neighborhood of this point
 for finding a better (more possible) point%
\footnote{
 Once a valid point,
 called the ``starting'' point with the value $a_i^{\ast}$,
 is found,
 we pick {\em randomly} one point in the range of
 \mbox{$\bbig{{\rm max}(a_i^{\ast} - \Delta_{a_i}, a_{i, {\rm min}}),
              {\rm min}(a_i^{\ast} + \Delta_{a_i}, a_{i, {\rm max}})}$},
 where
 $\Delta_{a_i} = \abrac{a_{i, {\rm max}} - a_{i, {\rm min}}} /$
 (the number of scanning points).
 We take this point if it is better and
 set this point as the next starting point
 for running this fine scanning process further;
 if this point is however {\em invalid},
 we through it away,
 use the old starting point to pick another checking point
 and run this fine scanning process.
 Note that,
 in the fine scanning process,
 we set {\em twice more} scanning points on one parameter--axis
 around the first starting point
 (see Sec.~5.4.2 for more details).
}.

 For a finer scanning,
 the third option above
 ``scan the whole parameter space roughly and
 then the neighborhood of the (almost) valid points more precisely''
 let \amidasii\ scan {\em immediately} and {\em randomly}
 the neighborhood of the point
 for finding a better point,
 once this point is valid or almost valid%
\footnote{
 For the Bayesian reconstruction
 of the one--dimensional WIMP velocity distribution function,
 we set the criterion of ``almost valid'' in \amidasii\ as that
 the value of the posterior probability distribution function
 at the {\em checking} point
 is {\em larger than 90\%} of the ``so far'' largest value,
 since we only need to find the point,
 in which the value of the posterior probability distribution function
 is the largest.
}.
\subsubsection{Number of scanning points for one fitting parameter}

 Users can set the number of scanning points
 for one fitting parameter:
\includefigure{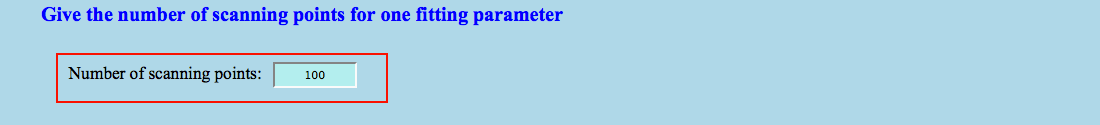}
 Note that
 this point number will be used for the regular scanning
 (option 1 in Sec.~5.4.1)
 as well as
 for the ``first step'' of the rough scanning
 (options 2 and 3).
 For the finer scanning
 around the (almost) valid points
 taken by the first step,
 {\em doubled} points
 (picked randomly around each (almost) valid point)
 for one parameter
 will be checked%
\footnote{
 This means that,
 once we set e.g.~100 scanning points for one fitting parameter,
 200 more points in one parameter--axis will be scanned
 in the neighborhood around each (almost) valid point.
 This means in turn that,
 for a fitting (velocity distribution) function
 with ``three'' parameters to scan,
 for ``each'' (almost) valid point,
 \amidasii\ needs to check $\sim$ 8,000,000 more points
 (and perhaps do 1 or 2 $\times$ $\sim$ 8,000,000 more times
  one--dimensional numerical integrations
  for the normalization constant of the fitting velocity distribution!).
}.
 Hence,
 the default number of scanning points
 for one fitting parameter
 needs to set different
 according to the choice of the scanning method.

 \subsection{Output results}

 In this subsection,
 we present particularly the output results
 of the (Bayesian) reconstruction of
 the one--dimensional WIMP velocity distribution function.
 Note that
 so far
 the \amidasii\ package can only offer
 the reconstruction results of
 the WIMP velocity distribution
 in form of plots.
 However,
 all important information,
 e.g.~the ($1\~(2)\~\sigma$ statistical uncertainty ranges of the)
 reconstructed characteristic Solar and Earth's Galactic velocities,
 will be provided in the TXT file(s).

 Remind that
 the default \amidas\ simulation setup
 shown in Secs.~2.2 to 2.4,
 Secs.~3.2 to 3.7 and Secs.~5.2 to 5.4
 has been generally used;
 the total number of WIMP--signal events
 is now set as 500 on average
 (Sec.~3.7.1)
 \cite{DMDDf1v}.

 \subsubsection{(Bayesian) reconstructed WIMP velocity distribution}

 For both of numerical simulations and data analyses,
 the \amidasii\ package provides one plot
 for the rough reconstructed velocity distribution
 \cite{DMDDf1v}
 and the ($1\~(2)\~\sigma$ statistical uncertainty bands of the)
 fitted velocity distribution function
 \cite{DMDDf1v-Bayesian}:
\includefigure{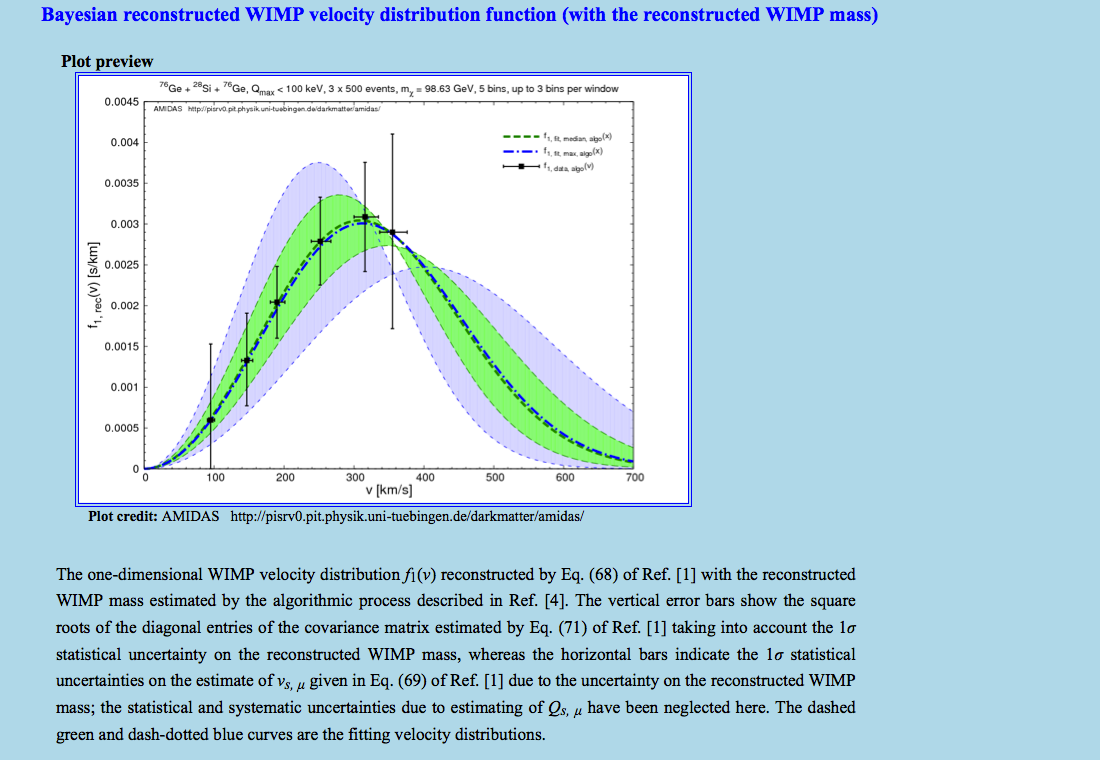}

 Meanwhile,
 for numerical simulations,
 two plots for users' comparisons will also be offered.
 One is only the comparison of
 the reconstructed rough velocity distribution
 with the input functional form
 \cite{DMDDf1v}:
\includefigure{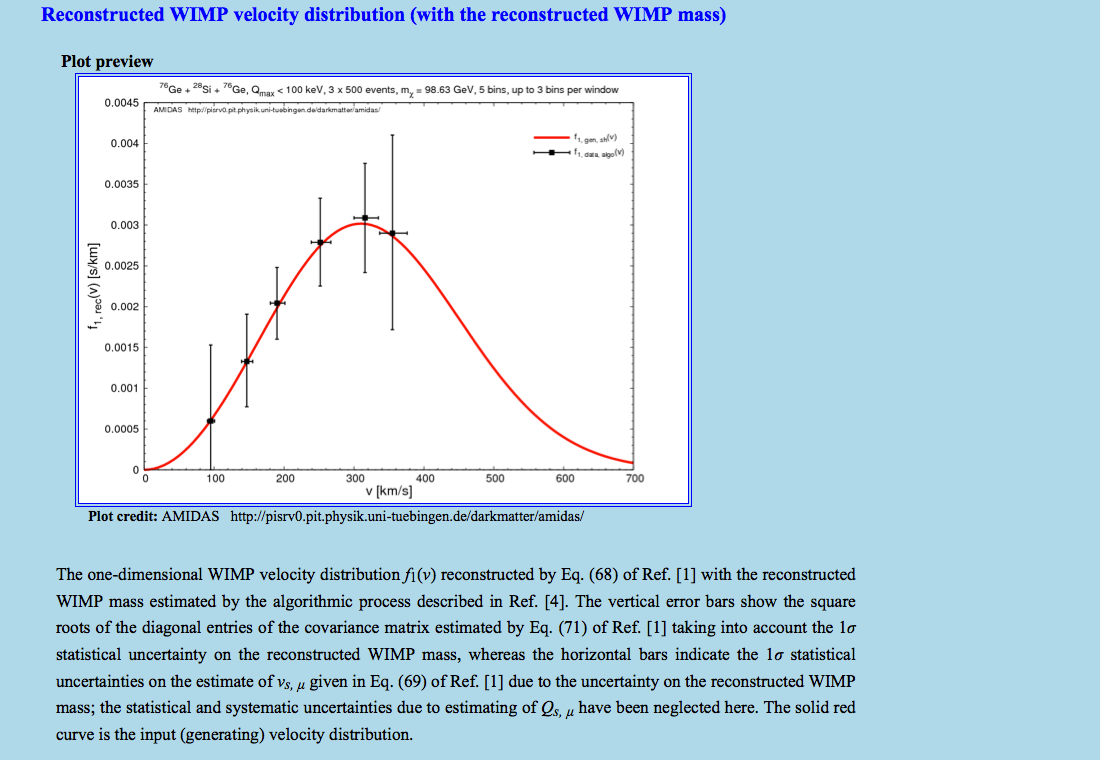}
 The other plot shows
 the input and
 the ($1\~(2)\~\sigma$ statistical uncertainty bands of the)
 fitted velocity distribution functions
 with the reconstructed rough distribution
 together:
\includefigure{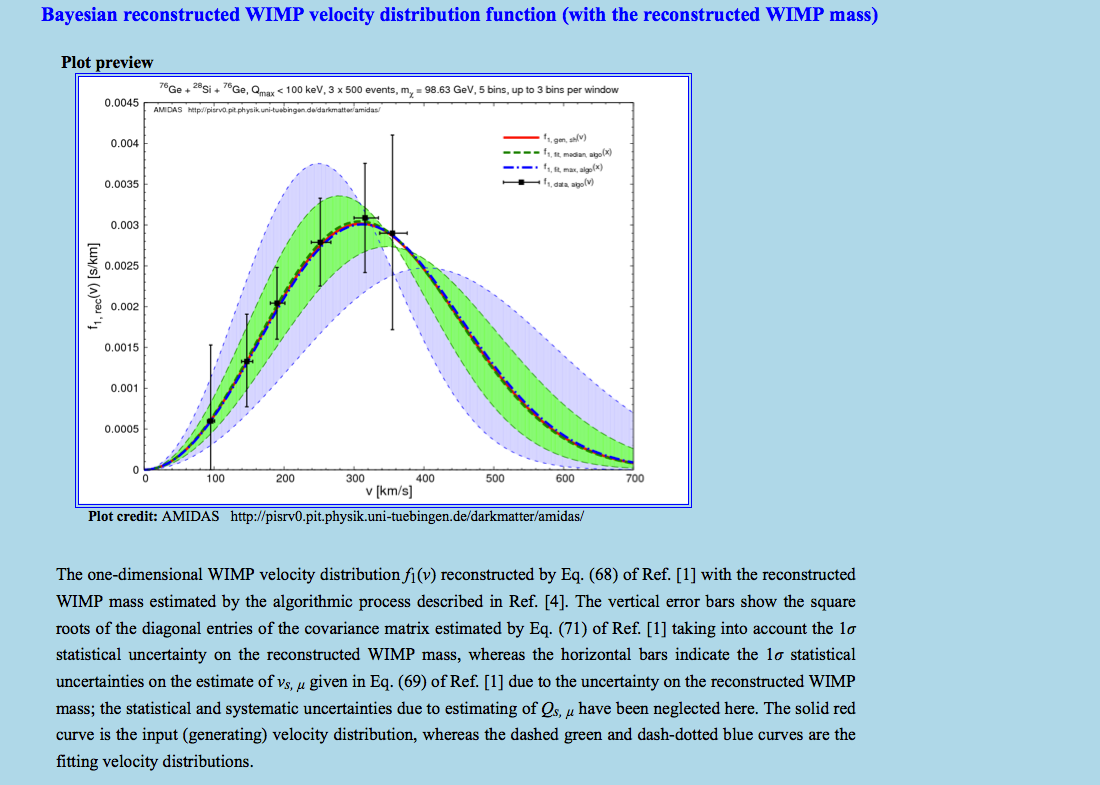}
 \subsubsection{Distribution(s) of the reconstructed fitting parameter(s)}

 The distribution of each Bayesian reconstructed fitting parameter,
 e.g.~the characteristic Solar and Earth's Galactic velocities,
 in all simulated experiments or analyzed data sets
 will also be provided by the \amidasii\ package:
\includefigure{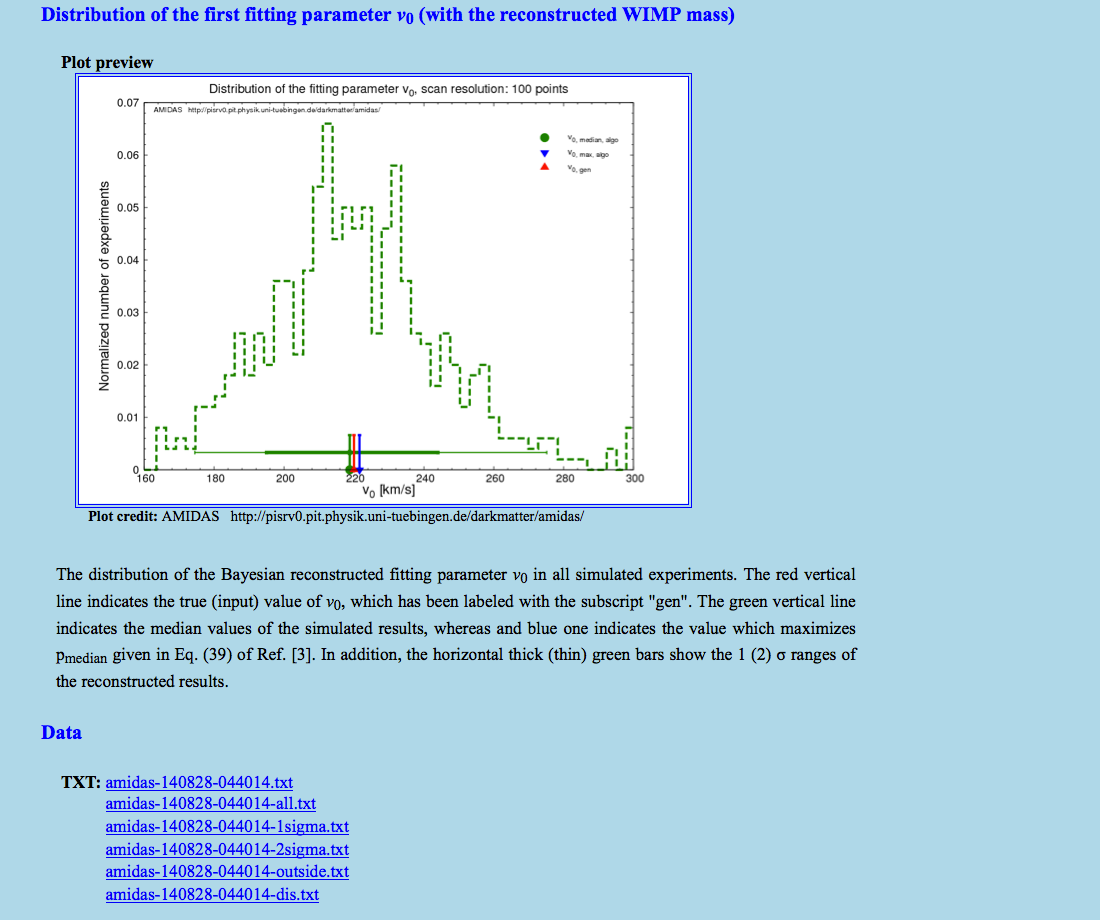}
 Note that,
 as shown in this figure,
 the Bayesian reconstructed result points
 which are in the $1\~(2)\~\sigma$ statistical uncertainty ranges/volumes
 or outside of them
 will be given separately in different TXT files.

 Moreover,
 for reconstructions with more than one fitting parameter,
 the projection of the result points
 on each 2-D plane of the parameter space
 will also be given:
\includefigure{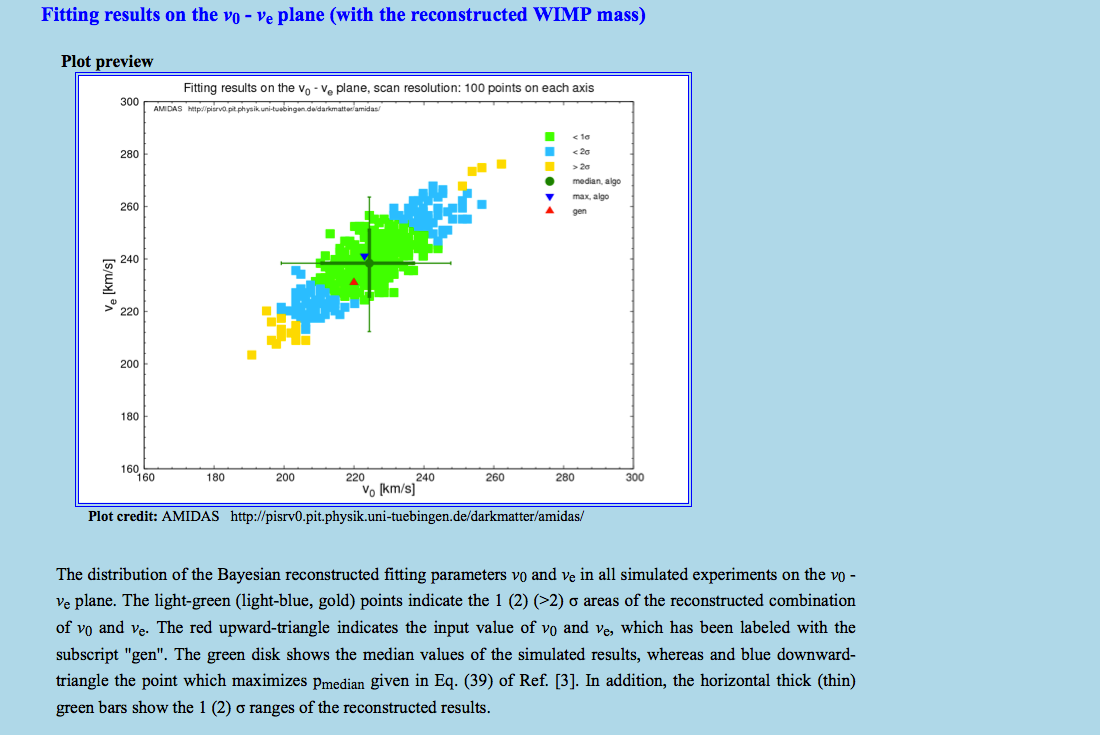}
 Here the shifted Maxwellian velocity distribution
 $f_{1, \sh}(v)$ given in Eq.~(\ref{eqn:f1v_sh})
 with two fitting parameters:
 $v_0$ and $\ve$
 has been used.

 \section{Summary}

 In this paper,
 we give a detailed user's guide
 to the \amidas\ (A Model--Independent Data Analysis System)
 package and website,
 which is developed for
 online simulations and data analyses
 for direct Dark Matter detection experiments and phenomenology.

 \amidas\ has the ability to do
 full Monte Carlo simulations
 as well as
 to analyze real/pseudo data sets
 either generated by another event generating programs 
 or
 recorded in direct DM detection experiments.
 Recently,
 the whole \amidas\ package and website system
 has been upgraded to the second phase:
 \amidasii,
 for including the new developed Bayesian analysis technique.

 Users can run all functions and
 adopt the default input setup used in our earlier works
 \cite{DMDDf1v, DMDDf1v-Bayesian,
       DMDDmchi, DMDDfp2, DMDDranap,
       DMDDbg-mchi, DMDDbg-f1v, DMDDbg-fp2, DMDDbg-ranap,
       AMIDAS-CYGNUS2011, AMIDASbg-DSU2011}
 for their simulations
 as well as
 analyzing their own real/pseudo data sets.
 The use of the \amidas\ website
 for users' simulations and data analyses
 has been explained step--by--step with plots
 in this paper.
 The preparations of function/data files to upload
 for simulations and data analyses
 have also been described.

 Moreover,
 for more flexible and user--oriented use,
 users have the option to set
 their own target nuclei
 as well as
 their favorite/needed (fitting) one--dimensional
 WIMP velocity distribution function,
 elastic nuclear form factors
 for the SI and SD WIMP--nucleus cross sections
 and different probability distribution functions
 needed in the Bayesian reconstruction procedure.
 As examples,
 the \amidasii\ codes
 for all user--uploadable functions are given
 in Secs.~3 and 5
 as well as
 Appendix B and C.

 In summary,
 up to now all basic functions of
 the \amidas\ package and website
 have been well established.
 Hopefully this new tool can
 help our theoretical as well as experimental colleagues
 to understand properties of halo WIMPs,
 offer useful information to indirect DM detection
 as well as collider experiments,
 and finally discover
 (the mystery of) Galactic DM particles.

\subsubsection*{Acknowledgments}

 The author would like to thank
 James Yi-Yu Liu
 for useful discussion about the basic idea
 of including user--defined functions
 into the source code.
 The author appreciates the ILIAS Project and
 the Physikalisches Institut der Universit\"at T\"ubingen
 for kindly providing the opportunity of the collaboration
 and the technical support of the \amidas\ website.
 The author would also like to thank
 the friendly hospitality of
 the Institute of Physics,
 National Chiao Tung University,
 the Graduate School of Science and Engineering for Research,
 University of Toyama,
 the Institute of Modern Physics,
 Chinese Academy of Sciences,
 the Center for High Energy Physics,
 Peking University,
 and
 the Xinjiang Astronomical Observatory,
 Chinese Academy of Sciences,
 where part of this work was completed.
 This work
 was partially supported
 by the National Science Council of R.O.C.~%
 under the contracts no.~NSC-98-2811-M-006-044 and
 no.~NSC-99-2811-M-006-031
 as well as
 the CAS Fellowship for Taiwan Youth Visiting Scholars
 under the grant no.~2013TW2JA0002.
\appendix
\setcounter{equation}{0}
\setcounter{figure}{0}
\setcounter{amidascode}{0}
\renewcommand{\theequation}       {A\arabic{equation}}
\renewcommand{\thefigure}         {A\arabic{figure}}
\renewcommand{\code}         [1]  {\refstepcounter{amidascode}
                                    \tabt{Code A\arabic{amidascode}: #1}}
%
%
%
\section{Intrinsically defined constants}

 In this section,
 we list all constants
 as well as
 the reduced mass $\mrN$ in Eq.~(\ref{eqn:mrN})
 and the transformation constant $\alpha$ in Eq.~(\ref{eqn:alpha})
 defined intrinsically in the \amidas\ package
 for users' reference.
 Users could use these intrinsically defined constants and functions directly
 and/or include them into their own defined functions.

\subsection{Constants and translators}

 The physical constants and needed translators
 defined in the \amidas\ package
 are in {\em natural units} as following:
\code{Physical constants and needed translators
      \label{code:constants_translators}}
\verb# double c   = 1.0;                                                       # \\
\verb#                                                                         # \\
\verb# double v_U = c / (2.9979246 * 1e5);                                     # \\
\verb#                                                                         # \\
\verb# double m_U = 1e6 / (c * c);                                             # \\
\verb# double m_p = 0.938272 * m_U;                                            # \\
\verb#                                                                         # \\
\verb# double G_F = 1.166379 * ( 1e-5 / (1e6 * 1e6) ) * (c * c * c);           # \\
\verb#                                                                         # \\
\verb# double fm_U = c / (0.197327 * 1e6);                                     # \\
\verb# double pb_U = 1e-36 * (1e13 * fm_U) * (1e13 * fm_U);                    # \\
\verb#                                                                         # \\
\verb# double s_U = 1.0 / (6.582119 * 1e-19);                                  # \\
\verb# double kg_day_U = (m_U / (1.78266 * 1e-27)) * (86400.0 * s_U);          # \\
\verb#                                                                         # \\
\verb# double rho_U = m_U / ( (1e13 * fm_U) * (1e13 * fm_U) * (1e13 * fm_U) ); # \\
\verb#                                                                         # \\
\verb# double omega = 2.0 * M_PI / 365.0;                                      # \\
\tabb
 The meanings of these constants and translators are
{\small
\begin{center}
\renewcommand{\arraystretch}{1.5}
\begin{tabular}{|| c | l | l ||}
\hline
\hline
 \makebox[ 1.4 cm][c]{Symbol}  &
 \makebox[12   cm][c]{Meaning} &
 \makebox[ 2.3 cm][c]{Remarks} \\
\hline
\hline
 \verb+v_U+      &
 The velocity translator from km/s to
 $c = 2.997~924~58 \times 10^5$ km/s  &
 \cite{RPP12P}   \\
\hline
 \verb+m_U+      &
 The mass translator from GeV/$c^2$ to keV/$c^2$ &
                 \\
 \verb+m_p+      &
 The proton mass
 $m_{\rm p} = 938.272~046(21)~{\rm MeV}/c^2$ &
 \cite{RPP12P}   \\
\hline
 \verb+G_F+      &
 The Fermi coupling constant
 $G_F = 1.166~378~7(6) \times 10^{-5}~{\rm GeV}^{-2} (\hbar c)^3$ &
 \cite{RPP12P}   \\
\hline
 \verb+fm_U+     &
 \begin{minipage}{12cm}
 \vspace{0.2 cm}
 The length translator from $1~{\rm fm} = 10^{-15}~{\rm m}$ to $c$ \\
 $\abrac{\hbar c = 0.197~326~9718(44)~{\rm GeV~fm}}$
 \vspace{0.25cm}
 \end{minipage}  &
 \cite{RPP12P}   \\
 \verb+pb_U+     &
 The area translator from $1~{\rm pb} = 10^{-36}~{\rm cm}^2$ to the natural units &
                 \\
\hline
 \verb+s_U+      &
 \begin{minipage}{12cm}
 \vspace{0.2 cm}
 The time translator from 1 s to the natural units \\
 $\abrac{\hbar = 6.582~119~28(15) \times 10^{-25}~{\rm GeV~s}}$
 \vspace{0.25cm}
 \end{minipage}  &
 \cite{RPP12P}   \\
 \verb+kg_day_U+ &
 \begin{minipage}{12cm}
 \vspace{0.2 cm}
 The exposure translator from 1 kg--day to the natural units \\
 $\abrac{1~{\rm GeV} / c^2 = 1.782~661~845(39) \times 10^{-27}~{\rm kg}}$
 \vspace{0.25cm}
 \end{minipage}  &
 \cite{RPP12P}   \\
\hline
 \verb+rho_U+    &
 The density translator from $1~{\rm GeV}/c^2/{\rm cm^3}$ to the natural units &
                 \\
\hline
 \verb+omega+    &
 $\omega = 2 \pi / 365$ needed in Eq.~(\ref{eqn:v_e}) &
                 \\
\hline
\hline
\end{tabular}
\end{center}
}
\subsection{The reduced mass \boldmath$\mrN$
            and the transformation constant $\alpha$}

 All masses in the \amidas\ code are in the unit of GeV$/c^2$.
 First,
 the mass of the target nucleus
 is defined as a function of the atomic mass number $A$ by
\code{Nuclear mass \boldmath$\mN(A)$
      \label{code:m_N}}
\verb# double m_N(int A)        # \\
\verb# {                        # \\
\verb#   return m_p * A * 0.99; # \\
\verb# }                        # \\
\tabb
 Here the mass difference between a proton and a neutron
 has been neglected.
 And the reduced mass between the WIMP mass
 and ``something'' is given by
\code{Reduced mass between the WIMP mass and ``something''
      \label{code:mchi_r}}
\verb# double mchi_r(double mchi, double mx) # \\
\verb# {                                     # \\
\verb#   return mchi * mx / (mchi + mx);     # \\
\verb# }                                     # \\
\tabb
 Hence,
 the reduced mass of the WIMP mass
 with the proton (nucleon) mass
 and
 with the mass of the target nucleus,
 $\mrN$ in Eq.~(\ref{eqn:mrN}),
 are defined as
\code{Reduced mass \boldmath$\mrp(\mchi)$
      \label{code:mchi_rp}}
\verb# double mchi_rp(double mchi)       # \\
\verb# {                                 # \\
\verb#   return mchi_r(mchi * m_U, m_p); # \\
\verb# }                                 # \\
\tabb
 and
\code{Reduced mass \boldmath$\mrN(\mchi, A)$
      \label{code:mchi_rN}}
\verb# double mchi_rN(double mchi, int A)   # \\
\verb# {                                    # \\
\verb#   return mchi_r(mchi * m_U, m_N(A)); # \\
\verb# }                                    # \\
\tabb
 Finally,
 the transformation constant $\alpha$
 defined in Eq.~(\ref{eqn:alpha})
 can also be given as a function of $\mchi$ and $A$ as:
\code{transformation constant \boldmath$\alpha(\mchi, A)$
      \label{code:alpha}}
\verb# double alpha(double mchi, int A)                # \\
\verb# {                                               # \\
\verb#   return sqrt(m_N(A) / 2.0) / mchi_rN(mchi, A); # \\
\verb# }                                               # \\
\tabb
\subsection{Declared variables and constants}

 Here
 we list the {\em short--name} variables and constants
 declared in the \amidasii\ package.
 Note that
 these names should be {\em avoided} to use
 in the user--defined function,
 except that
 you know their exact meanings and
 can apply them properly;
 otherwise,
 the \amidasii\ package might not work correctly.
\tabt{Declared short--name variables and constants
      \label{code:variables_constants}}
\verb# a_point, a_pointNo, a_subpoint, a_subpointNo, an_in, ap_in, Ax, AX, AX_max, AX_str,   # \\
\verb# b_point, b_pointNo, b_subpoint, b_subpointNo, bar_Q, bar_Q_win, bar_QQ, bar_QQ_win,   # \\
\verb# bgevent_ratio, bgevent_ratio_common, c_point, c_pointNo, c_subpoint, c_subpointNo,    # \\
\verb# calN, chi2_mchi_min, CnX, CnX_n, CnX_n_sh, CnX_sh, cov_InX, cp, cp_sh, CpX, CpX_p,    # \\
\verb# CpX_p_sh, CpX_sh, Delta_Qt, Delta_Qt_tmp, Delta_Qtp, Delta_Qtp_tmp, Delta_t,          # \\
\verb# Delta_t_tmp, delta_Xbinwidth, delta2_k, delta2_k_win, delta2_Q, delta2_Q_win,         # \\
\verb# Epsilon, error_mchi_in_fixed, error_sigmaSIp_in, event, eventNo_bg, eventNo_bg_ave,   # \\
\verb# eventNo_max, eventNo_Q, eventNo_Q_win, eventNo_sg, eventNo_sg_ave, eventNo_tot,       # \\
\verb# eventNo_tot_ave, exptNo_common, exptNo_max, fp2_algo, fp2_input, fp2_th, fp2_tmp,     # \\
\verb# FQmin_SD_TX, FQmin_SI_TX, FQmin_TX, FQ_TS_const, hn, hn_median, hn_tmp,               # \\
\verb# Int_f1v_Gau_k, Intf_dim, Intf_function_Simpson, Intf_pointNo, InX, JJ, k, k_f1v,      # \\
\verb# k_win, ka, kappa, kappa_hi, kappa_lo, kappa_win, kn, kn_median, kn_tmp, lambda, ln,   # \\
\verb# ln_median, ln_tmp, m_N_TX, mchi_assumed, mchi_in, mchi_in_fixed, mchi_solve,          # \\
\verb# mchi_solve_hi, mchi_solve_lo, mchi_tmp, mchiNo, mchiNo_max, mm, n_expt, n_mchi,       # \\
\verb# n_plot, n_point, n_ranap, NmX, nn, nn_max, nn_win, nTX, nTXh, nTXl, nTXx, nTXxn,      # \\
\verb# nTXxp, nTXy, nTXyn, nTXyp, nTXz, order_sigmaSIp_in, pm, pointNo, Qbin, Qbin_1_common, # \\
\verb# Qbin_hi, Qbin_lo, Qbin_max, QbinNo, QbinNo_common, QbinNo_max, Qbinwidth, Qmax,       # \\
\verb# Qmax_algo, Qmax_bg_str, Qmax_bgwindow, Qmax_bgwindow_common, Qmax_gen, Qmax_Int,      # \\
\verb# Qmax_InX_Int, Qmax_kin, Qmax_set, Qmax_set_common, Qmax_str, Qmax_win, Qmid, Qmid_sh, # \\
\verb# Qmid_win, Qmid_win_sh, Qmin, Qmin_bg_str, Qmin_bgwindow, Qmin_bgwindow_common,        # \\
\verb# Qmin_gen, Qmin_Int, Qmin_set, Qmin_set_common, Qmin_str, Qmin_win, QminXInX,          # \\
\verb# qqR_1_min, qqR_1_max, QQ_SD_max_TX, QQ_SD_min_TX, Qtp, Qvin, Qwin, Qwin_max, QwinNo,  # \\
\verb# QwinNo_max, Qwinwidth, R_Simpson, R_TX, r_win, ranap_in, ranap_in_fixed, ranap_SD,    # \\
\verb# ranap_SD_sh, ranap_tmp, ranapNo, ranapNo_max, ratio_eventNo_bg_ave,                   # \\
\verb# ratio_eventNo_sg_ave, ratio_eventNo_tot_ave, ratio_Xbinwidth, ratio_Ybinwidth,        # \\
\verb# ratio_Zbinwidth, rfnfp_in_fixed, rho, rho_0, rho0_input, rho0_tmp, RJ, RJ_sh, RJX,    # \\
\tabc
\verb# rlh, rmin_bgwindow, rmin_gen, rminXInX, Rn, Rn_sh, RnX, Rsigma, Rsigma_sh,            # \\
\verb# rsigmaSDnSI_th, rsigmaSDnSI_tmp, rsigmaSDpSI_th, rsigmaSDpSI_tmp, RsigmaX, run,       # \\
\verb# run_Intf, run_Intf_max, runi, runj, runk, scan_pointNo, scan_subpointNo, sh,          # \\
\verb# sigmaSIp_in, Sn, Snp, sol, Sp, Spn, subpointNo, sum_Q, sum_Q_win, sum_QQ, sum_QQ_win, # \\
\verb# t_end, t_expt, t_p, t_start, tbin, tbinNo, tbinNo_common, tbinNo_max, tbinwidth, tmax,# \\
\verb# tmax_Int, tmid, tmin, tmin_Int, TX, TXNo, TXNo_max, v_0, v_esc, v_max, Xbin, XbinNo,  # \\
\verb# XbinNo_max, Xbinwidth, Xbinwidth_1, Xbinwidth_common, Xmax, Xmax_binning, Xmax_Intf,  # \\
\verb# Xmax_plot, Xmid, Xmin, Xmin_binning, Xmin_Intf, Xmin_plot, Ybin, YbinNo, YbinNo_max,  # \\
\verb# Ybinwidth_1, Ymax_binning, Ymax_Intf, Ymax_plot, Ymin_binning, Ymin_Intf, Ymin_plot,  # \\
\verb# Zbin, ZbinNo, ZbinNo_max, Zbinwidth_1, Zmax_binning, Zmax_Intf, Zmin_binning,         # \\
\verb# Zmin_Intf, ZX                                                                         # \\
\tabb
%

%
%
%
\section{Intrinsically defined functions}

 In this section,
 we give all codes for
 intrinsically defined
 velocity distribution functions
 given in Sec.~3.4,
 elastic nuclear form factors
 given in Sec.~3.5
 as well as
 simple artificial background spectrum
 given in Sec.~3.6
 in the \amidasii\ package.

\subsection{One--dimensional WIMP velocity distribution function}

 In this subsection,
 we give first the codes for
 intrinsically defined
 velocity distribution functions
 given in Sec.~3.4
 in the \amidasii\ package.

\subsubsection{Simple Maxwellian velocity distribution
               \boldmath$f_{1, \Gau}(v)$}

 According to Eq.~(\ref{eqn:f1v_Gau}),
 we define
\code{Simple Maxwellian velocity distribution
      \boldmath$f_{1, \Gau}(v)$
      \label{code:f1v_Gau}}
\verb# double f1v_Gau(double vv)                     # \\
\verb# {                                             # \\
\verb#   return                                      # \\
\verb#     (4.0 / sqrt(M_PI))                      * # \\
\verb#     ( (vv * vv) / (v_0 * v_0 * v_0) ) / v_U * # \\
\verb#     exp(-(vv * vv) / (v_0 * v_0) );           # \\
\verb# }                                             # \\
\tabb
 Then,
 since
\beq
    \int_{\alpha \sqrt{Q}}^{\infty} \bfrac{f_{1, \Gau}(v)}{v} dv
 =  \frac{2}{\sqrt{\pi}} \afrac{1}{v_0} \~ e^{-\alpha^2 Q / v_0^2}
\~,
\label{eqn:Intf1v_v_Gau}
\eeq
 we can define
\code{Integral over the simple Maxwellian velocity distribution
      \boldmath$\int_{\alpha \sqrt{Q}}^{\infty} \bbig{f_{1, \Gau}(v) / v} dv$
      \label{code:Intf1v_v_Gau}}
\verb# double Intf1v_v_Gau(double mchi, int A, double QQ) # \\
\verb# {                                                  # \\
\verb#   return                                           # \\
\verb#     ( 2.0 / sqrt(M_PI) / (v_0 * v_U) ) *           # \\
\verb#     exp(-alpha(mchi, A) * alpha(mchi, A) * QQ /    # \\
\verb#          (  (v_0 * v_U) * (v_0 * v_U)  )       );  # \\
\verb# }                                                  # \\
\tabb

 On the other hand,
 for drawing output plots,
 we have
\code{Simple Maxwellian velocity distribution
      \boldmath$f_{1, \Gau}(v)$
      (for Gnuplot)
      \label{code:f1v_x_Gau}}
\verb#   M_PI = 3.141593                        # \\
\verb#                                          # \\
\verb#    f1v_Gau(x)                          \ # \\
\verb# =  (4.0 / sqrt(M_PI))              *   \ # \\
\verb#    ( (x * x) / (v_0 * v_0 * v_0) ) *   \ # \\
\verb#    exp(-(x * x) / (v_0 * v_0))           # \\
\tabb
 and
\code{Integral over the simple Maxwellian velocity distribution
      \boldmath$\int_{\alpha \sqrt{Q}}^{\infty} \bbig{f_{1, \Gau}(v) / v} dv$
      (for Gnuplot)
      \label{code:Intf1v_v_x_Gau}}
\verb# c   = 1.0                                                       # \\
\verb# m_U = 1e6 / (c * c)                                             # \\
\verb# v_U = c / (2.9979246 * 1e5)                                     # \\
\verb#                                                                 # \\
\verb# m_N  = (0.938272 * m_U) * AX * 0.99                             # \\
\verb# m_rN = (m_chi * m_U) * m_N / (m_chi * m_U + m_N)                # \\
\verb#                                                                 # \\
\verb# alpha = sqrt(m_N / (2.0 * m_rN * m_rN))                         # \\
\verb#                                                                 # \\
\verb#    Intf1v_v_Gau(x)                                            \ # \\
\verb# =  (2.0 / sqrt(M_PI) / (v_0 * v_U)) *                         \ # \\
\verb#    exp(-alpha * alpha * x / ( (v_0 * v_U) * (v_0 * v_U) ) )     # \\
\tabb
 Remind that
 firstly,
 ``\verb+\+'' (backslash) must be used
 in order to let the including of the definition(s) given in this file
 into the other intrinsic commands correctly.
 Secondly,
 three flexible--kept parameters:
 \verb+AX+,
 \verb+m_chi+ and \verb+v_0+
 will be read directly from
 the users' initial simulation/data analysis setup
 set earlier on the website.

\subsubsection{Modified Maxwellian velocity distribution
               \boldmath$f_{1, \Gau, k}(v)$}

 According to Eq.~(\ref{eqn:f1v_Gau_k}),
 we can obtain the integral over $f_{1, \Gau, k}(v)$
 and then define the normalization constant $N_f$
 in the \amidasii\ package
 as follows.

 For $k = 1$,
 we have
\beqn
     \int_0^{\vmax} f_{1, \Gau, k = 1}(v) \~ dv
 \=  \frac{1}{N_{f, k = 1}}
     \int_0^{\vmax} v^2 \abrac{e^{-v^2     / v_0^2} - e^{-\vmax^2 / v_0^2}} \~ dv
     \non\\
 \=  \frac{1}{N_{f, k = 1}}
     \bbrac{  \frac{\sqrt{\pi}}{4} \~
              v_0^3 \~
              \erf\afrac{\vmax}{v_0}
            - \vmax
              \abrac{\frac{v_0^2}{2} + \frac{\vmax^2}{3}}
              e^{-\vmax^2 / v_0^2}                         }
     \non\\
 \=  1
\~,
\label{eqn:Int_f1v_Gau_k_1}
\eeqn
 and then can define $N_f$ by
\newpage
\code{Normalization constant \boldmath$N_f$
      of $f_{1, \Gau, k}(v)$  for $k = 1$
      \label{code:Int_f1v_Gau_k_1}}
\verb#    Int_f1v_Gau_k                                # \\
\verb# =  (  (sqrt(M_PI) / 4.0)                   *    # \\
\verb#       (v_0 * v_0 * v_0)                    *    # \\
\verb#       erf(v_max / v_0)                          # \\
\verb#     - v_max                                *    # \\
\verb#      (  v_0   * v_0   / 2.0                     # \\
\verb#       + v_max * v_max / 3.0  )             *    # \\
\verb#       exp(-(v_max * v_max) / (v_0 * v_0) )  ) * # \\
\verb#    v_U;                                         # \\
\tabb
 For $k = 2$,
 we have
\beqn
 \conti
     \int_0^{\vmax} f_{1, \Gau, k = 2}(v) \~ dv
     \non\\
 \=  \frac{1}{N_{f, k = 2}}
     \int_0^{\vmax} v^2 \abrac{e^{-v^2     / 2 v_0^2} - e^{-\vmax^2 / 2 v_0^2}}^2 \~ dv
     \non\\
 \=  \frac{1}{N_{f, k = 2}}
     \bleft{   \frac{\sqrt{\pi}}{4} \~
                v_0^3 \~
               \erf\afrac{\vmax}{v_0}
             - \vmax
               \abrac{\frac{v_0^2}{2} - \frac{\vmax^2}{3} - 2 v_0^2}
               e^{-\vmax^2 / v_0^2}                                   }
     \non\\
 \conti ~~~~~~~~~~~~~~~ 
     \bright{- \sqrt{2 \pi} \~
               v_0^3 \~
               \erf\afrac{\vmax}{\sqrt{2} \~ v_0}
               e^{-\vmax^2 / 2 v_0^2}                                 }
     \non\\
 \=  1
\~,
\label{eqn:Int_f1v_Gau_k_2}
\eeqn
 and then can define $N_f$ by
\code{Normalization constant \boldmath$N_f$
      of $f_{1, \Gau, k}(v)$  for $k = 2$
      \label{code:Int_f1v_Gau_k_2}}
\verb#    Int_f1v_Gau_k                                      # \\
\verb# =  (  (sqrt(M_PI) / 4.0)                         *    # \\
\verb#       (v_0 * v_0 * v_0)                          *    # \\
\verb#       erf(v_max / v_0)                                # \\
\verb#     - v_max                                      *    # \\
\verb#       (  v_0   * v_0   / 2.0                          # \\
\verb#        - v_max * v_max / 3.0                          # \\
\verb#        - 2.0 * v_0 * v_0    )                    *    # \\
\verb#       exp(-(v_max * v_max) / (v_0 * v_0) )            # \\
\verb#     - sqrt(2.0 * M_PI)                           *    # \\
\verb#       (v_0 * v_0 * v_0)                          *    # \\
\verb#       erf( v_max / v_0 / sqrt(2.0) )             *    # \\
\verb#       exp(-(v_max * v_max) / (2.0 * v_0 * v_0) )  ) * # \\
\verb#    v_U;                                               # \\
\tabb
 For $k = 3$,
 we have
\beqn
 \conti
     \int_0^{\vmax} f_{1, \Gau, k = 3}(v) \~ dv
     \non\\
 \=  \frac{1}{N_{f, k = 3}}
     \int_0^{\vmax} v^2 \abrac{e^{-v^2     / 3 v_0^2} - e^{-\vmax^2 / 3 v_0^2}}^3 \~ dv
     \non\\
 \=  \frac{1}{N_{f, k = 3}}
     \cleft{   \frac{\sqrt{\pi}}{4} \~
                v_0^3 \~
               \erf\afrac{\vmax}{v_0}
             - \vmax
               \abrac{  \frac{v_0^2}{2}
                      + \frac{\vmax^2}{3}
                      + \frac{9 v_0^2}{4} }
               e^{-\vmax^2 / v_0^2}                                      }
     \non\\
 \conti ~~~~~~~~~~~ 
     \cright{- v_0^3
             \bbrac{  \frac{9 \sqrt{6 \pi}}{16} \~
                      \erf\afrac{\sqrt{2} \~ \vmax}{\sqrt{3} \~ v_0}
                      e^{-\vmax^2 / 3 v_0^2}
                    - \frac{9 \sqrt{3 \pi}}{4} \~
                      \erf\afrac{\vmax}{\sqrt{3} \~ v_0}
                      e^{-2 \vmax^2 / 3 v_0^2}                        }  }
     \non\\
 \=  1
\~,
\label{eqn:Int_f1v_Gau_k_3}
\eeqn
 and then can define $N_f$ by
\code{Normalization constant \boldmath$N_f$
      of $f_{1, \Gau, k}(v)$  for $k = 3$
      \label{code:Int_f1v_Gau_k_3}}
\verb#    Int_f1v_Gau_k                                                  # \\
\verb# =  (  (sqrt(M_PI) / 4.0)                                     *    # \\
\verb#       (v_0 * v_0 * v_0)                                      *    # \\
\verb#       erf(v_max / v_0)                                            # \\
\verb#     - v_max                                                  *    # \\
\verb#       (  v_0   * v_0   / 2.0                                      # \\
\verb#        + v_max * v_max / 3.0                                      # \\
\verb#        + (9.0 / 4.0) * v_0 * v_0  )                          *    # \\
\verb#       exp(-(v_max * v_max) / (v_0 * v_0) )                        # \\
\verb#     + (v_0 * v_0 * v_0)                                      *    # \\
\verb#       (- sqrt(6.0 * M_PI) * (9.0 / 16.0)                  *       # \\
\verb#          erf( v_max / v_0 * (sqrt(6.0) / 3.0) )           *       # \\
\verb#          exp(-(      v_max * v_max) / (3.0 * v_0 * v_0) )         # \\
\verb#        + sqrt(3.0 * M_PI) * (9.0 /  4.0)                  *       # \\
\verb#          erf( v_max / v_0 / sqrt(3.0) )                   *       # \\
\verb#          exp(-(2.0 * v_max * v_max) / (3.0 * v_0 * v_0) )  )  ) * # \\
\verb#    v_U;                                                           # \\
\tabb
 For $k = 4$,
 we have
\beqn
 \conti
     \int_0^{\vmax} f_{1, \Gau, k = 4}(v) \~ dv
     \non\\
 \=  \frac{1}{N_{f, k = 4}}
     \int_0^{\vmax} v^2 \abrac{e^{-v^2     / 4 v_0^2} - e^{-\vmax^2 / 4 v_0^2}}^4 \~ dv
     \non\\
 \=  \frac{1}{N_{f, k = 4}}
     \cleft{   \frac{\sqrt{\pi}}{4} \~
                v_0^3 \~
               \erf\afrac{\vmax}{v_0}
             - \vmax
               \abrac{  \frac{v_0^2}{2}
                      - \frac{\vmax^2}{3}
                      - \frac{14 v_0^2}{3} }
               e^{-\vmax^2 / v_0^2}                               }
     \non\\
 \conti ~~~~~~~~~~~ 
            {- v_0^3
               \bleft{   \frac{8 \sqrt{3 \pi}}{9} \~
                         \erf\afrac{\sqrt{3} \~ \vmax}{2 v_0}
                         e^{-\vmax^2 / 4 v_0^2}
                       - 3 \sqrt{2 \pi} \~
                         \erf\afrac{\vmax}{\sqrt{2} \~ v_0}
                         e^{-\vmax^2 / 2 v_0^2}                }  }
     \non\\
 \conti ~~~~~~~~~~~~~~~~~~~~ 
     \cBiggr{  \bBiggr{+ 8 \sqrt{\pi} \~
               \erf\afrac{\vmax}{2 v_0}
               e^{-3 \vmax^2 / 4 v_0^2}                        }  }
     \non\\
 \=  1
\~,
\label{eqn:Int_f1v_Gau_k_4}
\eeqn
 and then can define $N_f$ by
\newpage
\code{Normalization constant \boldmath$N_f$
      of $f_{1, \Gau, k}(v)$  for $k = 4$
      \label{code:Int_f1v_Gau_k_4}}
\verb#    Int_f1v_Gau_k                                                  # \\
\verb# =  (  (sqrt(M_PI) / 4.0)                                     *    # \\
\verb#       (v_0 * v_0 * v_0)                                      *    # \\
\verb#       erf(v_max / v_0)                                            # \\
\verb#     - v_max                                                  *    # \\
\verb#       (  v_0   * v_0   / 2.0                                      # \\
\verb#        - v_max * v_max / 3.0                                      # \\
\verb#        - (14.0 / 3.0) * v_0 * v_0  )                         *    # \\
\verb#       exp(-(v_max * v_max) / (v_0 * v_0) )                        # \\
\verb#     + (v_0 * v_0 * v_0)                                      *    # \\
\verb#       (- sqrt(3.0 * M_PI) * (8.0 / 9.0)                   *       # \\
\verb#          erf( v_max / v_0 * (sqrt(3.0) / 2.0) )           *       # \\
\verb#          exp(-(      v_max * v_max) / (4.0 * v_0 * v_0) )         # \\
\verb#        + sqrt(2.0 * M_PI) * 3.0                           *       # \\
\verb#          erf( v_max / v_0 / sqrt(2.0) )                   *       # \\
\verb#          exp(-(      v_max * v_max) / (2.0 * v_0 * v_0) )         # \\
\verb#        - sqrt(      M_PI) * 8.0                           *       # \\
\verb#          erf( v_max / v_0 / 2.0 )                         *       # \\
\verb#          exp(-(3.0 * v_max * v_max) / (4.0 * v_0 * v_0) )  )  ) * # \\
\verb#    v_U;                                                           # \\
\tabb
 Now we can define
 the modified Maxwellian velocity distribution
 given in Eq.~(\ref{eqn:f1v_Gau_k})
 for \mbox{$k = 1,~2,~3$,} and 4
 as
\code{Modified Maxwellian velocity distribution
      \boldmath$f_{1, \Gau, k}(v)$
      \label{code:f1v_Gau_k}}
\verb# double f1v_Gau_k(double vv)                                         # \\
\verb# {                                                                   # \\
\verb#   return                                                            # \\
\verb#     (  (vv * vv)                                            *       # \\
\verb#        pow(  exp(-(vv    * vv   ) / (k_f1v * v_0 * v_0) )           # \\
\verb#            - exp(-(v_max * v_max) / (k_f1v * v_0 * v_0) ),          # \\
\verb#            k_f1v                                          )  ) /    # \\
\verb#     Int_f1v_Gau_k;                                                  # \\
\verb# }                                                                   # \\
\tabb

 Meanwhile,
 for predicting the WIMP--signal spectrum
 one needs also that,
 for $k = 1$,
\beqn
     \int \bfrac{f_{1, \Gau, k = 1}(v)}{v} dv
 \=  \frac{1}{N_{f, k = 1}}
     \int v \abrac{e^{-v^2     / v_0^2} - e^{-\vmax^2 / v_0^2}} \~ dv
     \non\\
 \=  \frac{1}{N_{f, k = 1}}
     \abrac{- \frac{v_0^2}{2} \~
              e^{-v^2     / v_0^2}
            - \frac{v^2}{2} \~
              e^{-\vmax^2 / v_0^2}  }
\~,
\label{eqn:Int_f1v_v_Gau_k_1}
\eeqn
 to define
\newpage
\code{Integral over the modified Maxwellian velocity distribution
      \boldmath$\int \bbig{f_{1, \Gau, k}(v) / v} dv$ for $k = 1$
      \label{code:Int_f1v_v_Gau_k_1}}
\verb# double Int_f1v_v_Gau_k(double vv)                                    # \\
\verb# {                                                                    # \\
\verb#     return                                                           # \\
\verb#       (- (v_0 * v_0) / 2.0                    *                      # \\
\verb#          exp(-(vv    * vv   ) / (v_0 * v_0) )                        # \\
\verb#        - (vv  * vv ) / 2.0                    *                      # \\
\verb#          exp(-(v_max * v_max) / (v_0 * v_0) )  ) /                   # \\
\verb#       Int_f1v_Gau_k;                                                 # \\
\verb# }                                                                    # \\
\tabb
 For $k = 2$,
 one has
\beqn
     \int \bfrac{f_{1, \Gau, k = 2}(v)}{v} dv
 \=  \frac{1}{N_{f, k = 2}}
     \int v \abrac{e^{-v^2     / 2 v_0^2} - e^{-\vmax^2 / 2 v_0^2}}^2 \~ dv
     \non\\
 \=  \frac{1}{N_{f, k = 2}}
     \abrac{- \frac{v_0^2}{2} \~
              e^{-v^2     / v_0^2}
            + \frac{v^2}{2} \~
              e^{-\vmax^2 / v_0^2}
            + 2 v_0^2 \~
              e^{-\vmax^2 / 2 v_0^2} \~
              e^{-v^2     / 2 v_0^2}   }
\~,
\label{eqn:Int_f1v_v_Gau_k_2}
\eeqn
 and can define
\code{Integral over the modified Maxwellian velocity distribution
      \boldmath$\int \bbig{f_{1, \Gau, k}(v) / v} dv$ for $k = 2$
      \label{code:Int_f1v_v_Gau_k_2}}
\verb# double Int_f1v_v_Gau_k(double vv)                                    # \\
\verb# {                                                                    # \\
\verb#     return                                                           # \\
\verb#       (- (v_0 * v_0) / 2.0                          *                # \\
\verb#          exp(-(vv    * vv   ) / (v_0 * v_0) )                        # \\
\verb#        + (vv  * vv ) / 2.0                          *                # \\
\verb#          exp(-(v_max * v_max) / (v_0 * v_0) )                        # \\
\verb#        + 2.0                                        *                # \\
\verb#          (v_0 * v_0)                                *                # \\
\verb#          exp(-(v_max * v_max) / (2.0 * v_0 * v_0) ) *                # \\
\verb#          exp(-(vv    * vv   ) / (2.0 * v_0 * v_0) )  ) /             # \\
\verb#       Int_f1v_Gau_k;                                                 # \\
\verb# }                                                                    # \\
\tabb
 For $k = 3$,
 one has
\beqn
     \int \bfrac{f_{1, \Gau, k = 3}(v)}{v} dv
 \=  \frac{1}{N_{f, k = 3}}
     \int v \abrac{e^{-v^2     / 3 v_0^2} - e^{-\vmax^2 / 3 v_0^2}}^3 \~ dv
     \non\\
 \=  \frac{1}{N_{f, k = 3}}
     \bleft{ - \frac{v_0^2}{2} \~
               e^{-v^2     / v_0^2}
             - \frac{v^2}{2} \~
               e^{-\vmax^2 / v_0^2}              }
     \non\\
 \conti ~~~~~~~~~~~~ 
     \bBiggr{+ \frac{9}{4} \~ v_0^2
               \aBig{  e^{-  \vmax^2 / 3 v_0^2} \~
                       e^{-2 v^2     / 3 v_0^2}
                     - 2
                       e^{-2 \vmax^2 / 3 v_0^2} \~
                       e^{-  v^2     / 3 v_0^2}    }  }
\~,
\label{eqn:Int_f1v_v_Gau_k_3}
\eeqn
 and can define
\newpage
\code{Integral over the modified Maxwellian velocity distribution
      \boldmath$\int \bbig{f_{1, \Gau, k}(v) / v} dv$ for $k = 3$
      \label{code:Int_f1v_v_Gau_k_3}}
\verb# double Int_f1v_v_Gau_k(double vv)                                    # \\
\verb# {                                                                    # \\
\verb#     return                                                           # \\
\verb#       (- (v_0 * v_0) / 2.0                                      *    # \\
\verb#          exp(-(vv    * vv   ) / (v_0 * v_0) )                        # \\
\verb#        - (vv  * vv ) / 2.0                                      *    # \\
\verb#          exp(-(v_max * v_max) / (v_0 * v_0) )                        # \\
\verb#        + (v_0 * v_0)                                            *    # \\
\verb#          (  (9.0 / 4.0)                                      *       # \\
\verb#             exp(-(      v_max * v_max) / (3.0 * v_0 * v_0) ) *       # \\
\verb#             exp(-(2.0 * vv    * vv   ) / (3.0 * v_0 * v_0) )         # \\
\verb#           - (9.0 / 2.0)                                      *       # \\
\verb#             exp(-(2.0 * v_max * v_max) / (3.0 * v_0 * v_0) ) *       # \\
\verb#             exp(-(      vv    * vv   ) / (3.0 * v_0 * v_0) )  )  ) / # \\
\verb#       Int_f1v_Gau_k;                                                 # \\
\verb# }                                                                    # \\
\tabb
 For $k = 4$,
 one has
\beqn
 \conti
     \int \bfrac{f_{1, \Gau, k = 4}(v)}{v} dv
     \non\\
 \=  \frac{1}{N_{f, k = 4}}
     \int v \abrac{e^{-v^2     / 4 v_0^2} - e^{-\vmax^2 / 4 v_0^2}}^4 \~ dv
     \non\\
 \=  \frac{1}{N_{f, k = 4}}
     \bleft{ - \frac{v_0^2}{2} \~
               e^{-v^2     / v_0^2}
             + \frac{v^2}{2} \~
               e^{-\vmax^2 / v_0^2}              }
     \non\\
 \conti ~~~~~~~~~~~ 
     \bBiggr{+ v_0^2
               \abrac{  \frac{8}{3} \~
                        e^{-  \vmax^2 / 4 v_0^2} \~
                        e^{-3 v^2     / 4 v_0^2}
                      - 6 \~
                        e^{-  \vmax^2 / 2 v_0^2} \~
                        e^{-  v^2     / 2 v_0^2}
                      + 8 \~
                        e^{-3 \vmax^2 / 4 v_0^2} \~
                        e^{-  v^2     / 4 v_0^2}    }\!\!}
,
\label{eqn:Int_f1v_v_Gau_k_4}
\eeqn
 and can define
\code{Integral over the modified Maxwellian velocity distribution
      \boldmath$\int \bbig{f_{1, \Gau, k}(v) / v} dv$ for $k = 4$
      \label{code:Int_f1v_v_Gau_k_4}}
\verb# double Int_f1v_v_Gau_k(double vv)                                    # \\
\verb# {                                                                    # \\
\verb#     return                                                           # \\
\verb#       (- (v_0 * v_0) / 2.0                                      *    # \\
\verb#          exp(-(vv    * vv   ) / (v_0 * v_0) )                        # \\
\verb#        + (vv  * vv ) / 2.0                                      *    # \\
\verb#          exp(-(v_max * v_max) / (v_0 * v_0) )                        # \\
\verb#        + (v_0 * v_0)                                            *    # \\
\verb#          (  (8.0 / 3.0)                                      *       # \\
\verb#             exp(-(      v_max * v_max) / (4.0 * v_0 * v_0) ) *       # \\
\verb#             exp(-(3.0 * vv    * vv   ) / (4.0 * v_0 * v_0) )         # \\
\verb#           - 6.0                                              *       # \\
\verb#             exp(-(      v_max * v_max) / (2.0 * v_0 * v_0) ) *       # \\
\verb#             exp(-(      vv    * vv   ) / (2.0 * v_0 * v_0) )         # \\
\verb#           + 8.0                                              *       # \\
\verb#             exp(-(3.0 * v_max * v_max) / (4.0 * v_0 * v_0) ) *       # \\
\verb#             exp(-(      vv    * vv   ) / (4.0 * v_0 * v_0) )  )  ) / # \\
\verb#       Int_f1v_Gau_k;                                                 # \\
\verb# }                                                                    # \\
\tabb
 Hence,
 we can define the required integral
\beq
   \int_{\alpha \sqrt{Q}}^{\vmax} \bfrac{f_{1, \Gau, k}(v)}{v} dv
\label{eqn:Intf1v_Gau_k}
\eeq
 for $k = 1,~2,~3~{\rm and}~4$
 as
\code{Integral over the modified Maxwellian velocity distribution
      \boldmath$\int_{\alpha \sqrt{Q}}^{\vmax} \bbig{f_{1, \Gau, k}(v) / v} dv$
      \label{code:Intf1v_v_Gau_k}}
\verb# double Intf1v_v_Gau_k(double mchi, int A, double QQ)                 # \\
\verb# {                                                                    # \\
\verb#   return                                                             # \\
\verb#       Int_f1v_v_Gau_k(v_max)                                         # \\
\verb#     - Int_f1v_v_Gau_k(alpha(mchi, A) * sqrt(QQ) / v_U);              # \\
\verb# }                                                                    # \\
\tabb

 On the other hand,
 while
 for drawing the output
 generating velocity distribution function,
 we need to define,
 for $k = 1,~2,~3~{\rm and}~4$,
\code{Modified Maxwellian velocity distribution
      \boldmath$f_{1, \Gau, k}(v)$
      (for Gnuplot)
      \label{code:f1v_x_Gau_k}}
\verb#    f1v_Gau_k(x)                                                  # \\
\verb# =  (x * x) *                                                     # \\
\verb#    (  exp(-(x     * x    ) / (k_f1v * v_0 * v_0) )               # \\
\verb#     - exp(-(v_max * v_max) / (k_f1v * v_0 * v_0) )  ) ** k_f1v / # \\
\verb#    Int_f1v_Gau_k                                                 # \\
\tabb
 where the normalization constant \verb+Int_f1v_Gau_k+
 will be estimated directly
 by Codes A\ref{code:Int_f1v_Gau_k_1} to A\ref{code:Int_f1v_Gau_k_4},
 for drawing the output
 generating WIMP--signal spectrum,
 we need
\code{Integral over the modified Maxwellian velocity distribution
      \boldmath$\int \bbig{f_{1, \Gau, k}(v) / v} dv$ for $k = 1$
      (for Gnuplot)
      \label{code:Int_f1v_v_x_Gau_k_1}}
\verb#    Int_f1v_v_Gau_k(x)                             \ # \\
\verb# =  (- (v_0 * v_0) / 2.0                    *      \ # \\
\verb#       exp(-(x     * x    ) / (v_0 * v_0) )        \ # \\
\verb#     - (x   * x  ) / 2.0                    *      \ # \\
\verb#       exp(-(v_max * v_max) / (v_0 * v_0) )  ) /   \ # \\
\verb#    Int_f1v_Gau_k                                    # \\
\tabb
\code{Integral over the modified Maxwellian velocity distribution
      \boldmath$\int \bbig{f_{1, \Gau, k}(v) / v} dv$ for $k = 2$
      (for Gnuplot)
      \label{code:Int_f1v_v_x_Gau_k_2}}
\verb#    Int_f1v_v_Gau_k(x)                                   \ # \\
\verb# =  (- (v_0 * v_0) / 2.0                          *      \ # \\
\verb#       exp(-(x     * x    ) / (v_0 * v_0) )              \ # \\
\verb#     + (x   * x  ) / 2.0                          *      \ # \\
\verb#       exp(-(v_max * v_max) / (v_0 * v_0) )              \ # \\
\verb#     + 2.0                                        *      \ # \\
\verb#       (v_0 * v_0)                                *      \ # \\
\verb#       exp(-(v_max * v_max) / (2.0 * v_0 * v_0) ) *      \ # \\
\verb#       exp(-(x     * x    ) / (2.0 * v_0 * v_0) )  ) /   \ # \\
\verb#    Int_f1v_Gau_k                                          # \\
\tabb
\code{Integral over the modified Maxwellian velocity distribution
      \boldmath$\int \bbig{f_{1, \Gau, k}(v) / v} dv$ for $k = 3$
      (for Gnuplot)
      \label{code:Int_f1v_v_x_Gau_k_3}}
\verb#    Int_f1v_v_Gau_k(x)                                               \ # \\
\verb# =  (- (v_0 * v_0) / 2.0                                      *      \ # \\
\verb#       exp(-(x     * x    ) / (v_0 * v_0) )                          \ # \\
\verb#     - (x  * x ) / 2.0                                        *      \ # \\
\verb#       exp(-(v_max * v_max) / (v_0 * v_0) )                          \ # \\
\verb#     + (v_0 * v_0)                                            *      \ # \\
\verb#       (  (9.0 / 4.0)                                      *         \ # \\
\verb#          exp(-(      v_max * v_max) / (3.0 * v_0 * v_0) ) *         \ # \\
\verb#          exp(-(2.0 * x     * x    ) / (3.0 * v_0 * v_0) )           \ # \\
\verb#        - (9.0 / 2.0)                                      *         \ # \\
\verb#          exp(-(2.0 * v_max * v_max) / (3.0 * v_0 * v_0) ) *         \ # \\
\verb#          exp(-(      x     * x    ) / (3.0 * v_0 * v_0) )  )  ) /   \ # \\
\verb#    Int_f1v_Gau_k                                                      # \\
\tabb
\code{Integral over the modified Maxwellian velocity distribution
      \boldmath$\int \bbig{f_{1, \Gau, k}(v) / v} dv$ for $k = 4$
      (for Gnuplot)
      \label{code:Int_f1v_v_x_Gau_k_4}}
\verb#    Int_f1v_v_Gau_k(x)                                               \ # \\
\verb# =  (- (v_0 * v_0) / 2.0                                      *      \ # \\
\verb#       exp(-(x  * x ) / (v_0 * v_0) )                                \ # \\
\verb#     + (x  * x ) / 2.0                                        *      \ # \\
\verb#       exp(-(v_max * v_max) / (v_0 * v_0) )                          \ # \\
\verb#     + (v_0 * v_0)                                            *      \ # \\
\verb#       (  (8.0 / 3.0)                                      *         \ # \\
\verb#          exp(-(      v_max * v_max) / (4.0 * v_0 * v_0) ) *         \ # \\
\verb#          exp(-(3.0 * x     * x    ) / (4.0 * v_0 * v_0) )           \ # \\
\verb#        - 6.0                                              *         \ # \\
\verb#          exp(-(      v_max * v_max) / (2.0 * v_0 * v_0) ) *         \ # \\
\verb#          exp(-(      x     * x    ) / (2.0 * v_0 * v_0) )           \ # \\
\verb#        + 8.0                                              *         \ # \\
\verb#          exp(-(3.0 * v_max * v_max) / (4.0 * v_0 * v_0) ) *         \ # \\
\verb#          exp(-(      x     * x    ) / (4.0 * v_0 * v_0) )  )  ) /   \ # \\
\verb#    Int_f1v_Gau_k                                                      # \\
\tabb
 for finally defining
\code{Integral over the modified Maxwellian velocity distribution
      \boldmath$\int_{\alpha \sqrt{Q}}^{\vmax} \bbig{f_{1, \Gau, k}(v) / v} dv$
      (for Gnuplot)
      \label{code:Intf1v_v_x_Gau_k}}
\verb# c   = 1.0                                        # \\
\verb# m_U = 1e6 / (c * c)                              # \\
\verb# v_U = c / (2.9979246 * 1e5)                      # \\
\verb#                                                  # \\
\verb# m_N  = (0.938272 * m_U) * AX * 0.99              # \\
\verb# m_rN = (m_chi * m_U) * m_N / (m_chi * m_U + m_N) # \\
\verb#                                                  # \\
\verb# alpha = sqrt(m_N / (2.0 * m_rN * m_rN))          # \\
\verb#                                                  # \\
\verb#    Intf1v_v_Gau_k(x)                        \    # \\
\verb# =  Int_f1v_v_Gau_k(v_max)                   \    # \\
\verb#  - Int_f1v_v_Gau_k(alpha * sqrt(x) / v_U)        # \\
\tabb
 Remind that,
 firstly,
 ``\verb+\+'' (backslash) must be used
 in order to let the including of the definition(s) given in this file
 into the other intrinsic commands correctly.
 Secondly,
 all flexible--kept parameters:
 \verb+k_f1v+
 (stands for the power index $k$
 of the modified Maxwellian velocity distribution),
 \verb+v_0+,
 \verb+v_max+
 as well as
 \verb+AX+ and \verb+m_chi+
 will be read directly from
 the users' initial simulation/data analysis setup
 set earlier on the website.

\subsubsection{Shifted Maxwellian velocity distribution
               \boldmath$f_{1, \sh}(v)$}

 According to Eqs.~(\ref{eqn:f1v_sh}) and (\ref{eqn:v_e}),
 we define
\code{Shifted Maxwellian velocity distribution
      \boldmath$f_{1, \sh}(v)$
      \label{code:f1v_sh}}
\verb# double f1v_sh(double vv, double tt)                             # \\
\verb# {                                                               # \\
\verb#   return                                                        # \\
\verb#     (1.0 / sqrt(M_PI))         *                                # \\
\verb#     (vv / v_0 / v_e(tt)) / v_U *                                # \\
\verb#     (  exp(-(vv - v_e(tt)) * (vv - v_e(tt)) / (v_0 * v_0) )     # \\
\verb#      - exp(-(vv + v_e(tt)) * (vv + v_e(tt)) / (v_0 * v_0) )  ); # \\
\verb# }                                                               # \\
\tabb
 and
\code{Time--dependent Earth's velocity in the Galactic frame
      \boldmath$\ve(t)$
      \label{code:v_e}}
\verb# double v_e(double tt)                                     # \\
\verb# {                                                         # \\
\verb#   return v_0 * ( 1.05 + 0.07 * cos(omega * (tt - t_p)) ); # \\
\verb# }                                                         # \\
\tabb
 Then,
 since
\beq
    \int_{\alpha \sqrt{Q}}^{\infty} \bfrac{f_{1, \sh}(v)}{v} dv
 =  \frac{1}{2 \ve}
    \bbrac{  \erf\afrac{\alpha \sqrt{Q} + \ve}{v_0}
           - \erf\afrac{\alpha \sqrt{Q} - \ve}{v_0} }
\~,
\label{eqn:Intf1v_v_sh}
\eeq
 we can define
\code{Integral over the shifted Maxwellian velocity distribution
      \boldmath$\int_{\alpha \sqrt{Q}}^{\infty} \bbig{f_{1, \sh}(v) / v} dv$
      \label{code:Intf1v_v_sh}}
\verb# double Intf1v_v_sh(double mchi, int A, double QQ, double tt)     # \\
\verb# {                                                                # \\
\verb#   return                                                         # \\
\verb#     (1.0 / 2.0 / (v_e(tt) * v_U)) *                              # \\
\verb#     (  erf( (alpha(mchi, A) * sqrt(QQ) + (v_e(tt) * v_U)) /      # \\
\verb#             (v_0 * v_U)                                    )     # \\
\verb#      - erf( (alpha(mchi, A) * sqrt(QQ) - (v_e(tt) * v_U)) /      # \\
\verb#             (v_0 * v_U)                                    )  ); # \\
\verb# }                                                                # \\
\tabb

 On the other hand,
 for drawing output plots,
 we have
\newpage
\code{Shifted Maxwellian velocity distribution
      \boldmath$f_{1, \sh}(v)$
      (for Gnuplot)
      \label{code:f1v_x_sh}}
\verb# M_PI   = 3.141593                                         # \\
\verb# omega  = 2.0 * M_PI / 365.0                               # \\
\verb#                                                           # \\
\verb# v_e = v_0 * ( 1.05 + 0.07 * cos(omega * (t_expt - t_p)) ) # \\
\verb#                                                           # \\
\verb#    f1v_sh(x)                                         \    # \\
\verb# =  (1.0 / sqrt(M_PI)) *                              \    # \\
\verb#    (x / v_0 / v_e)    *                              \    # \\
\verb#    (  exp(-(x - v_e) * (x - v_e) / (v_0 * v_0))      \    # \\
\verb#     - exp(-(x + v_e) * (x + v_e) / (v_0 * v_0))  )        # \\
\tabb
 and
\code{Integral over the shifted Maxwellian velocity distribution
      \boldmath$\int_{\alpha \sqrt{Q}}^{\infty} \bbig{f_{1, \sh}(v) / v} dv$
      (for Gnuplot)
      \label{code:Intf1v_v_x_sh}}
\verb# c   = 1.0                                                           # \\
\verb# m_U = 1e6 / (c * c)                                                 # \\
\verb# v_U = c / (2.9979246 * 1e5)                                         # \\
\verb#                                                                     # \\
\verb# m_N  = (0.938272 * m_U) * AX * 0.99                                 # \\
\verb# m_rN = (m_chi * m_U) * m_N / (m_chi * m_U + m_N)                    # \\
\verb#                                                                     # \\
\verb# alpha = sqrt(m_N / (2.0 * m_rN * m_rN))                             # \\
\verb#                                                                     # \\
\verb#    Intf1v_v_sh(x)                                                 \ # \\
\verb# =  (1.0 / 2.0 / (v_e * v_U)) *                                    \ # \\
\verb#    (  erf( ( alpha * sqrt(x) + (v_e * v_U) ) / (v_0 * v_U) )      \ # \\
\verb#     - erf( ( alpha * sqrt(x) - (v_e * v_U) ) / (v_0 * v_U) )  )     # \\
\tabb
 Remind that,
 firstly,
 ``\verb+\+'' (backslash) must be used
 in order to let the including of the definition(s) given in this file
 into the other intrinsic commands correctly.
 Secondly,
 all flexible--kept parameters:
 \verb+t_p+,
 \verb+t_expt+,
 \verb+v_0+
 as well as
 \verb+AX+ and \verb+m_chi+
 will be read directly from
 the users' initial simulation/data analysis setup
 set earlier on the website.

\subsection{Elastic nuclear form factors}

 In this subsection,
 we give the codes for
 intrinsically defined
 elastic nuclear form factors
 given in Sec.~3.5
 in the \amidas\ package.

\subsubsection{Exponential form factor
               \boldmath$F_{\rm ex}(Q)$}

 For the exponential form factor $F_{\rm ex}(Q)$,
 the nuclear radius $R_0(A)$
 and the nuclear coherence energy $Q_0(A)$
 have been defined
 as functions of the atomic mass number $A$:
\code{Nuclear radius \boldmath$R_0(A)$
      \label{code:R_0}}
\verb# double R_0(int A)                                  # \\
\verb# {                                                  # \\
\verb#   return (0.3 + 0.91 * cbrt(m_N(A) / m_U)) * fm_U; # \\
\verb# }                                                  # \\
\tabb
 and
\code{Nuclear coherence energy \boldmath$Q_0(A)$
      \label{code:Q_0}}
\verb# double Q_0(int A)                          # \\
\verb# {                                          # \\
\verb#   return 1.5 / (m_N(A) * R_0(A) * R_0(A)); # \\
\verb# }                                          # \\
\tabb
 Then we can define
\code{Squared exponential form factor
      \boldmath$F_{\rm ex}^2(A, Q)$
      \label{code:FQ_ex}}
\verb# double FQ_ex(int A, double QQ) # \\
\verb# {                              # \\
\verb#   return exp(-QQ / Q_0(A));    # \\
\verb# }                              # \\
\tabb
 and
\code{Derivative of the squared exponential form factor
      \boldmath$d F_{\rm ex}^2(A, Q) / dQ$
      \label{code:dFQdQ_ex}}
\verb# double dFQdQ_ex(int A, double QQ)        # \\
\verb# {                                        # \\
\verb#   return -(1.0 / Q_0(A)) * FQ_ex(A, QQ); # \\
\verb# }                                        # \\
\tabb
 since
\beq
    \Dd{F_{\rm ex}^2(Q)}{Q}
 =  \dd{Q} \aBig{e^{-Q / Q_0}}
 =- \frac{1}{Q_0} \~ e^{-Q / Q_0}
\~.
\label{eqn:dFQdQ_ex}
\eeq

 On the other hand,
 for drawing the output
 generating WIMP--signal recoil spectrum,
 we need
\code{Squared exponential form factor
      \boldmath$F_{\rm ex}^2(Q)$
      (for Gnuplot)
      \label{code:FQ_x_ex}}
\verb# c    = 1.0                                               # \\
\verb# m_U  = 1e6 / (c * c)                                     # \\
\verb# fm_U = c / (0.197327 * 1e6)                              # \\
\verb#                                                          # \\
\verb# m_N  = (0.938272 * m_U) * AX * 0.99                      # \\
\verb# m_rN = (m_chi * m_U) * m_N / (m_chi * m_U + m_N)         # \\
\verb#                                                          # \\
\verb# alpha = sqrt(m_N / (2.0 * m_rN * m_rN))                  # \\
\verb#                                                          # \\
\verb# R_0 = ( 0.3 + 0.91 * (m_N / m_U) ** (1.0 / 3.0) ) * fm_U # \\
\verb# Q_0 = 1.5 / (m_N * R_0 * R_0)                            # \\
\verb#                                                          # \\
\verb#    FQ_ex(x)        \                                     # \\
\verb# =  exp(-x / Q_0)                                         # \\
\tabb
 Remind that,
 firstly,
 ``\verb+\+'' (backslash) must be used
 in order to let the including of the definition(s) given in this file
 into the other intrinsic commands correctly.
 Secondly,
 two flexible--kept parameters:
 \verb+AX+ and \verb+m_chi+
 will be read directly from
 the users' initial simulation/data analysis setup
 set earlier on the website.

\subsubsection{Woods--Saxon form factor
               \boldmath$F_{\rm WS}(Q)$}

 For the Woods--Saxon form factor $F_{\rm WS}(Q)$,
 the nuclear skin thickness $s$,
 the radius $R_A(A)$,
 and the effective nuclear radius $R_1(A)$
 are defined as
\code{Nuclear skin thickness \boldmath$s$
      \label{code:ss}}
\verb# double ss = fm_U; # \\
\tabb
 and
\code{\boldmath$R_A(A)$
      \label{code:R_A}}
\verb# double R_A(int A)              # \\
\verb# {                              # \\
\verb#   return 1.2 * cbrt(A) * fm_U; # \\
\verb# }                              # \\
\tabb
 Then we have
\code{Effective nuclear radius \boldmath$R_1(A)$
      \label{code:R_1}}
\verb# double R_1(int A)                               # \\
\verb# {                                               # \\
\verb#   return sqrt(R_A(A) * R_A(A) - 5.0 * ss * ss); # \\
\verb# }                                               # \\
\tabb
 Meanwhile,
 the transferred 3-momentum 
 $q$ given in Eq.~(\ref{eqn:qq})
 has been given
 as a function of the atomic mass number $A$ and recoil energy $Q$:
\code{Transferred 3-momentum \boldmath$q(A, Q)$
      \label{code:qq}}
\verb# double qq(int A, double QQ)       # \\
\verb# {                                 # \\
\verb#   return sqrt(2.0 * m_N(A) * QQ); # \\
\verb# }                                 # \\
\tabb
 Then we can define
\code{Squared Woods--Saxon form factor
      \boldmath$F_{\rm WS}^2(A, Q)$
      \label{code:FQ_WS}}
\verb# double FQ_WS(int A, double QQ)                                             # \\
\verb# {                                                                          # \\
\verb#   if (QQ == 0.0)                                                           # \\
\verb#   {                                                                        # \\
\verb#     return 1.0;                                                            # \\
\verb#   }                                                                        # \\
\verb#                                                                            # \\
\verb#   else                                                                     # \\
\verb#   {                                                                        # \\
\verb#     return                                                                 # \\
\verb#       ( 3.0 * sphBesselj(1, qq(A, QQ) * R_1(A)) / (qq(A, QQ) * R_1(A)) ) * # \\
\verb#       ( 3.0 * sphBesselj(1, qq(A, QQ) * R_1(A)) / (qq(A, QQ) * R_1(A)) ) * # \\
\verb#       exp(-(qq(A, QQ) * ss) * (qq(A, QQ) * ss));                           # \\
\verb#   }                                                                        # \\
\verb# }                                                                          # \\
\tabb

 For defining the derivative of the squared Woods--Saxon form factor,
 $d F_{\rm WS}^2(Q) / dQ$,
 first,
 we find that
\beqn
     \Dd{F_{\rm WS}^2(Q)}{Q}
 \=  \dd{Q} \cbrac{\bfrac{3 j_1(q R_1)}{q R_1}^2 e^{-(q s)^2}}
     \non\\
 \=  2 \bfrac{3 j_1(q R_1)}{q R_1} e^{-(q s)^2}
     \bfrac{3 j'_1(q R_1) (q' R_1) (q R_1) - 3 j_1(q R_1) (q' R_1)}{(q R_1)^2}
     \non\\
 \conti ~~~~~~~~~~~~ 
   + \bfrac{3 j_1(q R_1)}{q R_1}^2 e^{-(q s)^2}
     \bBig{\abrac{-2 q s^2} q'}
     \non\\
 \=  2
     \bfrac{3 j_1(q R_1)}{q R_1}^2 e^{-(q s)^2}
     \bbrac{\frac{j'_1(q R_1) (q R_1) - j_1(q R_1)}{j_1(q R_1) \~ q} - q s^2}
     \aDd{q}{Q}
     \non\\
 \=  \bbrac{\frac{j'_1(q R_1) \~ R_1}{j_1(q R_1)} - \frac{1}{q} - q s^2}
     \afrac{2 \mN}{q}
     F_{\rm WS}^2(Q)
\~,
\label{eqn:dFQdQ_WS}
\eeqn
 where,
 from the expression (\ref{eqn:qq}) for $q$,
 we have
\beq
    \Dd{q}{Q}
 =  \sqrt{2 \mN} \cdot \frac{1}{2 \sqrt{Q}}
 =  \sqrt{2 \mN} \cdot \frac{\sqrt{2 \mN}}{2 q}
 =  \frac{\mN}{q}
\~.
\label{eqn:dqdQ}
\eeq
 The expression (\ref{eqn:dFQdQ_WS}) can be used for $Q > 0$.
 However,
 for the limit case $Q = q = 0$,
 since
 \cite{SchaumMathHandbook}
\beq
    j_1(x)
 =  \sfrac{\pi}{2 x} \~ J_{3 / 2}(x)
 =  \frac{1}{x} \abrac{\frac{\sin x}{x} - \cos x}
\~,
\label{eqn:j_1}
\eeq
 we can get
\beq
    j'_1(x)
 =  \frac{\abrac{x^2 - 2} \sin x + 2 x \cos x}{x^3}
\~.
\label{eqn:j'_1}
\eeq
 Then,
 by letting $x = q R_1$,
 the first two terms in the bracket of
 the right--hand side of Eq.~(\ref{eqn:dFQdQ_WS})
 (multiply by $1 / q$)
 can be written
 and found that
\beqn
    \frac{j'_1(x)}{j_1(x)} \afrac{R_1^2}{x} - \frac{R_1^2}{x^2}
 =  \bbrac{  \frac{\abrac{x^2 - 2} \sin x + 2 x \cos x}{x^2 \abrac{\sin x - x \cos x}}
           - \frac{1}{x^2}                                                            }
    R_1^2
 \to
  - \frac{R_1^2}{5}
\~,
\eeqn
 as $x \to 0$.
 Hence,
 we can obtain that,
 for the limit case $Q = q = 0$,
\beq
    \vright{\Dd{F_{\rm WS}^2(Q)}{Q}}_{Q = 0}
 =  \abrac{- \frac{R_1^2}{5} - s^2}
    \abrac{2 \mN}
 =- \afrac{2}{5} R_A^2 \mN
\~,
\label{eqn:dFQdQ_WS_0}
\eeq
 where we use the expression (\ref{eqn:R1}) for $R_1$ and
\beq
    F_{\rm WS}^2(Q = 0)
  = 1
\~.
\label{eqn:FQ_WS_0}
\eeq
 Now,
 we can define
 the derivative of the squared Woods--Saxon form factor
 as
\newpage
\code{Derivative of the squared Woods--Saxon form factor
      \boldmath$d F_{\rm WS}^2(A, Q) / dQ$
      \label{code:dFQdQ_WS}}
\verb# double dFQdQ_WS(int A, double QQ)                         # \\
\verb# {                                                         # \\
\verb#   if (QQ == 0.0)                                          # \\
\verb#   {                                                       # \\
\verb#     return -0.4 * (R_A(A) * R_A(A)) * m_N(A);             # \\
\verb#   }                                                       # \\
\verb#                                                           # \\
\verb#   else                                                    # \\
\verb#   {                                                       # \\
\verb#     return                                                # \\
\verb#       (  dsphBesselj(1, qq(A, QQ) * R_1(A)) * R_1(A) /    # \\
\verb#           sphBesselj(1, qq(A, QQ) * R_1(A))               # \\
\verb#        - 1.0 / qq(A, QQ)                                  # \\
\verb#        - (qq(A, QQ) * ss * ss)                        ) * # \\
\verb#       FQ_WS(A, QQ) * ((2.0 * m_N(A)) / qq(A, QQ));        # \\
\verb#   }                                                       # \\
\verb# }                                                         # \\
\tabb
 Moreover,
 the needed spherical Bessel functions
\beq
    j_n(x)
 =  \sfrac{\pi}{2 x} \~ J_{n + 1/2}(x)
\~,
\label{eqn:jn_Jn_12}
\eeq
 and their derivatives
\beq
    j'_n(x)
 =  \sfrac{\pi}{2 x}
    \bbrac{-\frac{1}{2 x} \~ J_{n + 1/2}(x) + J_{n + 1/2}'(x)}
\~,
\label{eqn:j'n_Jn_12}
\eeq
 have also been defined
 in an intrinsic package of the \amidas\ code.
 We define at first
 the half--integer Bessel functions $J_{n + 1/2}(x)$,
 for $n = 0,~\pm 1,~\pm 2,~\cdots$,
 by
\code{Bessel functions \boldmath$J_{n / 2}(x)$
      for $n = \pm 1,~\pm 3,~\pm 5,~\cdots$
      \label{code:BesselJ}}
\verb# double BesselJ(int n, int m, double x)                                                # \\
\verb# {                                                                                     # \\
\verb#   if (m == 2)                                                                         # \\
\verb#   {                                                                                   # \\
\verb#     switch (n)                                                                        # \\
\verb#     {                                                                                 # \\
\verb#       case  1:                                                                        # \\
\verb#         return                                                                        # \\
\verb#           sqrt(2.0 / M_PI / x) * sin(x);                                              # \\
\verb#         break;                                                                        # \\
\verb#                                                                                       # \\
\verb#       case -1:                                                                        # \\
\verb#         return                                                                        # \\
\verb#           sqrt(2.0 / M_PI / x) * cos(x);                                              # \\
\verb#         break;                                                                        # \\
\verb#                                                                                       # \\
\verb#       case  3:                                                                        # \\
\verb#         return                                                                        # \\
\verb#           sqrt(2.0 / M_PI / x) * (sin(x) / x - cos(x));                               # \\
\verb#         break;                                                                        # \\
\tabc
%
\verb#       case -3:                                                                        # \\
\verb#         return                                                                        # \\
\verb#           sqrt(2.0 / M_PI / x) * (cos(x) / x + sin(x)) * (-1);                        # \\
\verb#         break;                                                                        # \\
\verb#                                                                                       # \\
\verb#       case  5:                                                                        # \\
\verb#         return                                                                        # \\
\verb#           sqrt(2.0 / M_PI / x) * ((3.0 / (x * x) - 1) * sin(x) - (3.0 / x) * cos(x)); # \\
\verb#         break;                                                                        # \\
\verb#                                                                                       # \\
\verb#       case -5:                                                                        # \\
\verb#         return                                                                        # \\
\verb#           sqrt(2.0 / M_PI / x) * ((3.0 / (x * x) - 1) * cos(x) + (3.0 / x) * sin(x)); # \\
\verb#         break;                                                                        # \\
\verb#                                                                                       # \\
\verb#       default:                                                                        # \\
\verb#         if (n >  5 &&  n % 2 == 1)                                                    # \\
\verb#         {                                                                             # \\
\verb#           return                                                                      # \\
\verb#             ((n - 2) / x) * BesselJ(n - 2, 2, x) - BesselJ(n - 4, 2, x);              # \\
\verb#         }                                                                             # \\
\verb#                                                                                       # \\
\verb#         else                                                                          # \\
\verb#         if (n < -5 && -n % 2 == 1)                                                    # \\
\verb#         {                                                                             # \\
\verb#           return                                                                      # \\
\verb#             ((n + 2) / x) * BesselJ(n + 2, 2, x) - BesselJ(n + 4, 2, x);              # \\
\verb#         }                                                                             # \\
\verb#     }                                                                                 # \\
\verb#   }                                                                                   # \\
\verb# }                                                                                     # \\
\tabb
 Here for the case $n \ge 3$ or $n \le -3$
 (i.e.~$\pm n \ge 3$),
 we use the following relation
 \cite{SchaumMathHandbook}:
\beq
    J_{n \pm 1}(x)
 =  \afrac{2 n}{x} J_{n}(x) - J_{n \mp 1}(x)
\~.
\eeq
 Then one can derive that
\beqn
     J_{m / 2}(x)
 \=  \bfrac{2 \abrac{m / 2 \mp 1}}{x} J_{m / 2 \mp 1}(x) - J_{m / 2 \mp 2}(x)
     \non\\
 \=  \afrac{m \mp 2}{x} J_{(m \mp 2) / 2}(x) - J_{(m \mp 4) / 2}(x)
\~.
\label{eqn:J_m2}
\eeqn
 For the derivatives of the Bessel functions,
 since
 \cite{SchaumMathHandbook}
\beqn
     J'_{n}(x)
 \=  \frac{1}{2} \bBig{J_{n - 1}(x) - J_{n + 1}(x)}
     \non\\
 \=- \afrac{n}{x} J_{n}(x) + J_{n - 1}(x)
     \non\\
 \=  \afrac{n}{x} J_{n}(x) - J_{n + 1}(x)
\~,
\label{eqn:J'n}
\eeqn
 we can find that
\beq
      J'_{m / 2}(x)
 =\mp \afrac{m}{2 x} J_{m / 2}(x) \pm J_{(m \mp 2) / 2}(x)
\~.
\label{eqn:J'_m2}
\eeq
 Therefore,
 according to Eq.~(\ref{eqn:jn_Jn_12}),
 we can first define
\code{Spherical Bessel functions \boldmath$j_{n}(x)$
      for $n = 0,~1,~2,~\cdots$
      \label{code:sphBesselj}}
\verb# double sphBesselj(int n, double x)                        # \\
\verb# {                                                         # \\
\verb#   return sqrt(M_PI / 2.0 / x) * BesselJ(2 * n + 1, 2, x); # \\
\verb# }                                                         # \\
\tabb
 Meanwhile,
 from Eqs.~(\ref{eqn:j'n_Jn_12}) and (\ref{eqn:J'n}),
 we can obtain that
\beqn
     j_n'(x)
 \=  \sfrac{\pi}{2 x}
     \bbrac{-\frac{1}{2 x} \~ J_{n + 1/2}(x) + J_{n + 1/2}'(x)}
     \non\\
 \=  \sfrac{\pi}{2 x}
     \cbrac{- \frac{1}{2 x} \~ J_{n + 1/2}(x)
            + \frac{1}{2}
              \bBig{  J_{n - 1/2}(x)
                    - J_{n + 3/2}(x)  }        }
     \non\\
 \=  \sfrac{\pi}{2 x} \cdot \frac{1}{2}
     \cbrac{- \frac{1}{x} \~ J_{n + 1/2}(x)
            +                J_{n - 1/2}(x)
            - \bbrac{  \afrac{2 n + 1}{x} \~ J_{n + 1/2}(x)
                     -                       J_{n - 1/2}(x)  }  }
     \non\\
 \=  \sfrac{\pi}{2 x}
     \bbrac{- \afrac{n + 1}{x} J_{n + 1/2}(x)
            +                  J_{n - 1/2}(x)  }
\~,
\eeqn
 and thus define
\code{Derivatives of the Spherical Bessel functions \boldmath$j'_{n}(x)$
      for $n = 0,~1,~2,~\cdots$
      \label{code:dsphBesselj}}
\verb# double dsphBesselj(int n, double x)             # \\
\verb# {                                               # \\
\verb#   return                                        # \\
\verb#     sqrt(M_PI / 2.0 / x) *                      # \\
\verb#     (- (n + 1) * BesselJ(2 * n + 1, 2, x) / x   # \\
\verb#      +           BesselJ(2 * n - 1, 2, x)    ); # \\
\verb# }                                               # \\
\tabb

 Finally,
 for drawing the output
 generating WIMP--signal recoil spectrum,
 we need
\newpage
\code{Squared Woods--Saxon form factor
      \boldmath$F_{\rm WS}^2(Q)$
      (for Gnuplot)
      \label{code:FQ_x_WS}}
\verb# c    = 1.0                                                       # \\
\verb# m_U  = 1e6 / (c * c)                                             # \\
\verb# fm_U = c / (0.197327 * 1e6)                                      # \\
\verb#                                                                  # \\
\verb# m_N  = (0.938272 * m_U) * AX * 0.99                              # \\
\verb#                                                                  # \\
\verb# ss  = fm_U                                                       # \\
\verb# R_A = 1.2 * (AX ** (1.0 / 3.0)) * fm_U                           # \\
\verb# R_1 = sqrt(R_A * R_A - 5.0 * ss * ss)                            # \\
\verb#                                                                  # \\
\verb#    qq(x)                \                                        # \\
\verb# =  sqrt(2.0 * m_N * x)                                           # \\
\verb#                                                                  # \\
\verb#    sphBesselj_1(x)                                             \ # \\
\verb# =  ( sin(qq(x) * R_1) / (qq(x) * R_1) - cos(qq(x) * R_1) ) /   \ # \\
\verb#    (qq(x) * R_1)                                                 # \\
\verb#                                                                  # \\
\verb#    FQ_WS(x)                                      \               # \\
\verb# =  ( 3.0 * sphBesselj_1(x) / (qq(x) * R_1) ) *   \               # \\
\verb#    ( 3.0 * sphBesselj_1(x) / (qq(x) * R_1) ) *   \               # \\
\verb#    exp(-(qq(x) * ss) * (qq(x) * ss))                             # \\
\tabb
 Remind that,
 firstly,
 ``\verb+\+'' (backslash) must be used
 in order to let the including of the definition(s) given in this file
 into the other intrinsic commands correctly.
 Secondly,
 the flexible--kept parameter:
 \verb+AX+
 will be read directly from
 the users' initial simulation/data analysis setup
 set earlier on the website.

\subsubsection{Modified Woods--Saxon form factor
               \boldmath$F_{\rm WS, Eder}(Q)$}

 First,
 the modified radius $R_A$ given in Eq.~(\ref{eqn:RA_Eder})
 has been defined as
\code{Modified \boldmath$R_A(A)$
      \label{code:R_A_Eder}}
\verb# double R_A_Eder(int A)                   # \\
\verb# {                                        # \\
\verb#   return (1.15 * cbrt(A) + 0.39) * fm_U; # \\
\verb# }                                        # \\
\tabb
 Then the effective nuclear radius $R_1$
 has to be redefined by
\code{Modified effective nuclear radius \boldmath$R_1(A)$
      \label{code:R_1_Eder}}
\verb# double R_1_Eder(int A)                                    # \\
\verb# {                                                         # \\
\verb#   return sqrt(R_A_Eder(A) * R_A_Eder(A) - 5.0 * ss * ss); # \\
\verb# }                                                         # \\
\tabb
 Therefore,
 similar to Codes A\ref{code:FQ_WS} and A\ref{code:dFQdQ_WS}
 for the (derivative of the) Woods--Saxon form factor $F_{\rm WS}^2(Q)$,
 we can define
\newpage
\code{Squared modified Woods--Saxon form factor
      \boldmath$F_{\rm WS, Eder}^2(A, Q)$
      \label{code:FQ_WS_Eder}}
\verb# double FQ_WS_Eder(int A, double QQ)                                                  # \\
\verb# {                                                                                    # \\
\verb#   if (QQ == 0.0)                                                                     # \\
\verb#   {                                                                                  # \\
\verb#     return 1.0;                                                                      # \\
\verb#   }                                                                                  # \\
\verb#                                                                                      # \\
\verb#   else                                                                               # \\
\verb#   {                                                                                  # \\
\verb#     return                                                                           # \\
\verb#       ( 3.0 * sphBesselj(1, qq(A, QQ) * R_1_Eder(A)) / (qq(A, QQ) * R_1_Eder(A)) ) * # \\
\verb#       ( 3.0 * sphBesselj(1, qq(A, QQ) * R_1_Eder(A)) / (qq(A, QQ) * R_1_Eder(A)) ) * # \\
\verb#       exp(-(qq(A, QQ) * ss) * (qq(A, QQ) * ss));                                     # \\
\verb#   }                                                                                  # \\
\verb# }                                                                                    # \\
\tabb
 and
\code{Derivative of the squared modified Woods--Saxon form factor
      \boldmath$d F_{\rm WS, Eder}^2(A, Q) / dQ$
      \label{code:dFQdQ_WS_Eder}}
\verb# double dFQdQ_WS_Eder(int A, double QQ)                              # \\
\verb# {                                                                   # \\
\verb#   if (QQ == 0.0)                                                    # \\
\verb#   {                                                                 # \\
\verb#     return -0.4 * (R_A_Eder(A) * R_A_Eder(A)) * m_N(A);             # \\
\verb#   }                                                                 # \\
\verb#                                                                     # \\
\verb#   else                                                              # \\
\verb#   {                                                                 # \\
\verb#     return                                                          # \\
\verb#       (  dsphBesselj(1, qq(A, QQ) * R_1_Eder(A)) * R_1_Eder(A) /    # \\
\verb#           sphBesselj(1, qq(A, QQ) * R_1_Eder(A))                    # \\
\verb#        - 1.0 / qq(A, QQ)                                            # \\
\verb#        - (qq(A, QQ) * ss * ss)                                  ) * # \\
\verb#       FQ_WS_Eder(A, QQ) * ((2.0 * m_N(A)) / qq(A, QQ));             # \\
\verb#   }                                                                 # \\
\verb# }                                                                   # \\
\tabb

 On the other hand,
 for drawing the output
 generating WIMP--signal recoil spectrum,
 we need
\newpage
\code{Squared modified Woods--Saxon form factor
      \boldmath$F_{\rm WS, Eder}^2(Q)$
      (for Gnuplot)
      \label{code:FQ_x_WS_Eder}}
\verb# c    = 1.0                                                                      # \\
\verb# m_U  = 1e6 / (c * c)                                                            # \\
\verb# fm_U = c / (0.197327 * 1e6)                                                     # \\
\verb#                                                                                 # \\
\verb# m_N  = (0.938272 * m_U) * AX * 0.99                                             # \\
\verb#                                                                                 # \\
\verb# ss       = fm_U                                                                 # \\
\verb# R_A_Eder = ( 1.15 * (AX ** (1.0 / 3.0)) + 0.39 ) * fm_U                         # \\
\verb# R_1_Eder = sqrt(R_A_Eder * R_A_Eder - 5.0 * ss * ss)                            # \\
\verb#                                                                                 # \\
\verb#    qq(x)                \                                                       # \\
\verb# =  sqrt(2.0 * m_N * x)                                                          # \\
\verb#                                                                                 # \\
\verb#    sphBesselj_1(x)                                                            \ # \\
\verb# =  ( sin(qq(x) * R_1_Eder) / (qq(x) * R_1_Eder) - cos(qq(x) * R_1_Eder) ) /   \ # \\
\verb#    (qq(x) * R_1_Eder)                                                           # \\
\verb#                                                                                 # \\
\verb#    FQ_WS_Eder(x)                                      \                         # \\
\verb# =  ( 3.0 * sphBesselj_1(x) / (qq(x) * R_1_Eder) ) *   \                         # \\
\verb#    ( 3.0 * sphBesselj_1(x) / (qq(x) * R_1_Eder) ) *   \                         # \\
\verb#    exp(-(qq(x) * ss) * (qq(x) * ss))                                            # \\
\tabb
 Remind that,
 firstly,
 ``\verb+\+'' (backslash) must be used
 in order to let the including of the definition(s) given in this file
 into the other intrinsic commands correctly.
 Secondly,
 the flexible--kept parameter:
 \verb+AX+
 will be read directly from
 the users' initial simulation/data analysis setup
 set earlier on the website.

\subsubsection{Helm form factor
               \boldmath$F_{\rm Helm}(Q)$}

 For the Helm form factor $F_{\rm Helm}(Q)$,
 the nuclear skin thickness $s$,
 the radii $r_0$ and $R_A(A)$,
 and the effective nuclear radius $R_1(A)$
 are defined as
\code{Nuclear skin thickness \boldmath$s$
      \label{code:ss_Helm}}
\verb# double ss_Helm = 0.9 * fm_U; # \\
\tabb
\code{\boldmath$r_0$
      \label{code:r_0_Helm}}
\verb# double r_0_Helm = 0.52 * fm_U; # \\
\tabb
\code{\boldmath$R_A(A)$
      \label{code:R_A_Helm}}
\verb# double R_A_Helm(int A)                  # \\
\verb# {                                       # \\
\verb#   return (1.23 * cbrt(A) - 0.6) * fm_U; # \\
\verb# }                                       # \\
\tabb
 Then the effective nuclear radius $R_1$
 has to be redefined by
\newpage
\code{Effective nuclear radius \boldmath$R_1(A)$
      \label{code:R_1_Helm}}
\verb# double R_1_Helm(int A)                                         # \\
\verb# {                                                              # \\
\verb#   return                                                       # \\
\verb#     sqrt(  R_A_Helm(A) * R_A_Helm(A)                           # \\
\verb#          + (7.0 / 3.0) * (M_PI * M_PI) * (r_0_Helm * r_0_Helm) # \\
\verb#          - 5.0 * ss_Helm * ss_Helm);                           # \\
\verb# }                                                              # \\
\tabb
 Therefore,
 similar to Codes A\ref{code:FQ_WS} and A\ref{code:dFQdQ_WS}
 for the (derivative of the) Woods--Saxon form factor $F_{\rm WS}^2(Q)$,
 we can define
\code{Squared Helm form factor
      \boldmath$F_{\rm Helm}^2(A, Q)$
      \label{code:FQ_Helm}}
\verb# double FQ_Helm(int A, double QQ)                                                     # \\
\verb# {                                                                                    # \\
\verb#   if (QQ == 0.0)                                                                     # \\
\verb#   {                                                                                  # \\
\verb#     return 1.0;                                                                      # \\
\verb#   }                                                                                  # \\
\verb#                                                                                      # \\
\verb#   else                                                                               # \\
\verb#   {                                                                                  # \\
\verb#     return                                                                           # \\
\verb#       ( 3.0 * sphBesselj(1, qq(A, QQ) * R_1_Helm(A)) / (qq(A, QQ) * R_1_Helm(A)) ) * # \\
\verb#       ( 3.0 * sphBesselj(1, qq(A, QQ) * R_1_Helm(A)) / (qq(A, QQ) * R_1_Helm(A)) ) * # \\
\verb#       exp(-(qq(A, QQ) * ss_Helm) * (qq(A, QQ) * ss_Helm));                           # \\
\verb#   }                                                                                  # \\
\verb# }                                                                                    # \\
\tabb
 and
\code{Derivative of the squared Helm form factor
      \boldmath$d F_{\rm Helm}^2(A, Q) / dQ$
      \label{code:dFQdQ_Helm}}
\verb# double dFQdQ_Helm(int A, double QQ)                                 # \\
\verb# {                                                                   # \\
\verb#   if (QQ == 0.0)                                                    # \\
\verb#   {                                                                 # \\
\verb#     return                                                          # \\
\verb#       -0.4                                                      *   # \\
\verb#       (  R_A_Helm(A) * R_A_Helm(A)                                  # \\
\verb#        + (7.0 / 3.0) * (M_PI * M_PI) * (r_0_Helm * r_0_Helm)  ) *   # \\
\verb#       m_N(A);                                                       # \\
\verb#   }                                                                 # \\
\verb#                                                                     # \\
\verb#   else                                                              # \\
\verb#   {                                                                 # \\
\verb#     return                                                          # \\
\verb#       (  dsphBesselj(1, qq(A, QQ) * R_1_Helm(A)) * R_1_Helm(A) /    # \\
\verb#           sphBesselj(1, qq(A, QQ) * R_1_Helm(A))                    # \\
\verb#        - 1.0 / qq(A, QQ)                                            # \\
\verb#        - (qq(A, QQ) * ss_Helm * ss_Helm)                        ) * # \\
\verb#       FQ_Helm(A, QQ) * ((2.0 * m_N(A)) / qq(A, QQ));                # \\
\verb#   }                                                                 # \\
\verb# }                                                                   # \\
\tabb
 Note here that
 the expression (\ref{eqn:R1_Helm}) for $R_1$
 has to be used.

 On the other hand,
 for drawing the output
 generating WIMP--signal recoil spectrum,
 we need
\code{Squared Helm form factor
      \boldmath$F_{\rm Helm}^2(Q)$
      (for Gnuplot)
      \label{code:FQ_x_Helm}}
\verb# M_PI = 3.141593                                                                 # \\
\verb#                                                                                 # \\
\verb# c    = 1.0                                                                      # \\
\verb# m_U  = 1e6 / (c * c)                                                            # \\
\verb# fm_U = c / (0.197327 * 1e6)                                                     # \\
\verb#                                                                                 # \\
\verb# m_N  = (0.938272 * m_U) * AX * 0.99                                             # \\
\verb#                                                                                 # \\
\verb# ss_Helm  = 0.9  * fm_U                                                          # \\
\verb# r_0_Helm = 0.52 * fm_U                                                          # \\
\verb#                                                                                 # \\
\verb# R_A_Helm = ( 1.23 * (AX ** (1.0 / 3.0)) - 0.6 ) * fm_U                          # \\
\verb#                                                                                 # \\
\verb#    R_1_Helm                                                        \            # \\
\verb# =  sqrt(  R_A_Helm * R_A_Helm                                      \            # \\
\verb#         + (7.0 / 3.0) * (M_PI ** 2.0) * (r_0_Helm * r_0_Helm)      \            # \\
\verb#         - 5.0 * ss_Helm * ss_Helm                              )                # \\
\verb#                                                                                 # \\
\verb#    qq(x)                \                                                       # \\
\verb# =  sqrt(2.0 * m_N * x)                                                          # \\
\verb#                                                                                 # \\
\verb#    sphBesselj_1(x)                                                            \ # \\
\verb# =  ( sin(qq(x) * R_1_Helm) / (qq(x) * R_1_Helm) - cos(qq(x) * R_1_Helm) ) /   \ # \\
\verb#    (qq(x) * R_1_Helm)                                                           # \\
\verb#                                                                                 # \\
\verb#    FQ_Helm(x)                                         \                         # \\
\verb# =  ( 3.0 * sphBesselj_1(x) / (qq(x) * R_1_Helm) ) *   \                         # \\
\verb#    ( 3.0 * sphBesselj_1(x) / (qq(x) * R_1_Helm) ) *   \                         # \\
\verb#    exp(-(qq(x) * ss_Helm) * (qq(x) * ss_Helm))                                  # \\
\tabb
 Remind that,
 firstly,
 ``\verb+\+'' (backslash) must be used
 in order to let the including of the definition(s) given in this file
 into the other intrinsic commands correctly.
 Secondly,
 the flexible--kept parameter:
 \verb+AX+
 will be read directly from
 the users' initial simulation/data analysis setup
 set earlier on the website.

\subsubsection{Thin--shell form factor
               \boldmath$F_{\rm TS}(Q)$}

 For the thin--shell form factor $F_{\rm TS}(Q)$,
 the lower and upper bounds of $q R_1$ given in Eq.~(\ref{eqn:FQ_SD_TS}),
 between which the form factor is a constant ($\simeq 0.047$),
 is defined by
\code{Lower and upper bounds of \boldmath$q R_1$
      \label{code:qqR_1_min_max}}
\verb# qqR_1_min = 2.55; # \\
\verb# qqR_1_max = 4.50; # \\
\tabb
 and the constant is redefined as
\code{Constant of the thin--shell form factort (\boldmath$\simeq 0.047$)
      \label{code:FQ_TS_const}}
\verb#    FQ_TS_const                # \\
\verb# =  sphBesselj(0, qqR_1_min) * # \\
\verb#    sphBesselj(0, qqR_1_min);  # \\
\tabb
 Meanwhile,
 the lower and upper energy bounds
 can be estimated by
\code{Lower and upper energy bounds
      \label{code:QQ_SD_min_max}}
\verb# double QQ_SD_min(int A)                                              # \\
\verb# {                                                                    # \\
\verb#   return (qqR_1_min * qqR_1_min) / (2.0 * m_N(A) * R_1(A) * R_1(A)); # \\
\verb# }                                                                    # \\
\verb#                                                                      # \\
\verb# double QQ_SD_max(int A)                                              # \\
\verb# {                                                                    # \\
\verb#   return (qqR_1_max * qqR_1_max) / (2.0 * m_N(A) * R_1(A) * R_1(A)); # \\
\verb# }                                                                    # \\
\tabb
 Then we can define
\code{Squared thin--shell form factor
      \boldmath$F_{\rm TS}^2(A, Q)$
      \label{code:FQ_TS}}
\verb# double FQ_TS(int A, double QQ)            # \\
\verb# {                                         # \\
\verb#   if (QQ == 0.0)                          # \\
\verb#   {                                       # \\
\verb#     return 1.0;                           # \\
\verb#   }                                       # \\
\verb#                                           # \\
\verb#   else                                    # \\
\verb#   if (QQ <= QQ_SD_min(A) ||               # \\
\verb#       QQ >= QQ_SD_max(A)   )              # \\
\verb#   {                                       # \\
\verb#     return                                # \\
\verb#       sphBesselj(0, qq(A, QQ) * R_1(A)) * # \\
\verb#       sphBesselj(0, qq(A, QQ) * R_1(A));  # \\
\verb#   }                                       # \\
\verb#                                           # \\
\verb#   else                                    # \\
\verb#   {                                       # \\
\verb#     return FQ_TS_const;                   # \\
\verb#   }                                       # \\
\verb# }                                         # \\
\tabb

 For defining the derivative of the squared thin--shell form factor,
 $d F_{\rm TS}^2(Q) / dQ$,
 first,
 we find that
\beq
     \Dd{F_{\rm TS}^2(Q)}{Q}
  =  \dd{Q} \bbigg{j_0^2(q R_1)}
  =  \bfrac{j'_0(q R_1) \~ R_1}{j_0(q R_1)}
     \afrac{2 \mN}{q}
     F_{\rm TS}^2(Q)
\~,
\label{eqn:dFQdQ_TS}
\eeq
 where we have used Eq.~(\ref{eqn:dqdQ}).
 For the limit case $Q = q = 0$,
 since
 \cite{SchaumMathHandbook}
\beq
    j_0(x)
 =  \sfrac{\pi}{2 x} \~ J_{1 / 2}(x)
 =  \frac{\sin x}{x}
\~,
\label{eqn:j_0}
\eeq
 we can get
\beq
    j'_0(x)
 =  \frac{x \cos x - \sin x}{x^2}
\~.
\label{eqn:j'_0}
\eeq
 Then,
 by letting $x = q R_1$,
 the term in the bracket of
 the right--hand side of Eq.~(\ref{eqn:dFQdQ_TS})
 (multiply by $1 / q$)
 can be written
 and found that
\beqn
    \frac{j'_0(x)}{j_0(x)} \afrac{R_1^2}{x}
 =  \afrac{x \cos x - \sin x}{x^2 \sin x}
    R_1^2
 \to
  - \frac{R_1^2}{3}
\~,
\eeqn
 as $x \to 0$.
 Hence,
 we can obtain that,
 for the limit case $Q = q = 0$,
\beq
    \vright{\Dd{F_{\rm TS}^2(Q)}{Q}}_{Q = 0}
 =- \afrac{2}{3} R_1^2 \mN
\~,
\label{eqn:dFQdQ_WS_0}
\eeq
 where we have used
\beq
    F_{\rm TS}^2(Q = 0)
  = 1
\~.
\label{eqn:FQ_WS_0}
\eeq
 Now,
 we can define
 the derivative of the squared thin--shell form factor
 as
\code{Derivative of the squared thin--shell form factor
      \boldmath$d F_{\rm TS}^2(A, Q) / dQ$
      \label{code:dFQdQ_TS}}
\verb# double dFQdQ_TS(int A, double QQ)                   # \\
\verb# {                                                   # \\
\verb#   if (QQ == 0.0)                                    # \\
\verb#   {                                                 # \\
\verb#     return                                          # \\
\verb#       -(2.0 / 3.0) * (R_1(A) * R_1(A)) * m_N(A);    # \\
\verb#   }                                                 # \\
\verb#                                                     # \\
\verb#   else                                              # \\
\verb#   if (QQ <= QQ_SD_min(A) ||                         # \\
\verb#       QQ >= QQ_SD_max(A)   )                        # \\
\verb#   {                                                 # \\
\verb#     return                                          # \\
\verb#       dsphBesselj(0, qq(A, QQ) * R_1(A)) * R_1(A) * # \\
\verb#        sphBesselj(0, qq(A, QQ) * R_1(A)) *          # \\
\verb#        ( (2.0 * m_N(A)) / qq(A, QQ) );              # \\
\verb#   }                                                 # \\
\verb#                                                     # \\
\verb#   else                                              # \\
\verb#   {                                                 # \\
\verb#     return 0.0;                                     # \\
\verb#   }                                                 # \\
\verb# }                                                   # \\
\tabb

 Finally,
 for drawing the output
 generating WIMP--signal recoil spectrum,
 we need
\newpage
\code{Squared thin--shell form factor
      \boldmath$F_{\rm TS}^2(Q)$
      (for Gnuplot)
      \label{code:FQ_x_TS}}
\verb# qqR_1_min = 2.55                                                      # \\
\verb# qqR_1_max = 4.50                                                      # \\
\verb#                                                                       #\\
\verb#    FQ_TS_const                      \                                 # \\
\verb# =  (sin(qqR_1_min) / qqR_1_min) *   \                                 # \\
\verb#    (sin(qqR_1_min) / qqR_1_min)                                       # \\
\verb#                                                                       # \\
\verb# c    = 1.0                                                            # \\
\verb# m_U  = 1e6 / (c * c)                                                  # \\
\verb# fm_U = c / (0.197327 * 1e6)                                           # \\
\verb#                                                                       # \\
\verb# m_N  = (0.938272 * m_U) * AX * 0.99                                   # \\
\verb#                                                                       # \\
\verb# ss  = fm_U                                                            # \\
\verb# R_A = 1.2 * (AX ** (1.0 / 3.0)) * fm_U                                # \\
\verb# R_1 = sqrt(R_A * R_A - 5.0 * ss * ss)                                 # \\
\verb#                                                                       # \\
\verb#    qq(x)                \                                             # \\
\verb# =  sqrt(2.0 * m_N * x)                                                # \\
\verb#                                                                       # \\
\verb#    sphBesselj_0(x)                    \                               # \\
\verb# =  sin(qq(x) * R_1) / (qq(x) * R_1)                                   # \\
\verb#                                                                       # \\
\verb#    Xi(x)                                                            \ # \\
\verb# =  ( (qq(x) * R_1 - qqR_1_min) / abs(qq(x) * R_1 - qqR_1_min) ) *   \ # \\
\verb#    ( (qq(x) * R_1 - qqR_1_max) / abs(qq(x) * R_1 - qqR_1_max) )       # \\
\verb#                                                                       # \\
\verb#    FQ_TS(x)                              \                            # \\
\verb# =  sphBesselj_0(x) * sphBesselj_0(x) *   \                            # \\
\verb#    (1.0 + Xi(x)) / 2.0                   \                            # \\
\verb#  + FQ_TS_const * (1.0 - Xi(x)) / 2.0                                  # \\
\tabb
 Here we define an auxiliary function
\beq
    \Xi(x)
 =  \frac{x - x_{\rm min}}{\vbrac{x - x_{\rm min}}} \cdot
    \frac{x - x_{\rm max}}{\vbrac{x - x_{\rm max}}}
 =  \cleft{\renewcommand{\arraystretch}{1.6}
           \begin{array}{l l l}
            +1 \~, & ~~~~~~ &
            {\rm for}~x \le x_{\rm min}~{\rm or}~x \ge x_{\rm max} \~, \\
            -1 \~, &        &
            {\rm for}~x_{\rm min} \le x \le x_{\rm max} \~.
           \end{array}}
\label{eqn:Xi}
\eeq
 Thus we can get
\cheqnXa{A}
\beq
    \frac{1}{2} \bBig{1 + \Xi(x)}
 =  \cleft{\renewcommand{\arraystretch}{1.6}
           \begin{array}{l l l}
             1 \~, & ~~~~~~ &
            {\rm for}~x \le x_{\rm min}~{\rm or}~x \ge x_{\rm max} \~, \\
             0 \~, &        &
            {\rm for}~x_{\rm min} \le x \le x_{\rm max} \~,
           \end{array}}
\label{eqn:Xi_outside}
\eeq
 and
\cheqnXb{A}
\beq
    \frac{1}{2} \bBig{1 - \Xi(x)}
 =  \cleft{\renewcommand{\arraystretch}{1.6}
           \begin{array}{l l l}
             0 \~, & ~~~~~~ &
            {\rm for}~x \le x_{\rm min}~{\rm or}~x \ge x_{\rm max} \~, \\
             1 \~, &        &
            {\rm for}~x_{\rm min} \le x \le x_{\rm max} \~.
           \end{array}}
\label{eqn:Xi_inside}
\eeq
\cheqnX{A}
 Remind that,
 firstly,
 ``\verb+\+'' (backslash) must be used
 in order to let the including of the definition(s) given in this file
 into the other intrinsic commands correctly.
 Secondly,
 the flexible--kept parameter:
 \verb+AX+
 will be read directly from
 the users' initial simulation/data analysis setup
 set earlier on the website.

\subsection{Background spectrum}

 In this subsection,
 we give the codes for
 intrinsically defined
 simple artificial background spectrum
 given in Sec.~3.6
 in the \amidas\ package.

\subsubsection{Signal and background windows}

 Similar to the function $\Xi(x)$
 defined in Eqs.~(\ref{eqn:Xi}),
 (\ref{eqn:Xi_outside}) and (\ref{eqn:Xi_inside}),
 we define the signal and background windows and vacua
 for drawing the predicted WIMP--signal
 and artificial background spectra
 on the output plots together
 as following:
\code{Signal window
      \label{code:sgwindow_x}}
\verb#    sgwindow(x)                                          \ # \\
\verb# =  (  1                                                 \ # \\
\verb#     - ( (x - Qmin_gen) / abs(x - Qmin_gen) ) *          \ # \\
\verb#       ( (x - Qmax_gen) / abs(x - Qmax_gen) )  ) / 2.0     # \\
\tabb
\code{Signal vacuum
      \label{code:sgvacuum_x}}
\verb#    sgvacuum(x)                                          \ # \\
\verb# =  (  1                                                 \ # \\
\verb#     + ( (x - Qmin_gen) / abs(x - Qmin_gen) ) *          \ # \\
\verb#       ( (x - Qmax_gen) / abs(x - Qmax_gen) )  ) / 2.0     # \\
\tabb
\code{Background window
      \label{code:bgwindow_x}}
\verb#    bgwindow(x)                                                    \ # \\
\verb# =  (  1                                                           \ # \\
\verb#     - ( (x - Qmin_bgwindow) / abs(x - Qmin_bgwindow) ) *          \ # \\
\verb#       ( (x - Qmax_bgwindow) / abs(x - Qmax_bgwindow) )  ) / 2.0     # \\
\tabb
\code{Background vacuum
      \label{code:bgvacuum_x}}
\verb#    bgvacuum(x)                                                    \ # \\
\verb# =  (  1                                                           \ # \\
\verb#     + ( (x - Qmin_bgwindow) / abs(x - Qmin_bgwindow) ) *          \ # \\
\verb#       ( (x - Qmax_bgwindow) / abs(x - Qmax_bgwindow) )  ) / 2.0     # \\
\tabb

 According to these definitions,
 \amidas\ will use the given analytic expressions
 to draw the WIMP--signal and background spectra
 exactly in the energy range
 between the minimal and maximal cut--offs
 $Q_{\rm (min,max)}$ and $Q_{\rm bg, (min,max)}$,
 respectively.
 Note here that
 the minimal and maximal cut--offs:
 \verb+Qmin_gen+, \verb+Qmax_gen+ and
 \verb+Qmin_bgwindow+, \verb+Qmax_bgwindow+
 will be read directly from
 the users' initial simulation/data analysis setup
 set earlier on the website.
 Remind als here that
 ``\verb+\+'' (backslash) must be used
 in order to let the including of the definition(s) given in this file
 into the other intrinsic commands correctly.

\subsubsection{Constant background spectrum
               \boldmath$\abrac{dR / dQ}_{\rm bg, const}$}

 The simplest considerable artificial background spectrum
 is the constant one
 $\abrac{dR / dQ}_{\rm bg, const}$
 given in Eq.~(\ref{eqn:dRdQ_bg_const})
 \cite{DMDDbg-mchi}:
\code{Constant background spectrum
      \boldmath$\abrac{dR / dQ}_{\rm bg, const}(Q)$
      \label{code:dRdQ_bg_const}}
\verb# double dRdQ_bg_const(double QQ) # \\
\verb# {                               # \\
\verb#   return 1.0;                   # \\
\verb# }                               # \\
\tabb
 For drawing the output
 generating background spectrum,
 we have
\code{Constant background spectrum
      \boldmath$\abrac{dR / dQ}_{\rm bg, const}(Q)$
      (for Gnuplot)
      \label{code:dRdQ_bg_x_const}}
\verb#    dRdQ_bg_const(x)   \ # \\
\verb# =  1.0                  # \\
\tabb
 combined with the integral over $\abrac{dR / dQ}_{\rm bg, const}(Q)$
 for {\em normalizing} 
 the background spectrum properly
 according to the user's required background ratio
 to the WIMP--induced signals:
\code{Integral over the constant background spectrum
      \boldmath$\int \abrac{dR / dQ}_{\rm bg, const}(Q) \~ dQ$
      (for Gnuplot)
      \label{code:IntdRdQ_bg_x_const}}
\verb#    IntdRdQ_bg_const(x)   \ # \\
\verb# =  x                       # \\
\tabb
 Remind that
 ``\verb+\+'' (backslash) must be used
 in order to let the including of the definition(s) given in this file
 into the other intrinsic commands correctly.

\subsubsection{Exponential background spectrum
               \boldmath$\abrac{dR / dQ}_{\rm bg, ex}$}

 More realistically,
 we considered in our simulations
 the ``target--dependent'' exponential background spectrum
 $\abrac{dR / dQ}_{\rm bg, ex}$
 given in Eq.~(\ref{eqn:dRdQ_bg_ex})
 \cite{DMDDbg-mchi}:
\code{Target--dependent exponential background spectrum
      \boldmath$\abrac{dR / dQ}_{\rm bg, ex}(A, Q)$
      \label{code:dRdQ_bg_ex}}
\verb#  double Q_0_bg(int A)                # \\
\verb#  {                                   # \\
\verb#    return pow(A, 0.6);               # \\
\verb#  }                                   # \\
\verb#                                      # \\
\verb#  double dRdQ_bg_ex(int A, double QQ) # \\
\verb#  {                                   # \\
\verb#    return exp(-QQ / Q_0_bg(A));      # \\
\verb#  }                                   # \\
\tabb
 For drawing the output
 generating background spectrum,
 one needs
\code{Target--dependent exponential background spectrum
      \boldmath$\abrac{dR / dQ}_{\rm bg, ex}(Q)$
      (for Gnuplot)
      \label{code:dRdQ_bg_x_ex}}
\verb#    dRdQ_bg_ex(x)         \ # \\
\verb# =  exp(-x / AX ** 0.6)     # \\
\tabb
 and
\code{Integral over the Target--dependent exponential background spectrum
      \boldmath$\int \abrac{dR / dQ}_{\rm bg, ex}(Q) \~ dQ$
      (for Gnuplot)
      \label{code:IntdRdQ_bg_x_ex}}
\verb#    IntdRdQ_bg_ex(x)                    \ # \\
\verb# =- (AX ** 0.6) * exp(-x / AX ** 0.6)     # \\
\tabb
 since
\beq
    \int \aDd{R}{Q}_{\rm bg, ex} dQ
 =  \int e^{-Q / A^{0.6}} \~ dQ
 = - A^{0.6} \~ e^{-Q / A^{0.6}}
\~.
\eeq
 Remind that,
 firstly,
 ``\verb+\+'' (backslash) must be used
 in order to let the including of the definition(s) given in this file
 into the other intrinsic commands correctly.
 Secondly,
 the flexible--kept parameter:
 \verb+AX+
 will be read directly from
 the users' initial simulation/data analysis setup
 set earlier on the website.

\subsubsection{Gaussian--excess background spectrum
               \boldmath$\abrac{dR / dQ}_{\rm bg, Gau}$}

 For describing background events induced by
 e.g.~radioactivity of some nuclei,
 we introduce in our simulations
 the Gaussian--excess background spectrum
 $\abrac{dR / dQ}_{\rm bg, Gau}$
 given in Eq.~(\ref{eqn:dRdQ_bg_Gau})
 \cite{AMIDASbg-DSU2011}:
\code{Gaussian--excess background spectrum
      \boldmath$\abrac{dR / dQ}_{\rm bg, Gau}(Q_{\rm bg, peak}, \sigma_{Q, {\rm bg}}, Q)$
      \label{code:dRdQ_bg_Gau}}
\verb# double dRdQ_bg_Gau(double dRdQ_bg_mu, double dRdQ_bg_sigma_mu, double QQ) # \\
\verb# {                                                                         # \\
\verb#   return                                                                  # \\
\verb#     exp(-(QQ - dRdQ_bg_mu)    *                                           # \\
\verb#          (QQ - dRdQ_bg_mu)    /                                           # \\
\verb#          (2.0              *                                              # \\
\verb#           dRdQ_bg_sigma_mu *                                              # \\
\verb#           dRdQ_bg_sigma_mu  )  ) /                                        # \\
\verb#     (dRdQ_bg_sigma_mu * sqrt(2.0 * M_PI));                                # \\
\verb# }                                                                         # \\
\tabb
 For drawing the output
 generating background spectrum,
 one needs
\code{Gaussian--excess background spectrum
      \boldmath$\abrac{dR / dQ}_{\rm bg, Gau}(Q)$
      (for Gnuplot)
      \label{code:dRdQ_bg_x_Gau}}
\verb# M_PI = 3.141593                              # \\
\verb#                                              # \\
\verb# dRdQ_bg_mu       = 50.0                      # \\
\verb# dRdQ_bg_sigma_mu =  2.0                      # \\
\verb#                                              # \\
\verb#    dRdQ_bg_Gau(x)                          \ # \\
\verb# =  exp(-(x - dRdQ_bg_mu)     *             \ # \\
\verb#         (x - dRdQ_bg_mu)     /             \ # \\
\verb#         (2.0              *                \ # \\
\verb#          dRdQ_bg_sigma_mu *                \ # \\
\verb#          dRdQ_bg_sigma_mu  )  ) /          \ # \\
\verb#    (dRdQ_bg_sigma_mu * sqrt(2.0 * M_PI))     # \\
\tabb
 and
\code{Integral over the Gaussian--excess background spectrum
      \boldmath$\int \abrac{dR / dQ}_{\rm bg, Gau}(Q) \~ dQ$
      (for Gnuplot)
      \label{code:IntdRdQ_bg_x_Gau}}
\verb#    IntdRdQ_bg_Gau(x)                        \ # \\
\verb# =  (1.0 / 2.0) *                            \ # \\
\verb#    erf( (x - dRdQ_bg_mu)               /    \ # \\
\verb#         (sqrt(2.0) * dRdQ_bg_sigma_mu)  )     # \\
\tabb
 since
\beqn
     \int \aDd{R}{Q}_{\rm bg, Gau} dQ
 \=  \int
     \frac{1}{\sqrt{2 \pi} \~ \sigma_{Q, {\rm bg}}} \~
     e^{-\abrac{Q - Q_{\rm bg, peak}}^2 / 2 \sigma_{Q, {\rm bg}}^2} \~ dQ
     \non\\
 \=  \frac{1}{2} \~
     \erf\afrac{Q - Q_{\rm bg, peak}}{\sqrt{2} \~ \sigma_{Q, {\rm bg}}}
\~.
\eeqn
 Remind that,
 firstly,
 ``\verb+\+'' (backslash) must be used
 in order to let the including of the definition(s) given in this file
 into the other intrinsic commands correctly.
 Secondly,
 the flexible--kept parameters:
 \verb+dRdQ_bg_mu+ and \verb+dRdQ_bg_sigma_mu+
 will be read directly from
 the users' initial simulation/data analysis setup
 set earlier on the website.

%
%
%
\section{Intrinsically defined functions for Bayesian analyses}

 In this section,
 we give all codes for
 intrinsically defined
 fitting velocity distribution functions
 given in Sec.~5.1,
 distribution function for describing the statistical uncertainty
 given in Sec.~5.2
 as well as
 probability distribution functions
 for each fitting parameter
 given in Sec.~5.3
 in the \amidasii\ package.

\subsection{Fitting one--dimensional velocity distribution function}

 In this subsection,
 we give first the codes for
 intrinsically defined
 fitting velocity distribution functions
 given in Sec.~5.1
 in the \amidasii\ package.

\subsubsection{Simple Maxwellian velocity distribution
               \boldmath$f_{1, \Gau}(v)$}

 From the expression (\ref{eqn:f1v_Gau})
 for the simple Maxwellian velocity distribution function
 $f_{1, \Gau}(v)$,
 we can obtain that
\beq
    \int f_{1, \Gau}(v) \~ dv
 =  \erf\afrac{v}{v_0} - \frac{2}{\sqrt{\pi}} \afrac{v}{v_0} \~ e^{-v^2 / v_0^2}
\~.
\label{eqn:Int_f1v_Gau}
\eeq
 Then we can define
\code{Integral over the simple Maxwellian velocity distribution
      \boldmath$\int f_{1, \Gau}(v) \~ dv$
      \label{code:Int_f1v_Bayesian_fit_Gau}}
\verb# double Int_f1v_Bayesian_fit_Gau(double vv, double aa, double bb, double cc) # \\
\verb# {                                                                           # \\
\verb#   return                                                                    # \\
\verb#       erf(vv / aa)                                                          # \\
\verb#     - (2.0 / sqrt(M_PI)) *                                                  # \\
\verb#       (vv / aa)          *                                                  # \\
\verb#       exp(-(vv * vv) / (aa * aa) );                                         # \\
\verb# }                                                                           # \\
\tabb
 and the fitting simple Maxwellian velocity distribution,
 which satisfies the normalization condition:
\beq
    \int_0^{\vmax} f_{1, {\rm fit}}(v) \~ dv
 =  1
\~,
\label{eqn:f1v_fit_normalization}
\eeq
\code{Fitting simple Maxwellian velocity distribution
      \boldmath$f_{1, \Gau}(v)$
      \label{code:f1v_Bayesian_fit_Gau}}
\verb# double f1v_Bayesian_fit_Gau(double vv, double aa, double bb, double cc, double N_f) # \\
\verb# {                                                                                   # \\
\verb#   return                                                                            # \\
\verb#     (  (4.0 / sqrt(M_PI))                   *                                       # \\
\verb#        ( (vv * vv) / (aa * aa * aa) ) / v_U *                                       # \\
\verb#        exp(-(vv * vv) / (aa * aa) )          ) /                                    # \\
\verb#     (  Int_f1v_Bayesian_fit_Gau(v_max, aa, bb, cc)                                  # \\
\verb#      - Int_f1v_Bayesian_fit_Gau(0.0,   aa, bb, cc)  );                              # \\
\verb# }                                                                                   # \\
\tabb

 On the other hand,
 for drawing output plots,
 we have
\code{Fitting simple Maxwellian velocity distribution
      \boldmath$f_{1, \Gau}(v)$
      (for Gnuplot)
      \label{code:f1v_Bayesian_fit_x_Gau}}
\verb# M_PI = 3.141593                                                       # \\
\verb#                                                                       # \\
\verb#    f1v_Bayesian_fit_Gau(x)                                          \ # \\
\verb# =  (4.0 / sqrt(M_PI))                                             * \ # \\
\verb#    ((x * x) / (a_Bayesian_fit * a_Bayesian_fit * a_Bayesian_fit)) * \ # \\
\verb#    exp(-(x * x) / (a_Bayesian_fit * a_Bayesian_fit))              / \ # \\
\verb#    N_f                                                                # \\
\tabb
 Here the normalization constant \verb+N_f+
 will be estimated by using the function \\
 \verb+Int_f1v_Bayesian_fit_Gau+
 defined in Code A\ref{code:Int_f1v_Bayesian_fit_Gau} as
\beqn
     \verb+N_f+
 \=  \verb+Int_f1v_Bayesian_fit_Gau+(\vmax, \verb+a_Bayesian_fit+, 0, 0)
     \non\\
 \conti ~~~~~~~~~~~~ 
   - \verb+Int_f1v_Bayesian_fit_Gau+(0,     \verb+a_Bayesian_fit+, 0, 0)
\~,
\eeqn
 and the value of \verb+a_Bayesian_fit+
 will be given from the fitting result.
 Remind that
 ``\verb+\+'' (backslash) must be used
 in order to let the including of the definition(s) given in this file
 into the other intrinsic commands correctly.

\subsubsection{Modified Maxwellian velocity distribution
               \boldmath$f_{1, \Gau, k}(v)$}

 Since the power index $k$
 in the modified Maxwellian velocity distribution function $f_{1, \Gau, k}(v)$
 is one of our fitting parameter
 and thus unfixed,
 an analytic form for the integral over $f_{1, \Gau, k}(v)$
 is in general ``unknown''.
 For this need,
 in the \amidasii\ code
 one can simply define that
\code{Integral over the modified Maxwellian velocity distribution
      \boldmath$\int f_{1, \Gau, k}(v) \~ dv$
      \label{code:Int_f1v_Bayesian_fit_Gau_k}}
\verb# double Int_f1v_Bayesian_fit_Gau_k(double vv, double aa, double bb, double cc) # \\
\verb# {                                                                             # \\
\verb#   return -1.0;                                                                # \\
\verb# }                                                                             # \\
\tabb
 and
\newpage
\code{Fitting modified Maxwellian velocity distribution
      \boldmath$f_{1, \Gau, k}(v)$
      \label{code:f1v_Bayesian_fit_Gau_k}}
\verb# double f1v_Bayesian_fit_Gau_k(double vv, double aa, double bb, double cc, double N_f) # \\
\verb# {                                                                                     # \\
\verb#   return                                                                              # \\
\verb#     (  (vv * vv)                                       *                              # \\
\verb#        pow(  exp(-(vv    * vv   ) / (cc * aa * aa) )                                  # \\
\verb#            - exp(-(v_max * v_max) / (cc * aa * aa) ),                                 # \\
\verb#            cc                                        )  ) /                           # \\
\verb#     N_f;                                                                              # \\
\verb# }                                                                                     # \\
\tabb
 Here the normalization constant \verb+N_f+
 will be estimated {\em every time}
 by numerical integral over the function
 \verb+f1v_Bayesian_fit_Gau_k+ itself as
 as
\beqn
     \verb+N_f+
 \=  \left[\int_0^{\vmax}
           \verb+f1v_Bayesian_fit_Gau_k+(v, \verb+aa+, \verb+bb+, \verb+cc+, 1) \~ dv
           \right]^{-1}
\~,
     \non\\
\eeqn
 for different scanning points $(\verb+aa+, \verb+bb+, \verb+cc+)$.

 Similarly,
 for drawing the output plots,
 we have
\code{Fitting modified Maxwellian velocity distribution
      \boldmath$f_{1, \Gau, k}(v)$
      (for Gnuplot)
      \label{code:f1v_Bayesian_fit_x_Gau_k}}
\verb#    f1v_Bayesian_fit_Gau_k(x)                            \ # \\
\verb# =  (x * x)                                            * \ # \\
\verb#    (  exp(-(x     * x    )                   /          \ # \\
\verb#            (c_Bayesian_fit *                            \ # \\
\verb#             a_Bayesian_fit * a_Bayesian_fit)  )         \ # \\
\verb#     - exp(-(v_max * v_max)                   /          \ # \\
\verb#            (c_Bayesian_fit *                            \ # \\
\verb#             a_Bayesian_fit * a_Bayesian_fit)  )  ) **   \ # \\
\verb#    c_Bayesian_fit                                     / \ # \\
\verb#    N_f                                                    # \\
\tabb
 Here the normalization constant \verb+N_f+
 will be estimated by numerical integral over the function
 \verb+f1v_Bayesian_fit_Gau_k+
 defined in Code A\ref{code:f1v_Bayesian_fit_Gau_k} as
\beqn
\hspace{-0.6cm}
     \verb+N_f+
 \=  \left[\int_0^{\vmax}
           \verb+f1v_Bayesian_fit_Gau_k+(x, \verb+a_Bayesian_fit+, 0, \verb+c_Bayesian_fit+, 1) \~ dx
           \right]^{-1}
\~,
     \non\\
\eeqn
 and the values of \verb+a_Bayesian_fit+ and \verb+c_Bayesian_fit+
 will be given from the fitting result.
 Remind that
 ``\verb+\+'' (backslash) must be used
 in order to let the including of the definition(s) given in this file
 into the other intrinsic commands correctly.

\subsubsection{One--parameter shifted Maxwellian velocity distribution
               \boldmath$f_{1, \sh, v_0}(v)$}

 From the expression (\ref{eqn:f1v_sh})
 for the shifted Maxwellian velocity distribution function
 $f_{1, \sh}(v)$,
 we can obtain that
\beqn
     \int f_{1, \sh}(v) \~ dv
 \=  \frac{1}{2}
     \bbrac{  \erf\afrac{v + \ve}{v_0}
            + \erf\afrac{v - \ve}{v_0} }
     \non\\
 \conti ~~~~~~~~~~~~ 
   + \frac{1}{2 \~ \sqrt{\pi}} \afrac{v_0}{\ve}
     \bBig{  e^{-(v + \ve)^2 / v_0^2}
           - e^{-(v - \ve)^2 / v_0^2}  }
\~.
\label{eqn:Int_f1v_sh}
\eeqn
 Then,
 by using Eq.~(\ref{eqn:v_e_ave}),
 we can define
\code{Integral over the one--parameter shifted Maxwellian velocity distribution
      \boldmath$\int f_{1, \sh, v_0}(v) \~ dv$
      \label{code:Int_f1v_Bayesian_fit_sh_v0}}
\verb# double Int_f1v_Bayesian_fit_sh_v0(double vv, double aa, double bb, double cc) # \\
\verb# {                                                                             # \\
\verb#   return                                                                      # \\
\verb#       (1.0 / 2.0) *                                                           # \\
\verb#       (  erf((vv + (aa * 1.05)) / aa)                                         # \\
\verb#        + erf((vv - (aa * 1.05)) / aa)  )                                      # \\
\verb#     + (1.0 / 2.0 / sqrt(M_PI)) *                                              # \\
\verb#       (aa / (aa * 1.05))       *                                              # \\
\verb#       (  exp(-(vv + (aa * 1.05)) * (vv + (aa * 1.05)) / (aa * aa) )           # \\
\verb#        - exp(-(vv - (aa * 1.05)) * (vv - (aa * 1.05)) / (aa * aa) )  );       # \\
\verb# }                                                                             # \\
\tabb
 and the fitting one--parameter shifted Maxwellian velocity distribution,
 which satisfies the normalization condition
 (\ref{eqn:f1v_fit_normalization}),
\code{Fitting one--parameter shifted Maxwellian velocity distribution
      \boldmath$f_{1, \sh, v_0}(v)$
      \label{code:f1v_Bayesian_fit_sh_v0}}
\verb# double f1v_Bayesian_fit_sh_v0(double vv, double aa, double bb, double cc, double N_f) # \\
\verb# {                                                                                     # \\
\verb#   return                                                                              # \\
\verb#     (  (1.0 / sqrt(M_PI))                                               *             # \\
\verb#        (vv / aa / (aa * 1.05)) / v_U                                    *             # \\
\verb#        (  exp(-(vv - (aa * 1.05)) * (vv - (aa * 1.05)) / (aa * aa) )                  # \\
\verb#         - exp(-(vv + (aa * 1.05)) * (vv + (aa * 1.05)) / (aa * aa) )  )  ) /          # \\
\verb#     (  Int_f1v_Bayesian_fit_sh_v0(v_max, aa, bb, cc)                                  # \\
\verb#      - Int_f1v_Bayesian_fit_sh_v0(0.0,   aa, bb, cc)  );                              # \\
\verb# }                                                                                     # \\
\tabb

 On the other hand,
 for drawing output plots,
 we have
\code{Fitting one--parameter shifted Maxwellian velocity distribution
      \boldmath$f_{1, \sh, v_0}(v)$
      (for Gnuplot)
      \label{code:f1v_Bayesian_fit_x_sh_v0}}
\verb# M_PI = 3.141593                                       # \\
\verb#                                                       # \\
\verb#    f1v_Bayesian_fit_sh_v0(x)                        \ # \\
\verb# =  (1.0 / sqrt(M_PI))                             * \ # \\
\verb#    (x / a_Bayesian_fit / (a_Bayesian_fit * 1.05)) * \ # \\
\verb#    (  exp(-(x - (a_Bayesian_fit * 1.05)) *          \ # \\
\verb#            (x - (a_Bayesian_fit * 1.05)) /          \ # \\
\verb#            (a_Bayesian_fit * a_Bayesian_fit) )      \ # \\
\verb#     - exp(-(x + (a_Bayesian_fit * 1.05)) *          \ # \\
\verb#            (x + (a_Bayesian_fit * 1.05)) /          \ # \\
\verb#            (a_Bayesian_fit * a_Bayesian_fit) )  ) / \ # \\
\verb#    N_f                                                # \\
\tabb
 Here the normalization constant \verb+N_f+
 will be estimated by using the function \\
 \verb+Int_f1v_Bayesian_fit_sh_v0+
 defined in Code A\ref{code:Int_f1v_Bayesian_fit_sh_v0} as
\beqn
     \verb+N_f+
 \=  \verb+Int_f1v_Bayesian_fit_sh_v0+(\vmax, \verb+a_Bayesian_fit+, 0, 0)
     \non\\
 \conti ~~~~~~~~~~~~ 
   - \verb+Int_f1v_Bayesian_fit_sh_v0+(0,     \verb+a_Bayesian_fit+, 0, 0)
\~,
\eeqn
 and the value of \verb+a_Bayesian_fit+
 will be given from the fitting result.
 Remind that
 ``\verb+\+'' (backslash) must be used
 in order to let the including of the definition(s) given in this file
 into the other intrinsic commands correctly.

\subsubsection{Shifted Maxwellian velocity distribution
               \boldmath$f_{1, \sh}(v)$}

 According to Eq.~(\ref{eqn:Int_f1v_sh}),
 we can define
\code{Integral over the shifted Maxwellian velocity distribution
      \boldmath$\int f_{1, \sh}(v) \~ dv$
      \label{code:Int_f1v_Bayesian_fit_sh}}
\verb# double Int_f1v_Bayesian_fit_sh(double vv, double aa, double bb, double cc) # \\
\verb# {                                                                          # \\
\verb#   return                                                                   # \\
\verb#       (1.0 / 2.0) *                                                        # \\
\verb#       (  erf((vv + bb) / aa)                                               # \\
\verb#        + erf((vv - bb) / aa)  )                                            # \\
\verb#     + (1.0 / 2.0 / sqrt(M_PI)) *                                           # \\
\verb#       (aa / bb)                *                                           # \\
\verb#       (  exp(-(vv + bb) * (vv + bb) / (aa * aa) )                          # \\
\verb#        - exp(-(vv - bb) * (vv - bb) / (aa * aa) )  );                      # \\
\verb# }                                                                          # \\
\tabb
 and the fitting shifted Maxwellian velocity distribution,
 which satisfies the normalization condition
 (\ref{eqn:f1v_fit_normalization}),
\code{Fitting shifted Maxwellian velocity distribution
      \boldmath$f_{1, \sh}(v)$
      \label{code:f1v_Bayesian_fit_sh}}
\verb# double f1v_Bayesian_fit_sh(double vv, double aa, double bb, double cc, double N_f) # \\
\verb# {                                                                                  # \\
\verb#   return                                                                           # \\
\verb#     (  (1.0 / sqrt(M_PI))                             *                            # \\
\verb#        (vv / aa / bb) / v_U                           *                            # \\
\verb#        (  exp(-(vv - bb) * (vv - bb) / (aa * aa) )                                 # \\
\verb#         - exp(-(vv + bb) * (vv + bb) / (aa * aa) )  )  ) /                         # \\
\verb#     (  Int_f1v_Bayesian_fit_sh(v_max, aa, bb, cc)                                  # \\
\verb#      - Int_f1v_Bayesian_fit_sh(0.0,   aa, bb, cc)  );                              # \\
\verb# }                                                                                  # \\
\tabb

 On the other hand,
 for drawing output plots,
 we have
\code{Fitting shifted Maxwellian velocity distribution
      \boldmath$f_{1, \sh}(v)$
      (for Gnuplot)
      \label{code:f1v_Bayesian_fit_x_sh}}
\verb# M_PI = 3.141593                                       # \\
\verb#                                                       # \\
\verb#    f1v_Bayesian_fit_sh(x)                           \ # \\
\verb# =  (1.0 / sqrt(M_PI))                             * \ # \\
\verb#    (x / a_Bayesian_fit / b_Bayesian_fit)          * \ # \\
\verb#    (  exp(-(x - b_Bayesian_fit) *                   \ # \\
\verb#            (x - b_Bayesian_fit) /                   \ # \\
\verb#            (a_Bayesian_fit * a_Bayesian_fit) )      \ # \\
\verb#     - exp(-(x + b_Bayesian_fit) *                   \ # \\
\verb#            (x + b_Bayesian_fit) /                   \ # \\
\verb#            (a_Bayesian_fit * a_Bayesian_fit) )  ) / \ # \\
\verb#    N_f                                                # \\
\tabb
 Here the normalization constant \verb+N_f+
 will be estimated by using the function \\
 \verb+Int_f1v_Bayesian_fit_sh+
 defined in Code A\ref{code:Int_f1v_Bayesian_fit_sh} as
\beqn
     \verb+N_f+
 \=  \verb+Int_f1v_Bayesian_fit_sh+(\vmax, \verb+a_Bayesian_fit+, \verb+b_Bayesian_fit+, 0)
     \non\\
 \conti ~~~~
   - \verb+Int_f1v_Bayesian_fit_sh+(0,     \verb+a_Bayesian_fit+, \verb+b_Bayesian_fit+, 0)
\~,
\eeqn
 and the values of \verb+a_Bayesian_fit+ and \verb+b_Bayesian_fit+
 will be given from the fitting result.
 Remind that
 ``\verb+\+'' (backslash) must be used
 in order to let the including of the definition(s) given in this file
 into the other intrinsic commands correctly.

\subsubsection{Variated shifted Maxwellian velocity distribution
               \boldmath$f_{1, \sh, \Delta v}(v)$}

 According to Eq.~(\ref{eqn:Int_f1v_sh})
 and the relation given in Eq.~(\ref{eqn:Delta_v}),
 we can define
\code{Integral over the variated shifted Maxwellian velocity distribution
      \boldmath$\int f_{1, \sh, \Delta v}(v) \~ dv$
      \label{code:Int_f1v_Bayesian_fit_sh_Dv}}
\verb# double Int_f1v_Bayesian_fit_sh_Dv(double vv, double aa, double bb, double cc) # \\
\verb# {                                                                             # \\
\verb#   return                                                                      # \\
\verb#       (1.0 / 2.0) *                                                           # \\
\verb#       (  erf((vv + aa + bb) / aa)                                             # \\
\verb#        + erf((vv - aa - bb) / aa)  )                                          # \\
\verb#     + (1.0 / 2.0 / sqrt(M_PI)) *                                              # \\
\verb#       (aa / (aa + bb))         *                                              # \\
\verb#       (  exp(-(vv + bb) * (vv + aa + bb) / (aa * aa) )                        # \\
\verb#        - exp(-(vv - bb) * (vv - aa - bb) / (aa * aa) )  );                    # \\
\verb# }                                                                             # \\
\tabb
 and the fitting variated shifted Maxwellian velocity distribution,
 which satisfies the normalization condition
 (\ref{eqn:f1v_fit_normalization}),
\code{Fitting variated shifted Maxwellian velocity distribution
      \boldmath$f_{1, \sh, \Delta v}(v)$
      \label{code:f1v_Bayesian_fit_sh_Dv}}
\verb# double f1v_Bayesian_fit_sh_Dv(double vv, double aa, double bb, double cc, double N_f) # \\
\verb# {                                                                                     # \\
\verb#   return                                                                              # \\
\verb#     (  (1.0 / sqrt(M_PI))                                       *                     # \\
\verb#        (vv / aa / (aa + bb)) / v_U                              *                     # \\
\verb#        (  exp(-(vv - aa - bb) * (vv - aa - bb) / (aa * aa) )                          # \\
\verb#         - exp(-(vv + aa + bb) * (vv + aa + bb) / (aa * aa) )  )  ) /                  # \\
\verb#     (  Int_f1v_Bayesian_fit_sh_Dv(v_max, aa, bb, cc)                                  # \\
\verb#      - Int_f1v_Bayesian_fit_sh_Dv(0.0,   aa, bb, cc)  );                              # \\
\verb# }                                                                                     # \\
\tabb

 On the other hand,
 for drawing output plots,
 we have
\newpage
\code{Fitting variated shifted Maxwellian velocity distribution
      \boldmath$f_{1, \sh, \Delta v}(v)$
      (for Gnuplot)
      \label{code:f1v_Bayesian_fit_x_sh_Dv}}
\verb# M_PI = 3.141593                                                 # \\
\verb#                                                                 # \\
\verb#    f1v_Bayesian_fit_sh_Dv(x)                                  \ # \\
\verb# =  (1.0 / sqrt(M_PI))                                       * \ # \\
\verb#    (x / a_Bayesian_fit / (a_Bayesian_fit + b_Bayesian_fit)) * \ # \\
\verb#    (  exp(-(x - a_Bayesian_fit - b_Bayesian_fit) *            \ # \\
\verb#            (x - a_Bayesian_fit - b_Bayesian_fit) /            \ # \\
\verb#            (a_Bayesian_fit * a_Bayesian_fit)      )           \ # \\
\verb#     - exp(-(x + a_Bayesian_fit + b_Bayesian_fit) *            \ # \\
\verb#            (x + a_Bayesian_fit + b_Bayesian_fit) /            \ # \\
\verb#            (a_Bayesian_fit * a_Bayesian_fit) )    )         / \ # \\
\verb#    N_f                                                          # \\
\tabb
 Here the normalization constant \verb+N_f+
 will be estimated by using the function \\
 \verb+Int_f1v_Bayesian_fit_sh_Dv+
 defined in Code A\ref{code:Int_f1v_Bayesian_fit_sh_Dv} as
\beqn
     \verb+N_f+
 \=  \verb+Int_f1v_Bayesian_fit_sh_Dv+(\vmax, \verb+a_Bayesian_fit+, \verb+b_Bayesian_fit+, 0)
     \non\\
 \conti ~
   - \verb+Int_f1v_Bayesian_fit_sh_Dv+(0,     \verb+a_Bayesian_fit+, \verb+b_Bayesian_fit+, 0)
\~,
\eeqn
 and the values of \verb+a_Bayesian_fit+ and \verb+b_Bayesian_fit+
 will be given from the fitting result.
 Remind that
 ``\verb+\+'' (backslash) must be used
 in order to let the including of the definition(s) given in this file
 into the other intrinsic commands correctly.

\subsection{Distribution function for describing the statistical uncertainty}

 In this subsection,
 we give the codes for intrinsically defined
 distribution functions for describing the statistical uncertainty
 given in Sec.~5.2 in the \amidasii\ package,
 which will in turn be used in the likelihood function
 in Bayesian analyses.

\subsubsection{Poisson statistical--uncertainty distribution
               \boldmath${\rm Poi}(x_i, y_i;
                                   y_{i, {\rm th}})$}

 For describing,
 e.g.~the predicted signal/background event rates/numbers,
 one needs usually the Poisson statistical--uncertainty distribution
 ${\rm Poi}(x_i, y_i; a_j, j = 1,~2,~\cdots,~N_{\rm Bayesian})$
 given in Eq.~(\ref{eqn:Bayesian_DF_Poi}).
 For this use,
 we define
\newpage
\code{Poisson statistical--uncertainty distribution
      \boldmath${\rm Poi}(x_i, y_i; y_{i, {\rm th}})$
      \label{code:Bayesian_DF_Poi}}
\verb# double Bayesian_DF_Poi(double y_th, double y_rec)   # \\
\verb# {                                                   # \\
\verb#   if (y_rec == 0.0)                                 # \\
\verb#   {                                                 # \\
\verb#     return                                          # \\
\verb#       exp(-y_th);                                   # \\
\verb#   }                                                 # \\
\verb#                                                     # \\
\verb#   else                                              # \\
\verb#   if (y_rec >  0.0 &&                               # \\
\verb#       y_rec <= 1.0   )                              # \\
\verb#   {                                                 # \\
\verb#     return                                          # \\
\verb#       pow(y_th, y_rec) / exp(lgamma(y_rec + 1.0)) * # \\
\verb#       exp(-y_th);                                   # \\
\verb#   }                                                 # \\
\verb#                                                     # \\
\verb#   else                                              # \\
\verb#   if (y_rec >  1.0)                                 # \\
\verb#   {                                                 # \\
\verb#     return                                          # \\
\verb#       (y_th / y_rec) *                              # \\
\verb#       Bayesian_DF_Poi(y_th, y_rec - 1.0);           # \\
\verb#   }                                                 # \\
\verb# }                                                   # \\
\tabb
 Here we have used
 the expression for the Poisson distribution
 with $\mu$ as the expectation value:
\beqn
     {\rm Poi}(x; \mu)
 \=  \frac{\mu^x \~ e^{-\mu}}{x!}
\~,
     ~~~~~~~~~~~~~~~~~~~~~~~~~~~~\~ 
     {\rm for}~x \ge 0
\~,
     \non\\
 \=  \cleft{\renewcommand{\arraystretch}{0.6}
            \begin{array}{l l l}
             & & \\
             e^{-\mu}                                   \~, & ~~~~~~ & {\rm for}~x = 0       \~, \\
             & & \\
             \D \frac{\mu^x \~ e^{-\mu}}{\Gamma(x + 1)} \~, &        & {\rm for}~0 < x \le 1 \~, \\
             & & \\
             \D \afrac{\mu}{x} {\rm Poi}(x - 1; \mu)    \~, &        & {\rm for}~1 < x       \~, \\
             & & \\
            \end{array}}
\label{eqn:x_factorial}
\eeqn
 where the gamma function $\Gamma(x)$ is defined by
\beq
         \Gamma(x)
 \equiv  \int_0^\infty t^{x - 1} e^{-t} dt
 =       \frac{\Gamma(x + 1)}{x}
\~,
\label{eqn:Gamma_x}
\eeq
 and \verb+lgamma(x)+ is the intrinsic C mathematical function
 defined as
\beq
         \verb+lgamma+(x)
 \equiv  \ln \Gamma(x)
\~.
\label{eqn:lgamma}
\eeq
\subsubsection{(Double--)Gaussian statistical--uncertainty distribution
               \boldmath${\rm Gau}(x_i, y_i, y_{i, {\rm (lo, hi)}};
                                   y_{i, {\rm th}})$}

 For the most commonly needed Gaussian statistical--uncertainty distribution,
 we define in the \amidasii\ package
 an {\em asymmetric} form:
\code{(Double--)Gaussian statistical--uncertainty distribution
      \boldmath${\rm Gau}(x_i, y_i, y_{i, {\rm (lo, hi)}};
                          y_{i, {\rm th}})$
      \label{code:Bayesian_DF_Gau}}
\verb# double Bayesian_DF_Gau(double y_th, double y_rec, double y_lo, double y_hi) # \\
\verb# {                                                                           # \\
\verb#   if (y_rec >= y_th)                                                        # \\
\verb#   {                                                                         # \\
\verb#     return                                                                  # \\
\verb#       exp(-(y_rec - y_th) * (y_rec - y_th)         /                        # \\
\verb#            (2.0 * (y_rec - y_lo) * (y_rec - y_lo))  ) /                     # \\
\verb#       (sqrt(2.0 * M_PI) * (y_rec - y_lo));                                  # \\
\verb#   }                                                                         # \\
\verb#                                                                             # \\
\verb#   else                                                                      # \\
\verb#   if (y_rec <  y_th)                                                        # \\
\verb#   {                                                                         # \\
\verb#     return                                                                  # \\
\verb#       exp(-(y_rec - y_th) * (y_rec - y_th)         /                        # \\
\verb#            (2.0 * (y_hi - y_rec) * (y_hi - y_rec))  ) /                     # \\
\verb#       (sqrt(2.0 * M_PI) * (y_hi - y_rec));                                  # \\
\verb#   }                                                                         # \\
\verb# }                                                                           # \\
\tabb
 Remind that,
 for the case that
 the analyzed/fitted data point is larger (smaller)
 than the theoretical value
 estimated from the fitting (WIMP velocity distribution) function,
 we take the 1$\sigma$ lower (upper) (statistical) uncertainty
 as the uncertainty $\sigma(y_i)$
 in Eq.~(\ref{eqn:Bayesian_DF_Gau}).

\subsection{Distribution function of each fitting parameter}

 In the \amidasii\ package,
 we offer so far three probability distribution functions
 for each fitting parameter
 for describing the prior knowledge about it.

\subsubsection{Flat probability distribution function
               \boldmath${\rm p}_{i, {\rm flat}}(a_i)$}

 For the case that
 a prior knowledge about one of the fitting parameters
 is ``unknown'',
 one can use the flat probability distribution function
 ${\rm p}_{i, {\rm flat}}(a_i)$
 given in Eq.~(\ref{eqn:Bayesian_DF_a_flat}):
\code{Flat probability distribution function
      \boldmath${\rm p}_{i, {\rm flat}}(a_i)$
      \label{code:Bayesian_DF_a_flat}}
\verb# double Bayesian_DF_a_flat(double aa) # \\
\verb# {                                    # \\
\verb#   return 1.0;                        # \\
\verb# }                                    # \\
\tabb
\subsubsection{Poisson probability distribution function
               \boldmath${\rm p}_{i, {\rm Poi}}(a_i; \mu_{a, i})$}

 For the case that
 one of the fitting parameters
 is expected to be with an uncertainty
 obeying the ``Poisson'' statistics,
 one can use the Poisson probability distribution function
 ${\rm p}_{i, {\rm Poi}}(a_i; \mu_{a, i})$
 given in Eq.~(\ref{eqn:Bayesian_DF_a_Poi}):
\newpage
\code{Poisson probability distribution function
      \boldmath${\rm p}_{i, {\rm Poi}}(a_i; \mu_{a, i})$
      \label{code:Bayesian_DF_a_Poi}}
\verb# double Bayesian_DF_a_Poi(double aa)                     # \\
\verb# {                                                       # \\
\verb#   if (aa == 0.0)                                        # \\
\verb#   {                                                     # \\
\verb#     return                                              # \\
\verb#       exp(-a_ave_Bayesian);                             # \\
\verb#   }                                                     # \\
\verb#                                                         # \\
\verb#   else                                                  # \\
\verb#   if (aa >  0.0 &&                                      # \\
\verb#       aa <= 1.0   )                                     # \\
\verb#   {                                                     # \\
\verb#     return                                              # \\
\verb#       pow(a_ave_Bayesian, aa) / exp(lgamma(aa + 1.0)) * # \\
\verb#       exp(-a_ave_Bayesian);                             # \\
\verb#   }                                                     # \\
\verb#                                                         # \\
\verb#   else                                                  # \\
\verb#   if (aa >  1.0)                                        # \\
\verb#   {                                                     # \\
\verb#     return                                              # \\
\verb#       (a_ave_Bayesian / aa) *                           # \\
\verb#       Bayesian_DF_a_Poi(a_ave_Bayesian, aa - 1.0);      # \\
\verb#   }                                                     # \\
\verb# }                                                       # \\
\tabb
\subsubsection{Gaussian probability distribution function
               \boldmath${\rm p}_{i, {\rm Gau}}(a_i; \mu_{a, i}, \sigma_{a, i})$}

 For the most common case that
 one of the fitting parameters
 is expected to be with an uncertainty
 obeying the ``Gaussian/normal'' distribution,
 one can use the Gaussian probability distribution function
 ${\rm p}_{i, {\rm Gau}}(a_i; \mu_{a, i}, \sigma_{a, i})$
 given in Eq.~(\ref{eqn:Bayesian_DF_a_Gau}):
\code{Gaussian probability distribution function
      \boldmath${\rm p}_{i, {\rm Gau}}(a_i; \mu_{a, i}, \sigma_{a, i})$
      \label{code:Bayesian_DF_a_Gau}}
\verb# double Bayesian_DF_a_Gau(double aa)            # \\
\verb# {                                              # \\
\verb#   return                                       # \\
\verb#     exp(-(aa - a_ave_Bayesian)    *            # \\
\verb#          (aa - a_ave_Bayesian)    /            # \\
\verb#          (2.0                  *               # \\
\verb#           sigma_a_ave_Bayesian *               # \\
\verb#           sigma_a_ave_Bayesian  )  ) /         # \\
\verb#     (sigma_a_ave_Bayesian * sqrt(2.0 * M_PI)); # \\
\verb# }                                              # \\
\tabb
\end{document}